\documentclass[12pt,a4paper]{report}
\usepackage[titletoc]{appendix} 
\usepackage{enumerate}
\usepackage{epsfig}
\usepackage{subfigure}
\usepackage{algorithm} 
\usepackage{algorithmic}
\usepackage{amsthm,amssymb} 
\usepackage{graphicx}
\usepackage{enumerate}
\usepackage{hyperref} 
\usepackage{url}
\usepackage{multirow}
\usepackage{fancyhdr}
\usepackage{rotating}
\usepackage{textcomp}
\usepackage{graphicx, graphics,epstopdf}
\usepackage[printonlyused]{acronym}
\usepackage{listings}
\usepackage{color}
\usepackage{balance}
\usepackage[utf8]{inputenc}
\usepackage[T1]{fontenc}
\usepackage{microtype}
\usepackage{cite}
\usepackage{url}
\usepackage{tabularx}
\usepackage{booktabs}
\usepackage{multirow}
\usepackage{tikz}
\usepackage{gensymb}
\newcommand{\fakeparagraph}[1]{\smallskip\noindent\textbf{#1.}}

\newcommand*\circled[1]{\tikz[baseline=(char.base)]{
            \node[shape=circle,draw,inner sep=0.5pt, minimum size=0.5pt] (char) {#1};}}

\usepackage[bottom=1.0in,top=1.2in, headheight=3mm,  headsep=12mm,bindingoffset=6mm]{geometry}

\usepackage{longtable}
\usepackage[center]{caption}
\usepackage[singlespacing]{setspace}

\newcommand{\pic}[2]{\setlength{\epsfysize}{#1} \epsffile{#2}}

\begin{document}

\pagenumbering{roman}

\thispagestyle{empty}
\begin{center}
\vspace{1cm}
{\Huge \bf Programming Internet of Things, Service, and
People (IoTSP) Applications}\vspace{.05in}\\
\vspace{2.0cm}
{\large \bf Submitted by}\\
\vspace{0.2cm}
{\large \bf Saurabh Chauhan} \\
\vspace{0.2cm}
{\bf 1421015} \\
\vspace{2cm}
\textit{In the partial fulfillment of the requirements
for the degree of \\ Master of Technology \\ in \\ 
Computer Science and Engineering (CSE) \\ to}
\\
\vspace{15mm}
\begin{center}
\pic{70pt}{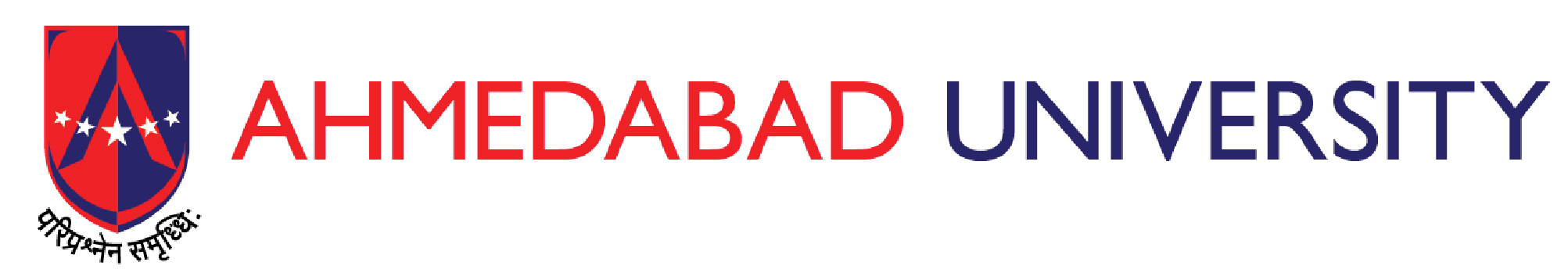}\\ ~\\
{\bf School of Engineering and Applied Science, Ahmedabad University \\Ahmedabad, India}\\ ~\\ 
{\bf June 2016}
\end{center}
\end{center}

\newpage 
\addcontentsline{toc}{chapter}{Acknowledgements}
\begin{center}
{\Large \bf Acknowledgments}
\end{center}
\vspace{10pt}
 On the very edge of my master's at Ahmedabad University, it is my radiant sentiment to express deepest sense of gratitude to Dr. Pankesh Patel and Prof. Sanjay Chaudhary for their careful and precious guidance which are extremely valuable for my study as well as for development. Without their active guidance, help, cooperation, encouragement, and constant support, I could not made headway in master's thesis and research career also. \\
\hspace*{0.2 in} I express my deepest thanks to Dr. Pankesh Patel and Prof. Sanjay Chaudhary for providing me the global exposure at ABB Corporate Research, India. I would like to express my deepest gratitude and special thanks to Dr. Pankesh Patel and Dr. Ashish Sureka who in spite of industrious, took time out to hear, and keep me on correct path. \\
\hspace*{0.2 in} I extended my gratitude to ABB Corporate Research, India for giving me edge in the era of research. I will strive to use gained skills and knowledge during internship in the best possible way, and I will continue to work on improvement, in order to attain desired career goals and place myself in the research community. \\
\hspace*{0.2 in} The research issues addressed and development process narrated in PhD thesis on \textit{Enabling High-Level Application Development for the Internet of Things} submitted to  Universit\'{e} Pierre et Marie Curie- Paris VI by Dr. Pankesh Patel  has provided me the motivation as well as the direction to initiate my research work.

\vspace{1.5cm}
\hfill{Saurabh Chauhan}

\newpage
\addcontentsline{toc}{chapter}{Declaration}

\begin{center}
{\Large \bf Declaration}\\
\end{center}
\vspace{10pt}

\textit{I hereby declare that this submission is my own work and that, to the best of my knowledge and belief, contains no material previously published or written by another person nor material which has been accepted for the award of any other degree or diploma of the university or other institute of higher learning, except where due acknowledgment has been made in the text.}

\vspace{2cm}
\hfill{{Signature of Student}}\\
\vspace{0.2cm}
\hfill{Saurabh Chauhan}\\
\vspace{0.2cm}
\hfill{1421015}

\newpage
\addcontentsline{toc}{chapter}{Certificate}
\begin{center}
{\Large \bf Certificate}\\
\end{center}
\vspace{10pt}

\noindent
\textit{ 
This is to certify that the thesis entitled \textbf{Programming Internet of Things, Service, and
People (IoTSP) Applications} submitted by Mr. Saurabh Chauhan (1421015) to the School of Engineering and Applied Science (SEAS), Ahmedabad University towards partial fulfillment of the requirements for the award of the Degree of Master of Technology in Computer Science Engineering, is a bonafide record of the work carried out by him under our supervision and guidance.}

\vspace{1.5cm}
\begin{center}
\hspace*{2.0in}Thesis Supervisor \hspace*{1in} Thesis Co-Advisor 
\newline
\hspace*{1.7in}Prof. Sanjay Chaudhary \hspace*{0.7in} Dr. Pankesh Patel \\ 
\end{center}
\vspace{0.5cm}

Date: $30^{th}$ June, 2016

Place: Ahmedabad \\

\newpage
\tableofcontents

\chapter*{Abstract}
\addcontentsline{toc}{chapter}{Abstract}
\vspace{10pt}
\hspace*{0.2in}Application development for Internet of Things, Service, and People~(IoTSP) is challenging because it
involves dealing with the heterogeneity that exists both in Physical and Internet worlds.
Second, stakeholders involved in the application development have to address 
issues pertaining to different life-cycles ranging from design, implementation 
to deployment. Given these, a critical challenge is to enable an application development 
for IoTSP applications with effectively and efficiently from various stakeholders. 

Several approaches to tackling this challenge have been proposed in the fields of 
Wireless Sensor Networks~(WSN) and Pervasive Computing, regarded 
as precursors to the modern day of IoTSP. However, existing approaches only cover limited 
subsets of the above mentioned challenges when applied to the IoTSP. 
In view of this, we have built
upon existing framework and evolved it into a framework for developing
IoTSP applications, with substantial additions and enhancements in high-level modeling languages
and their integration into the framework, and we present a comparative 
evaluation results with existing approaches. This provides 
the IoTSP community for further benchmarking. The evaluation is carried out on real devices 
exhibiting heterogeneity. Our experimental analysis and results demonstrate that our approach 
drastically reduces development effort for IoTSP applications compared to existing approaches.

\chapter*{List of Abbreviations}
\addcontentsline{toc}{chapter}{List of Abbreviations}
\begin{minipage}{0.3\textwidth}
\begin{flushleft} 
 \textbf{\hspace{0.1in} IoTSP} \\
\textbf{\hspace{0.1in} EoI} \\
\textbf{\hspace{0.1in} MDD} \\
\textbf{\hspace{0.1in} WSN} \\
\textbf{\hspace{0.1in} DL} \\
\textbf{\hspace{0.1in} AL} \\
\textbf{\hspace{0.1in} UIL} \\
\textbf{\hspace{0.1in} UI} \\
\textbf{\hspace{0.1in} DL} \\
\textbf{\hspace{0.1in} GPL} \\
\textbf{\hspace{0.1in} LoC} \\
\end{flushleft}
\end{minipage}
~
\begin{minipage}{0.6\textwidth}
\begin{flushleft}
 Internet of Things, Service and People \\
Entity of Interest\\
Model-driven development\\
 Wireless sensor networks\\
 Domain language\\
Architecture language\\
User interaction language\\
 User interface\\
 Deployment language\\
 General-purpose programming language\\
Lines of Code\\
\end{flushleft}
\end{minipage}

\vspace{10pt}

\newpage
\listoffigures
\addcontentsline{toc}{chapter}{List of Figures}
\listoftables
\addcontentsline{toc}{chapter}{List of Tables}
\newpage

\pagenumbering{arabic}

\chapter{Introduction}\label{c1}

\hspace*{0.2in} Internet of Things, Service and People~(IoTSP) is composed of  highly heterogeneous interconnections of 
elements from both the Physical as well as Internet worlds. IoTSP include WSN, RFID technologies, 
smart phones, and smart appliances as well as the elements of the traditional Internet such as Web 
and database servers, exposing their functionalists as Web services. In the \textit{IoTSP}, things formed a network, things could be devices, vehicles, buildings, and other embedded objects that enable exchange of the data. A \textit{thing} in the IoTSP is an entity which measures the \textit{Entity of Interest~(EoI)}, service could be element of the traditional Internet such as Web services~(e.g., Yahoo weather service, cloud services etc.), and people is an entity which interact with entity of interest. For example, smart watch is a thing and measures the person's body temperature, heart rate etc. Here person's body temperature and heart rate are entity of interest for the entity smart watch. IoTSP applications will involve interactions among large numbers of devices, many of them directly interacting with their physical surroundings.   
\section{Background and Motivation}
\hspace*{0.2in}While IoTSP have shown a great 
potential of usage into several fields such as smart home, personal health, energy usage monitoring and others, 
the heterogeneity and diversity involved in IoTSP represent a difficult challenge to deal with.  An important
challenge that needs to be addressed is to enable rapid development of IoTSP applications with minimal effort by various stakeholders\footnote{We use the term \textbf{stakeholders} as used in software engineering to mean- people, who are involved in the application development. Examples of stakeholders are software engineer, developer, domain expert, technologist etc.} involved in the application development process. To address above issues, several challenges have already been addressed in the closely related fields of the Wireless Sensor Networks~(WSNs)~\cite[p.~65]{vasseur2010interconnecting}and pervasive computing~\cite[p.~65]{vasseur2010interconnecting} to the modern day of IoTSP. \textbf{The goal of work\cite{Patel201562}  is to enable the development of applications for such complex systems}. In the following, 
we discuss one of such application.

\subsection{Application example: Smart Home}\label{sec:case-study}
To illustrate characteristics of IoTSP application, we consider smart home application. A home consists of several rooms, each one is instrumented with several heterogeneous entities~(physical devices) for providing residents' comfort, safety, and optimizing resources. Many applications can be developed on top of these devices, one of which we discuss below:
\begin{figure}[!ht]
\centering
\includegraphics[width=0.9\textwidth]{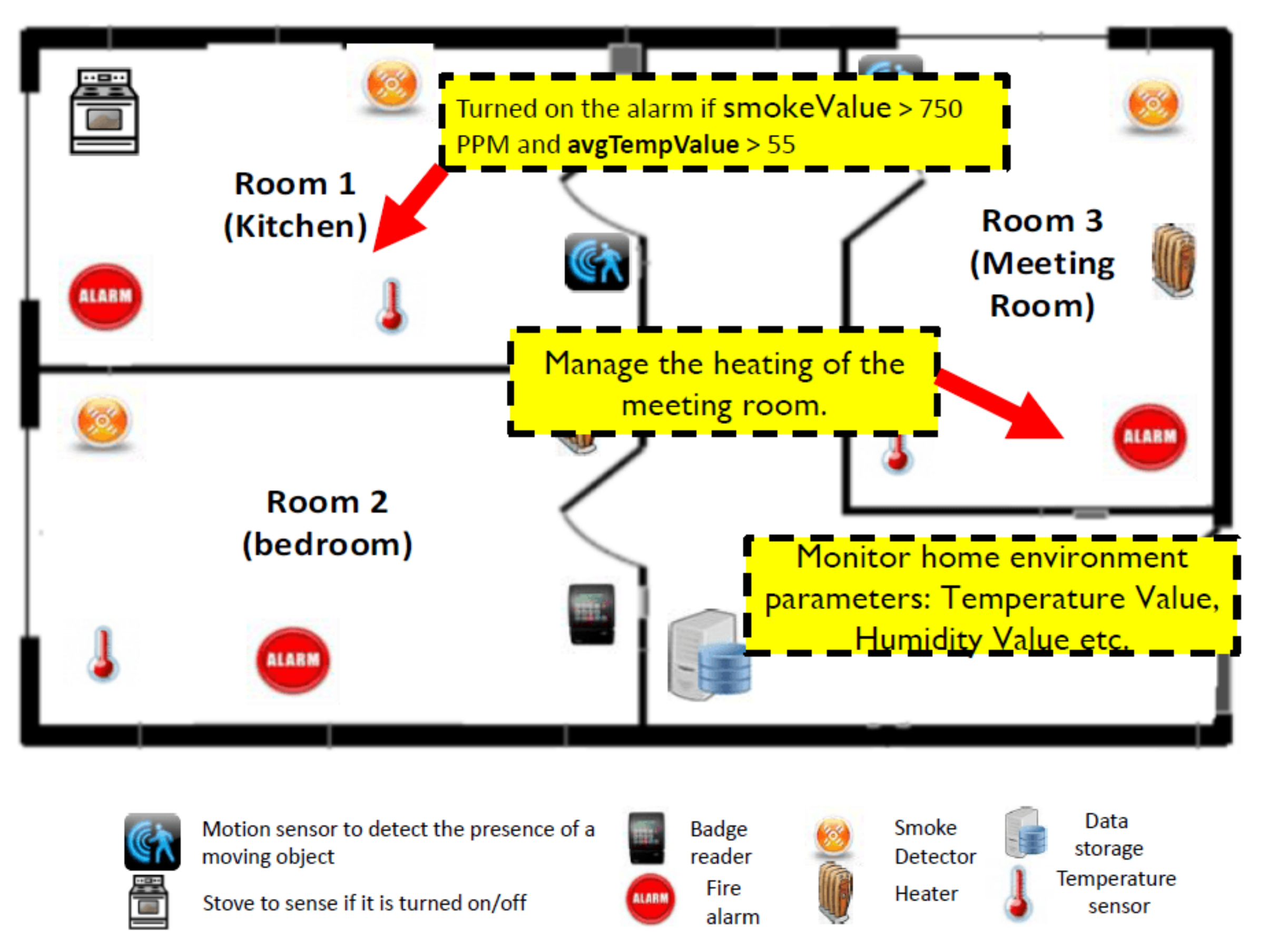}
\caption{A smart home with deployed devices (1) Temperature sensor, (2) Heater, (3) Fire alarm, (4) Smoke detector, (5) Badge reader, (6) Data storage, (7) Motion sensor, and (8) Smoke sensor.}
\label{homeautomation}
\end{figure}

To \textit{accommodate a resident's preference} in a room, a database is used to keep the profile of each resident, including his/her preferred temperature level. An RFID reader in the room detects the resident's entry and queries the database service. Based on this, the thresholds used by the room devices are updated. 

To ensure the \textit{safety of residents}, a fire detection application is installed. It  aims to detect fire by analyzing data  from smoke and temperature sensors. When fire occurs, residences are notified on their smart phone by an installed application. Additionally, residents and their neighbors are informed through a set of alarms. Moreover, the system generates the current environment status on dashboard (e.g., humidity, temperature, outside temperature by interacting with external web services) for the \textit{situation awareness}. 
 
\section{IoTSP application development challenges}\label{sec:challenges}

\begin{figure}[!ht]
\centering
\includegraphics[width=0.9\textwidth]{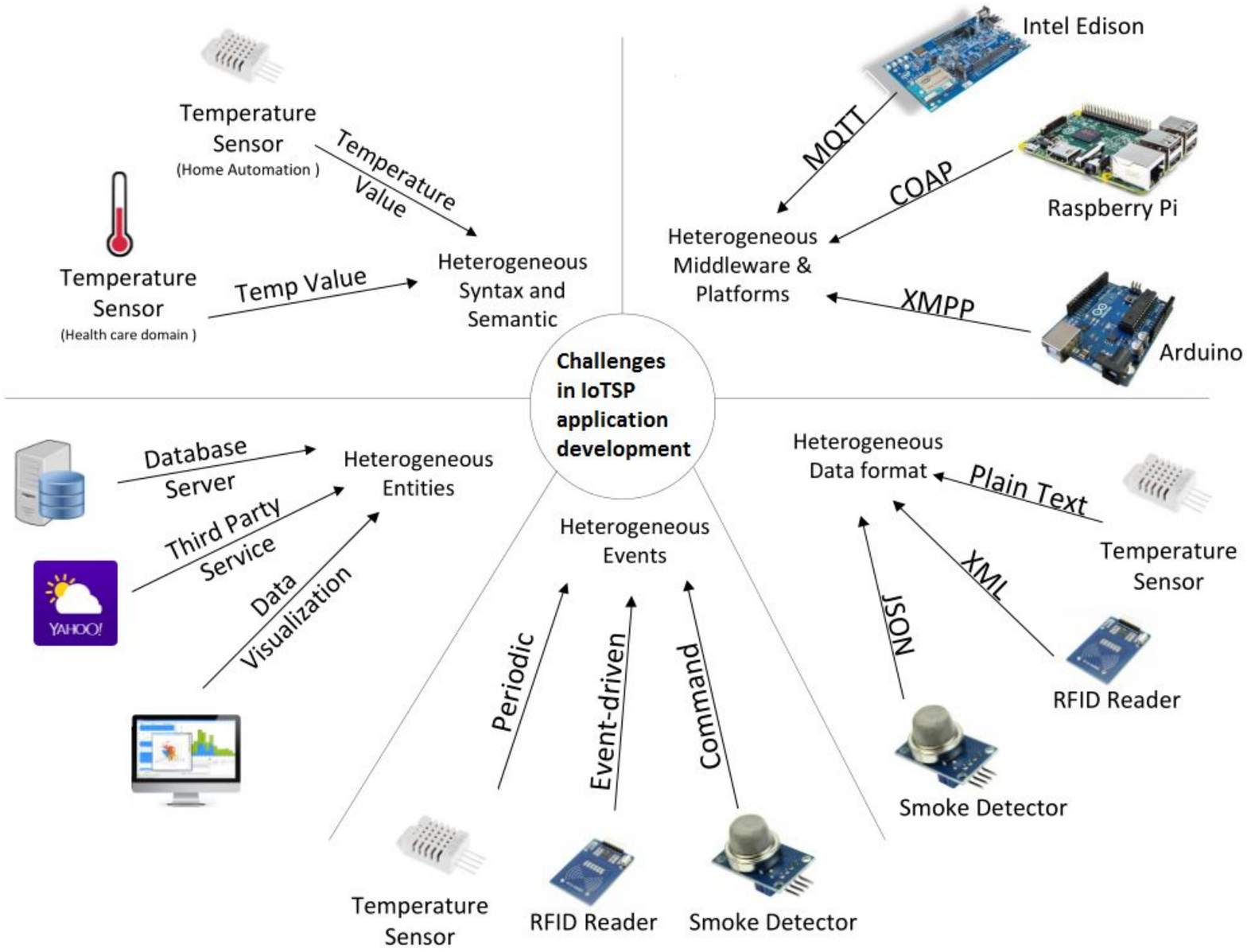}
\caption{Challenges in IoTSP application development}
\label{fig:challenges}
\end{figure}

The development of application~(discussed in Section~\ref{sec:case-study}) is difficult because IoTSP applications exhibit the following challenges~(Refer Figure~\ref{fig:challenges}): 

\textit{ Heterogeneous entities}: An IoTSP may execute on a network consisting of different types of entities. For example, a homeautomatione consists of entities encompassing, {\em sensors}~(e.g., temperature sensor), {\em tags}~(e.g., badgeReader to read a user's badge), {\em actuators}~(e.g., heater), {\em user interfaces}~(e.g., user receives notification in case of fire), {\em storage}~(e.g., profile storage to store users' data),  and 
elements of the ``traditional'' Internet such as Web and database servers, exposing their functionality as {\em Web services}~(e.g., Yahoo Weather service). 


\textit{Heterogeneous events and data exchange format }: An entity in IoTSP may exhibit heterogeneous events to communicate with other things/entities exist in the surrounding. The above mentioned entities exhibit different interaction modes such as {\em event-driven}~(e.g., a smoke sensor fires when a smoke is detected), 
{\em request-response}~(e.g., profile data is retrieved through request and response), 
{\em periodic}~(e.g., temperature sensor sense data periodically), {\em command}\cite{andrews1991paradigms}~(e.g., a heater fires command to regulate the temperature), and
{\em notify}\cite{Balland-2013}(e.g., a user is notified when the fire is detected). An entity also exhibits heterogeneous data exchange format to sense, affect, store and identify an EoI~(Entity of Interest). A temperature sensor may publish data in JSON format \{tempValue: 35, unitofMeasurement: "C"\}, and smoke sensor may publish data in XML format $\langle$ parameter name="smokeValue", value=550 ppm $\rangle$.

\textit{Heterogeneous middleware and platforms}: Unlike WSN, an IoTSP applications may involve entities running on different 
platforms. For instance, a temperature sensor may be attached with Raspberry 
Pi publish data using MQTT communication protocol, smoke sensor runs on a resource constrained 
micro-controller with no OS~(e.g., Arduino) publish data using CoAP communication protocol, an end-user 
application is deployed on Android Mobile OS, and a dashboard is implemented using JavaScript and HTML interaction is handled by WebSocket. 

\textit{Heterogeneous Syntax and Semantic}: The heterogeneous environment leverages IoT technology in multiple domains such as home automation, health care, transportation, energy monitoring etc. The cross domain application involves heterogeneous
semantic and syntax. For instance, in home automation domain temperature sensor is used to sense the temperature value (environment
parameter), where in health care domain temperature sensor is used to measure person’s body temperature.

\section{Contributions}\label{sec:contri}

In view of the above, we have built upon existing framework~\cite{Patel201562}  
and evolved it into a framework for IoTSP. The proposed development framework segregates IoTSP development 
concerns, provides a set of modeling languages to specify them and integrates automation techniques 
to parse these modeling languages. We present an enhanced and extended version of modeling languages and their integration into a development framework: (1) Domain language~(DL) that models IoTSP characteristics such as describing heterogeneous entities such as tags, external third-party services, and different types of sensors. (2) Architecture Language~(AL) to describe functionality of an application, (3) User Interaction Language~(UIL) to model interaction between an application and a user, and (4) Deployment Language~(DL) to describe deployment-specific features consisting information about a physical environment where devices are deployed. 

We present a comparative evaluation results with existing approaches. This further provides the IoTSP community for benchmarking.  The evaluation is carried out on {\em real devices} exhibiting IoTSP heterogeneity. 
The evaluation shows that our approach drastically reduces development effort as compare to existing approaches. We have created user manual of IoTSuite\footnote{IoTSuite user manual is available at http://www.slideshare.net/pankeshlinux/iotsuite-user-manual} which provides step by step guide to develop IoTSP application using IoTSuite and prepared video prototype\footnote{Video prototype is available at https://www.youtube.com/watch?v=nS Je7IzPvM} which demonstrates step by step process to develop smart home application using IoTSuite.

\section{Organization of the Thesis} 
\begin{itemize}
	\item Chapter~\ref{literature} focuses on the existing approaches available for programming IoTSP applications. The state of art represents comparative study of existing approaches with prototypes. 
	\item Chapter~\ref{appdev} presents development framework for programming IoTSP applications. It includes high-level programming constructs, separates different aspects of the system to ease application development and provides automation at different phases of development process to reduce development effort.
	\item Chapter~\ref{IoTSuite} presents user manual of IoTSuite for developing IoTSP applications. It provides step by step guide to develop IoTSP application.
	\item Chapter~\ref{evaluation} evaluates comparative analysis to develop smart home application using existing approaches. The experimental  results used by IoTSP community for further benchmarking.
\end{itemize}

\chapter{Literature Review}
\label{literature}
\section{State of the art: Programming Internet of Things, Service and People}\label{sec:sota}
To address the above mentioned challenges~(discussed in Section~\ref{sec:challenges}), several approaches exist to develop IoTSP application as described in following sections. 

\section{General Purpose Programming Languages}~\label{GPL approach} In this approach, stakeholders use General Purpose Programming Languages (GPLs) such as C, JavaScript, Python, and Android which encode activities of individual devices and target middleware to encode interaction with other devices~\cite{patel:tel-00927150}.  This approach requires more development efforts to develop IoTSP application using GPLs. The reasons are: 1) stakeholders have to write a program that reads data from sensors, 2) aggregate data to perform computational tasks, and 3) communicate with surrounding entities such as yahoo weather service and actuators to affect an entity of interest. The advantage of such approach is stakeholders have complete control over the system, and it also captures the heterogeneity that exists in the modern day IoTSP applications. 

\begin{figure}[!ht]
\centering
\includegraphics[width=0.5\textwidth]{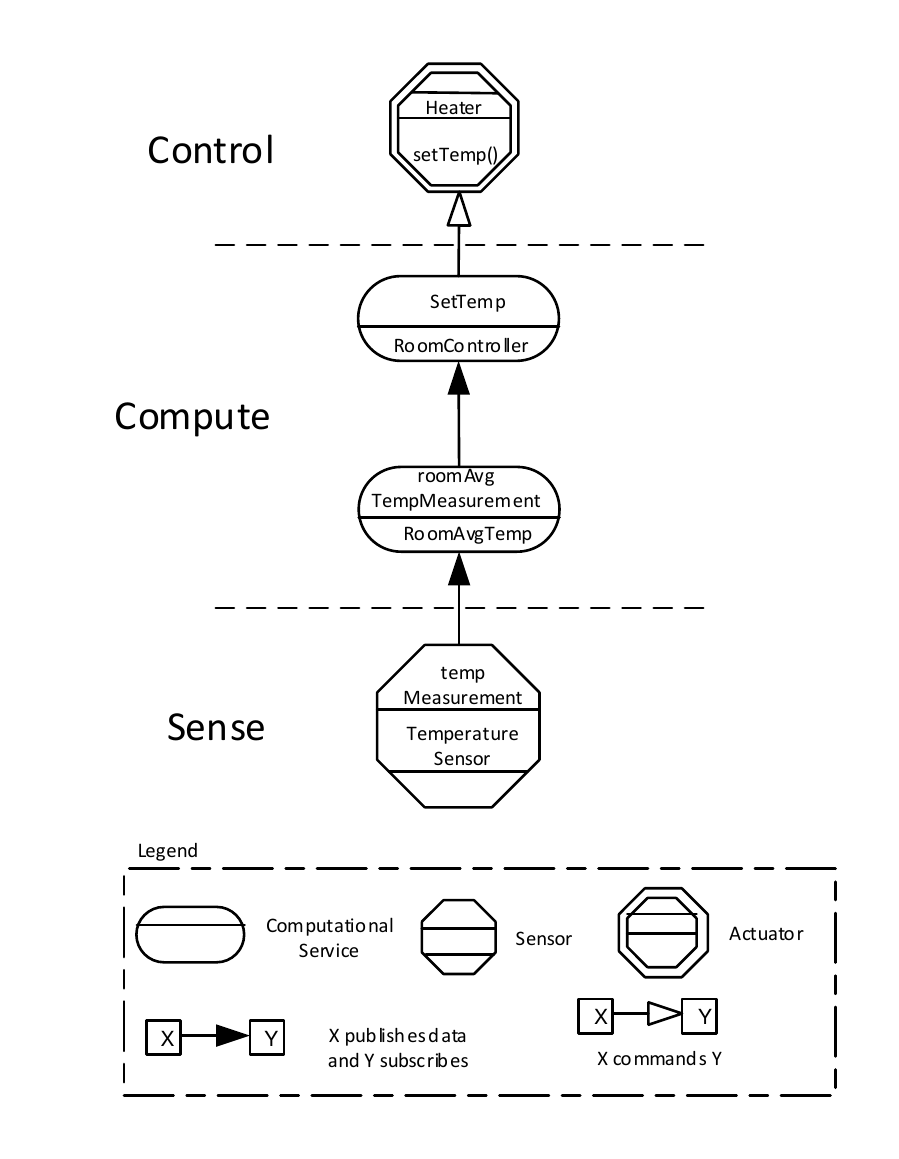}
\caption{Dataflow diagram of HVAC app.}
\label{fig:HVAC}
\end{figure}

To thoroughly understand the development effort required by this approach, let’s develop Heating, Ventilation and Air Conditioning (HVAC) application using GPLs\footnote{Code to develop HVAC application using GPL is available at https://github.com/chauhansaurabhb/Regulate-Temperature-HVAC-}. HVAC application consists of Temperature Sensor, RoomAvgTemp, Room Controller, and Heater as per data flow diagram shown in Figure~\ref{fig:HVAC}. The aim of HVAC application is to regulate temperature of room.
The code snippet of AM 2302 temperature sensor that is connected to Raspberry Pi is illustrated in~Listing~\ref{AM2302}. In order to read tempValue using AM2302 sensor, stakeholders have to connect this sensor to Pi  using GPIO pin.  Stakeholders have to write a program that periodically reads tempValue through GPIO pin 4~(Listing~\ref{AM2302}, line~\ref{GPIO}), and publish data using middleware.

\lstset{emph={var, function (), console, log, warn, if, else, require, return}, emphstyle={\color{blue}\bfseries\emph}, caption={Code snippet to read tempValue using AM2302}, label=AM2302, escapechar=\#}	
 
\begin{lstlisting}
1. var sensorLib = require('node-dht-sensor'); #\label{sensorLib}#
2. var mqtt=require('mqtt'); #\label{mqttlibstart}#
3. var client=mqtt.connect('mqtt://test.mosquitto.org:1883'); #\label{mqttlibend}#
4. var sensor = 
5. {
6.  initialize: function ()
7.  {
8.	 return sensorLib.initialize(22, 4); #\label{GPIO}#
9.  },
10. read: function () 
11.  {
12.	 var readout = sensorLib.read(); #\label{readout}#
13.	 var value={"tempValue":readout.temperature.toFixed(2),#\label{valuestart}#
14.	 "unitOfMeasurement":"C"};   #\label{valueend}#
15.	 client.publish('tempMeasurement',JSON.stringify(value)); #\label{publish}#
16.	 console.log('Current temperature value: ' + readout. #\label{consolestart}#
17.	 temperature.toFixed(2)+ 'C'+');  #\label{consoleend}#
18.			
19.	 setTimeout(function ()
20.	 {
21.	  sensor.read();
22.	 }, 5000);  #\label{periodically}#
23.  }
24. }; 
25.
26. if (sensor.initialize()) {  #\label{initializestart}#
27.	   sensor.read();
28.	 }
29. else {
30.		console.warn('Failed to initialize sensor'); #\label{initializeend}#
31.	 }
\end{lstlisting}

Here stakeholders used node.js module node-dht-sensor~(Listing~\ref{AM2302}, line~\ref{sensorLib}) query tempValue from DHT sensor, also used node.js client library for mqtt protocol (with mosquitto broker) written in JavaScript as illustrated in Listing~\ref{AM2302}, lines~\ref{mqttlibstart}-\ref{mqttlibend}. If the initialization of  AM2302 sensor succeeds than it reads value using readout (which contains temperature value)~(Listing~\ref{AM2302}, line~\ref{readout}) else display warning message to console~(Listing~\ref{AM2302}, line~\ref{initializeend}). On the successful initialization of sensor, it publishes JSON formatted data with topic tempMeasurement~(Listing~\ref{AM2302}, line~\ref{publish}). \\

We identify that stakeholders have complete control and freedom over the code/system  but it incurs more development efforts to build IoTSP application using this approach. The reason is stakeholders are responsible for end to end development. The code is difficult to reuse as it offers platform-dependent design (stakeholders have to think in terms of platform). 

\section{Macro Programming}
Macro prog. provides programming abstraction to specify global behavior of  distributed application to develop sense-compute-control application~\cite{mottola2011programming}. The programming abstraction allows the stakeholders to specify local behavior of nodes which encode the global behavior of distributed systems~\cite{gummadi2005macro}. It provides an abstraction of distributed application as  collection of nodes  that are bound together to perform a task within a single program. It provides abstraction to specify high-level constructs: configuration of GPIO pins in order to read/write values using nodes of sensors and actuators, passing these values to other nodes which allow the stakeholders to perform computational tasks by manipulating with these values at nodes. Macro prog. is  a viable approach compared to general-purpose programming approach as it provides flexibility to write custom-application logic and reduce development effort (Refer Section~\ref{LOCandCyclomatic})~\cite{Chauhan:2016:DFP:2897035.2897039}. Examples of macro prog. approach to develop IoTSP app. are Node-RED\footnote{http://nodered.org/}, WoTKit~\cite{wotkit} etc.

\textbf{Node-RED} is a drag-and-drop based editor for  connecting  hardware devices, services, and interfaces under the browser based flow editor. The main advantage of Node-RED is light-weight tool and built on Node.js which makes it easy to run on low-cost hardware and resource constraint device such as Raspberry Pi, Intel Edison etc. The primary reason for selecting Node-RED~(as example of macro prog.) is open source tool~(available at https://github.com/node-red with more than 2000 commits, last commit before 20 hours) allows stakeholders to create own node~(in case node doesn't exist) and contribute to Node-RED library. Node-RED is equipped with well defined documentations which 
\begin{figure}[!ht]
\centering
\includegraphics[width=1.0\textwidth, height=0.45\textwidth]{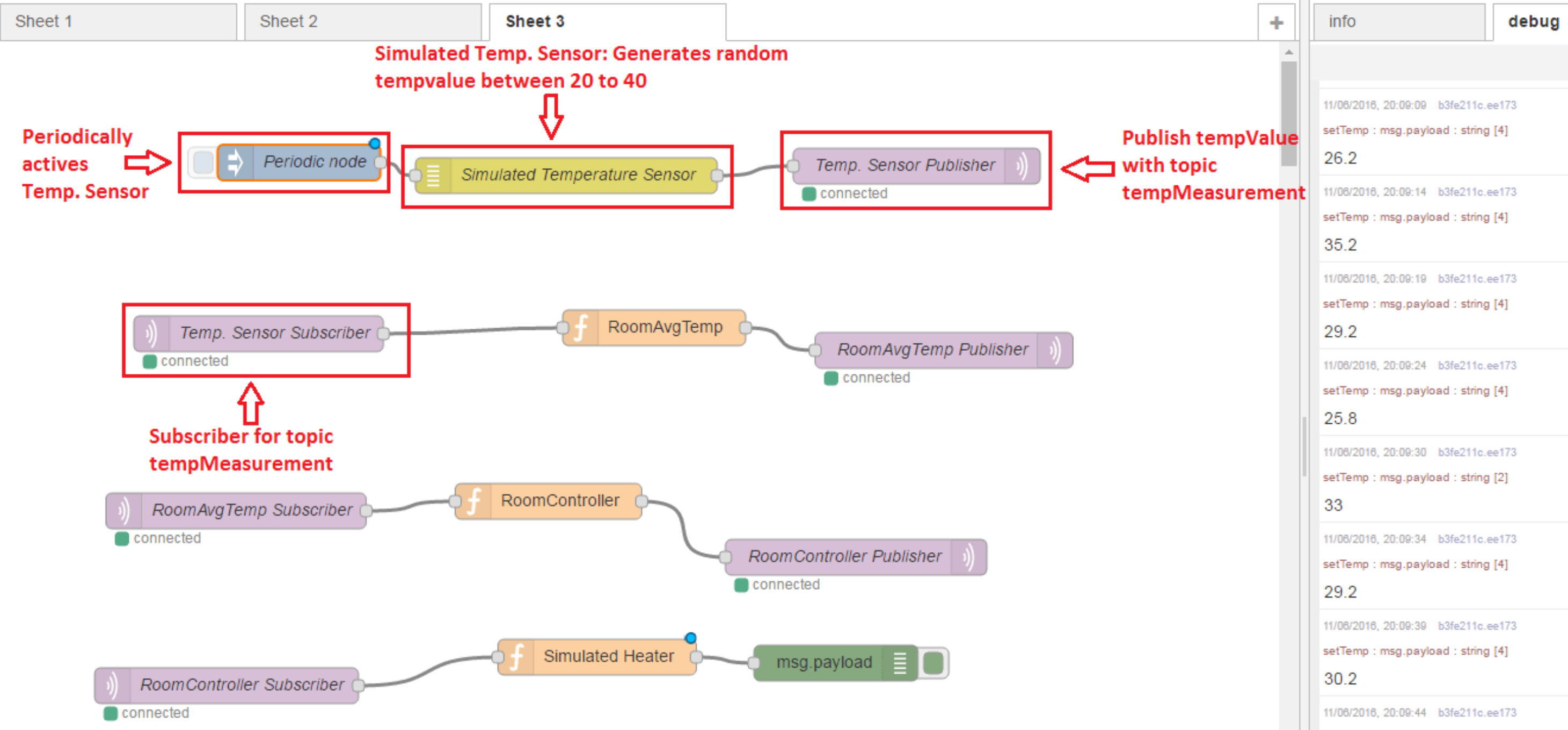}
\caption{Node-RED flow to develop HVAC app. (1/2)}
\label{fig:noderedflow}
\end{figure}
guide stakeholder to create flow, add nodes to Node-RED library etc. It also helps stakeholders to write custom application logic~(e.g., maintain the room temperature between $25^{\circ} C$ to $36^{\circ} C$) using JavaScript. The community is wide spread by providing support through stack overflow~(more than 248 questions posted with tag node-red).
Let's develop HVAC application discussed in \ref{GPL approach} using Node-RED\footnote{Node-RED flow to develop HVAC application is available at https://github.com/chauhansaurabhb/Regulate-Temperature-HVAC-} to understand development efforts required by this approach. Node-RED flow to develop HAVC application is shown Figure~\ref{fig:noderedflow}. Here, stakeholder used simulated
\begin{figure}[!ht]
\centering
\includegraphics[width=1.0\textwidth, height=0.45\textwidth]{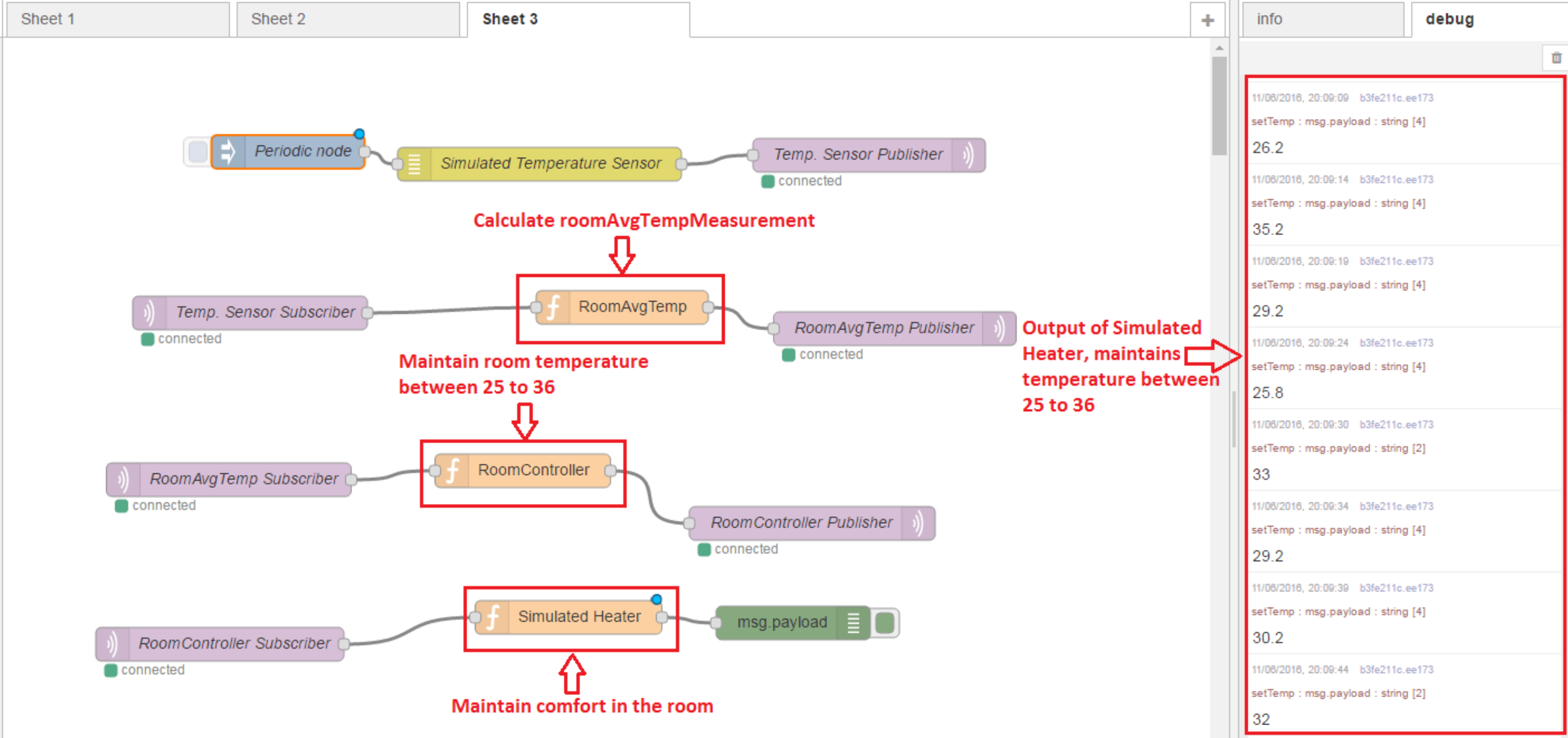}
\caption{Node-RED flow to develop HVAC app. (2/2)}
\label{fig:noderedflow1}
\end{figure} temperature sensor which publish tempValue (random number between 20 to 40) periodically (every 5 seconds), RoomAvgTemp subscribe this tempValue and calculate roomAvgTempMeasurement (it takes five sample of tempValue and calculate avgtempValue), and publish it. The RoomController component subscribe for roomAvgTempMeasurement and generates command setTemp() based on application logic (application logic is to maintain room temperature between 25 to 36) and Heater is used to maintain temperature of room as shown in debug console (refer Figure~\ref{fig:noderedflow1}). 
This approach reduces development efforts as compared to GPLs by providing abstractions which hide complexity to implement common functionality such as read/write values using node of sensor/actuator. Stakeholders have to write code to implement application specific logic~(node may not be available in library) so it incurs more development efforts to build IoTSP application.

\section{Cloud Based Platform}
Cloud based platform reduce development efforts compared to macro programming approach by providing  APIs or visual programming constructs which is used to  implement common functionality.  The resources are located in the cloud (centrally) so it offers ease of deployment and evolution. This approach sacrifices node-to-node communication because all communication happens through Cloud. 

\begin{figure}[!ht]
\centering
\includegraphics[width=0.99\textwidth]{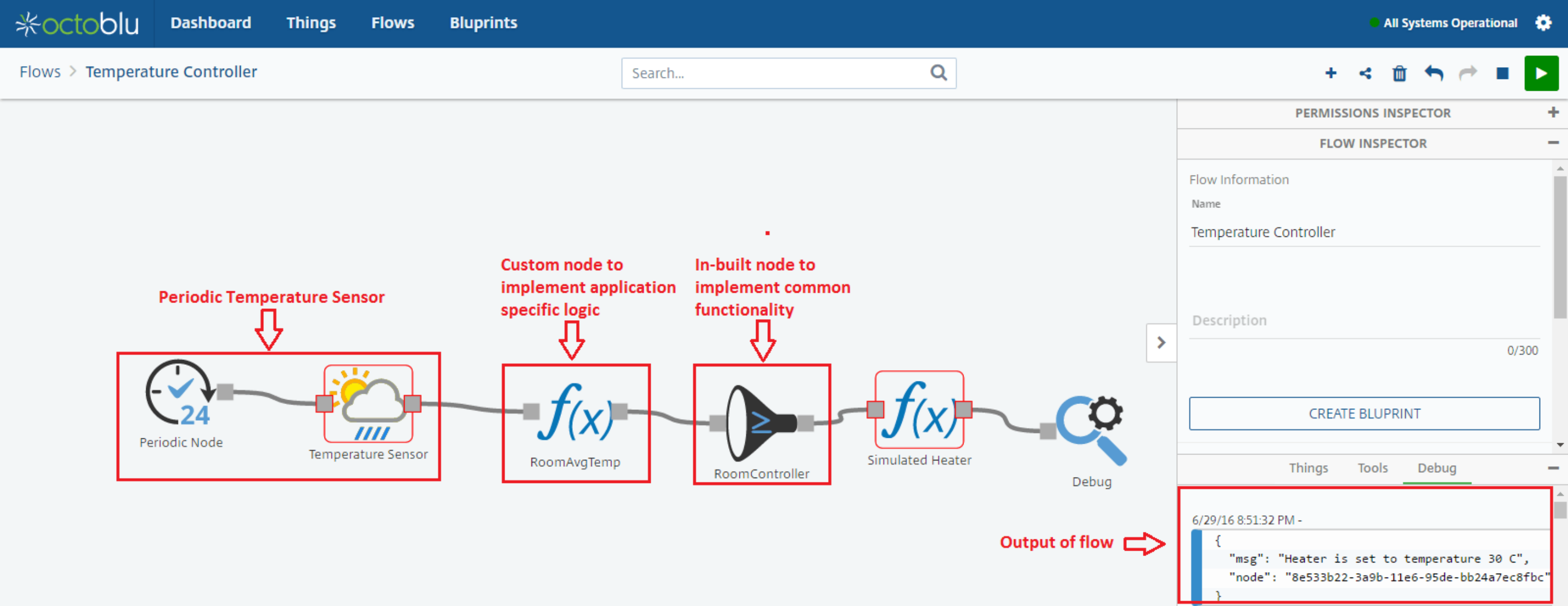}
\caption{Octoblu flow to develop HVAC app.}
\label{fig:Octablu}
\end{figure}

\textbf{Octoblu}\footnote{https://developer.octoblu.com/} is a cloud based approach enables communication between sensors, actuators, and resource constraint devices through Meshblu platform\footnote{Meshblu is the backbone of the Octoblu platform offering secure, cross-protocol cloud based communication between smart devices, cloud resources, and software API.}. The advantage of this approach is: it offers wide range of communication protocols such as CoAP, HTTP, MQTT, WebSocket etc. to enable communication between smart devices~(also enables communication across discrete protocols, and IoT platforms). The key benefit of this approach is, 1) collection of open-source libraries and services that allow micro-controllers such as Arduino to communicate with Octoblu platform, 2) communication libraries are written in Node.js runs on resource constraint device, 3) provides mobile app. for iOS, android device, and 4) integration of Octoblu with android, iOS, etc. using SDK. It's drag-and-drop design and one click automation~(using play button as shown in Figure~\ref{fig:Octablu}) reduce development efforts. It is open source cloud based approach (available at https://github.com/octoblu with more than 500 commits, and supported by active community). Octoblu available with well defined documentations and guide which allow stakeholders to develop IoTSP application.
Let's develop HVAC application discussed in \ref{GPL approach} using Octoblu. The flow for HVAC app. using Octoblu is shown in Figure~\ref{fig:Octablu}. The benefit of Octoblu compared to Node-RED is it doesn't require communication node to communicate with other devices thus it reduces development efforts~(in terms of LoC). It is integrated with tools which simplify task of writing custom application logic~(e.g., min, max, less than, equal to etc.). It allows stakeholders to implement application specific logic~(e.g., maintain the room temperature between $25^{\circ} C$ to $36^{\circ} C$) using JavaScript (in case in-built tool/node may not exist).
\begin{sidewaystable}
\centering 
\scriptsize
\begin{tabular}{|p{1.5cm} | p{1.2cm} | p{1.5cm} |p{1.5cm} |p{1.5cm} |p{2.2cm} | p{2.2cm} | p{2.3cm} | p{1.3cm} | p{1.3cm} | p{2.3cm} |}
 \toprule
\textbf{Approach}  &\textbf{Open Source}&\textbf{IDE Support}& \textbf{Operating Interface} & \textbf{Database Support}  &\textbf{Interaction Modes}&\textbf{Communication Medium}&\textbf{Prog. Interface}&\textbf{Types of app. Development}&\textbf{Data Exchange Format} & \textbf{Documentation Guideline/Support} \\ \midrule 
 Macro prog. (Node-RED) & Yes (GitHub commits more than 2000) & Browser-based IDE & Visual editor~(Drag-and-Drop) & MongoDB, MySQL & Periodic, Cmd, Req/Res, Event-driven, Notify & MQTT, HTTP, WebSocket & JavaScript & SCC, DV, Data collection, End User app. & JSON & Support on stack-overflow with node-red tag (more than 240 questions), wide range of documentation, and active community support    \\ \midrule
Cloud based platform (Octoblu) & Yes (GitHub commits more than 500) & Web-based IDE & Visual editor~(Drag-and-Drop) & MongoDB, AzureDB & Periodic, Cmd, Event-driven, Notify  & HTTP, WebSocket, MQTT, CoAP, XMPP etc. & JavaScript & SCC, DV, Data collection, End User app. & JSON &  Documentation available with examples and codes, Support from community and development team  \\ \midrule
													 
\end{tabular}
\caption{A brief comparison of existing approaches and prototypes. \textbf{SCC}~(Sense-Compute-Control), \textbf{DV}~(Data Visualization)} 
\label{table:comparisionofapproach} 
\end{sidewaystable} 

\chapter{IoTSP application development}
\label{appdev}
This chapter presents our development framework that separates IoTSP application development into different concerns, namely \emph{domain}, \emph{platform}, \emph{functional}, and \emph{deployment}. It integrates a set of high-level modeling languages to specify such concerns.  These languages are supported by automation techniques at various phases of application development process. Stakeholders carry out the following steps in order to develop an IoTSP application  using our approach.

\begin{figure*}[!ht]
\centering
\includegraphics[width=0.8\textwidth]{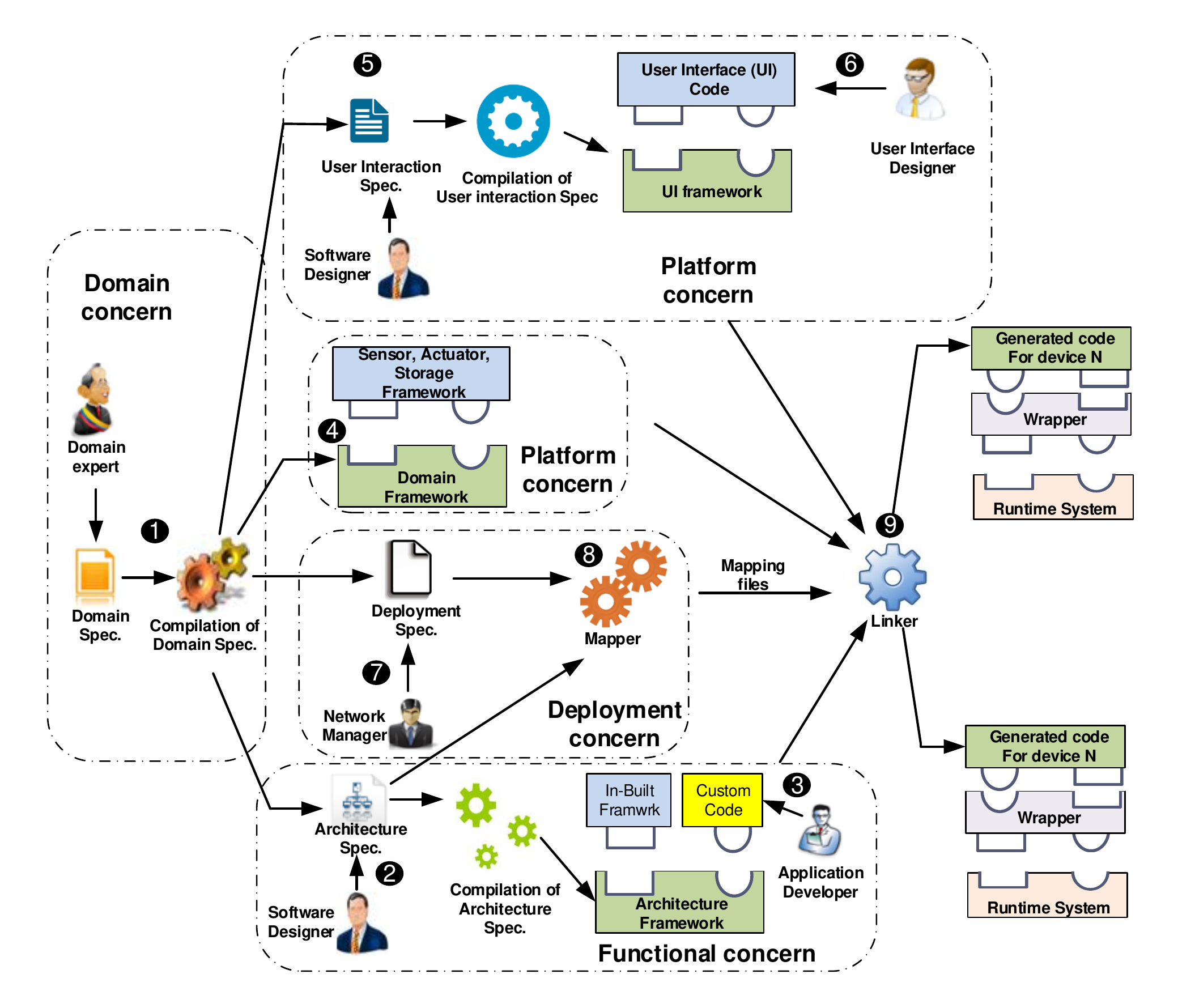}
\caption{IoTSP application  development: the overall process}
\label{developmentcycle}
\end{figure*}

\section{Domain Concern}
This concern is related to concepts that are specific to a domain (e.g., building automation, transport etc.) of an IoTSP. The stakeholders task regarding such 
concern consists of the following step:

\fakeparagraph{\em Specifying and compiling domain specification}
The domain expert specifies a domain specification using the Domain Language~(DL) 
(Step~\circled{1} in Figure~\ref{developmentcycle}). The domain specification includes specification of resources, which are responsible for interacting with Entities of Interest~(EoI). This includes \emph{tags}~(identify EoI), \emph{sensors}~(sense EoI), \emph{actuators}~(affect EoI), and \emph{storage}~(store information about EoI). In the domain specification, resources  are specified in a high-level manner to abstract low-level details from the domain expert~(detail in Section~\ref{sec:specdomain}).

\section{Functional Concern}
This concern is related to concepts that are specific to functionality of an IoTSP. An example of functionality is to open a window when an average temperature value of a room is greater than $30^{\circ} C$. The stakeholders task regarding such 
concern consists of the following steps:

\fakeparagraph{\em Specifying and compiling application architecture}
Referring the domain specification, the software designer specifies an application architecture using the Architecture Language~(AL)(Step \circled{2} in Figure~\ref{developmentcycle}). It consists of specification of computational 
services and interaction among them~(detail in Section~\ref{sec:arch}). A computational service is fueled by sensors and storage defined in the domain specification. They process inputs data and take appropriate decisions by triggering actuators defined in the domain specification. The architecture specification consists of two types of computational services: (1) {\em common} specifies common operations such as average, count, sum in the application logic, (2) {\em custom} specifies an application-specific logic (for instance, coordinating events from BadgeReader with the content of Profile data storage).

\section{Platform Concern}
This concern specifies the concepts that fall into computer 
programs that act as a translator between a hardware device and an application. 
The stakeholders task regarding such concern consists of the following steps:
		
\fakeparagraph{\em Generating device drivers}
The compilation of domain specification generates a domain framework~(Step~\circled{4} in Figure~\ref{developmentcycle}). 
It contains {\em concrete classes} corresponding to concepts defined in the domain specification. The concrete classes contain concrete
methods to interact with other software components and platform-specific device drivers,
described in our work~\cite[p.~75]{Patel201562}. We have integrated existing open-source sensing framework\footnote{\url{http://www.funf.org/}} 
for Android devices. Moreover, we have implemented sensing and actuating framework
for Raspberry Pi and storage framework for MongoDB, MySQL, and Microsoft AzureDB. So, the device developers do not have to implement platform-specific sensor, actuator, and storage code.

\fakeparagraph{\em Specifying user interactions}
To define user interactions, we present a set of {\em abstract interactors}, similar to work~\cite{Balland-2013}, that denotes information exchange between an application and a user. The software designer specifies user interactions using User Interaction Language~(UIL)~(Step~\circled{5} in Figure~\ref{developmentcycle}). The UIL provides three abstract interactors: (1) {\em command} denotes information flow from a user to an application~(e.g., controlling a heater according to a temperature preference), (2) {\em notify} denotes information flow from an application to a user~(e.g., fire notification in case of emergency), (3) {\em request} denotes information flows round-trip between an application and a user, initiated from the user~(e.g., requesting preference information to a database server). More detail is in Section~\ref{sec:spec-user-interactions}. 
The Implementation of user-interaction code is divided into two parts: (1) compilation of user-interaction specification, and (2) Writing user-interaction code. These two steps are described below:

\begin{itemize}
	\item \textbf{Compilation of user interaction spec}: Leveraging the user interaction specification, the development framework generates a User Interface~(UI) framework to aid the user interface designer~(step~\circled{6} in Figure~\ref{developmentcycle}). The UI framework contains a set of {\em interfaces} and {\em concrete classes} corresponding to resources defined in the user interaction specification.  The concrete classes contain concrete methods for interacting with other software components. For instance, the compilation of {\em command} interactors generates \texttt{sendCommand()} method to send command to other component. Similar way, the compilation of {\em notify} interaction generates \texttt{notifyReceived()} method to receive notifications. 
	
 \item \textbf{Writing user-interaction code}: Leveraging the UI framework, the user interface designer 
implements {\em interfaces}. These interfaces implements code that connects appropriate UI elements 
and concrete methods. For instance, a user initiates a command to heater by pressing UI elements such as \texttt{button} and \texttt{sendCommand()} method, or the application notifies a temperature value on \texttt{textlabel} through
\texttt{notifyReceived()} method. 
\end{itemize}

\section{Deployment Concern}
This concern is related to deployment-specific concepts that describe the
information about a device and its properties placed in the target deployment. 
It consists of the following steps:

\fakeparagraph{\em Specifying target deployment} 
Referring the domain specification, the network manager describes a deployment specification using the Deployment Language~(DL)~(Step~\circled{7} in Figure~\ref{developmentcycle}). The deployment specification 
includes the details of each device, resources hosted by each device, and the type of device. Ideally, the IoTSP application can be deployed on different deployments. This requirement is dictated by separating a deployment specification from other specifications.

\fakeparagraph{\em Mapping}  The mapper produces a mapping from a set of computational services to a set of devices. It takes a set of devices defined in the deployment specification and  a set of computation components defined in the architecture specification(Step~\circled{8} in Figure~\ref{developmentcycle}). The mapper devices devices where each computational services will be deployed. The current version of mapper algorithm~\cite{Patel201562} selects devices randomly and allocates computational services to the selected devices.

\fakeparagraph{\em Linking} The linker combines the code generated by various stages and creates packages that can be deployed on devices~(Step~\circled{9} in Figure~\ref{developmentcycle}).  It merges generated architecture framework, UI framework, domain framework, and mapping files. This stage supports the application deployment phase by producing device-specific code to result in a distributed software system collaboratively hosted by individual devices, thus providing automation at the deployment phase.

The final output of linker is composed of three parts: (1) a \emph{runtime-system} runs on each individual device and provides a support for executing distributed tasks, (2) a \emph{device-specific code} generated by the linker module, and (3) a \emph{wrapper} separates generated code from the linker module and underlying runtime system by implementing interfaces.

\section{Domain Specification}\label{sec:specdomain}
The Domain Language~(DL) offers high-level constructs to specify the domain-specific concepts. We describe these constructs as follows:

{Sensors}  A set of sensors is declared using the \texttt{sensors} keyword~(Listing~\ref{vocabDSL},  line~\ref{vocab:sensors-start}). Each sensor produces one or more sensor measurements along with the data-types specified in the data structure~(Listing~\ref{vocabDSL}, lines~\ref{vocab:tempstruct-start}-\ref{vocab:structs-end}), declared using the \texttt{generate} keyword~(Listing~\ref{vocabDSL}, line~\ref{vocab:periodicsensors-generate}). We categorize sensors into the following three types:
\begin{itemize}
\item \textbf{Periodic sensor:} It samples results every \texttt{d} seconds for a duration of \texttt{k} seconds. For instance, a temperature sensor generates a temperature measurement of \texttt{TempStruct} type~(Listing~\ref{vocabDSL}, lines~\ref{vocab:tempstruct-start}-\ref{vocab:structs-end}). It samples data every \texttt{1} second  for next \texttt{6} minutes~(Listing~\ref{vocabDSL}, line~\ref{vocab:periodicsensors-sampleperiod}).

\item \textbf{Event-driven sensor:} It produces data when the event condition is met. For instance, a smoke sensor generates \texttt{smokeMeasurement} when \texttt{smokeValue > 650 PPM} (Listing~\ref{vocabDSL}, lines~\ref{vocab:eventdrivensensor-generate-start}-\ref{vocab:eventdrivensensor-onCondition-end}). 

\item \textbf{Request-based sensor:} It responds its results only if it is requested. For instance, the
\texttt{YahooWeatherService} provides temperature value of a location given by a \texttt{locationID}~(Listing~\ref{vocabDSL}, lines~\ref{vocab:requestbasedsensor-start}-\ref{vocab:requestbasedsensor-end}).

\end{itemize}

{Tag} It is a physical object that can be applied to or incorporated into 
Entities of Interest for the purpose of identification. It is read through a reader. 
For instance, a \texttt{BadgeReader} read a user's badge and generates \texttt{badgeDetectedStruct} measurement~(Listing~\ref{vocabDSL}, lines~\ref{vocab:badgereader-tag-start}-\ref{vocab:badgereader-tag-end}).

{Actuators}  A set of actuators is declared using the \texttt{actuators} keyword~(Listing~\ref{vocabDSL}, line~\ref{vocab:actuators-start}). Each actuator has one or more actions declared using the \texttt{action} keyword. An action may take inputs specified as parameters (Listing~\ref{vocabDSL}, line~\ref{vocab:actuators-end}). For instance, a heater may have two actions (e.g., switch off, set heater), illustrated in Listing~\ref{vocabDSL}, lines~\ref{vocab:heater}-\ref{vocab:actuators-end}. 

{Storage} A set of storage is declared using the \texttt{storages} keyword~(Listing~\ref{vocabDSL}, line~\ref{vocab:storages-start}). 	A retrieval from  the storage requires a parameter, specified using the  \texttt{accessed-by}  keyword~(Listing~\ref{vocabDSL}, line~\ref{vocab:storages-generate-end}). The data insertion into the storage is performed by the \texttt{action} keyword with parameter. For instance,  a user's profile is accessed from storage by a \texttt{badgeID} and inserted by invoking an action~(Listing~\ref{vocabDSL}, line~\ref{vocab:storages-end}).

\lstset{emph={requestBasedSensors, sensors, String, generate, actuators, structs, double,  action, resources, long, storages, accessed, by, periodicSensors, sample, period, for, eventDrivenSensors, onCondition, true, tags, =}, emphstyle={\color{blue}\bfseries\emph}, caption={Code snippet of domain spec.}, escapechar=\#, 
 label=vocabDSL}	
 
\begin{lstlisting}
1 structs:   #\label{vocab:structs-start}#
2 TempStruct               #\label{vocab:tempstruct-start}#
3    tempValue : double;
4    unitOfMeasurement : String;	 #\label{vocab:structs-end}#
5 resources:
6 sensors:    #\label{vocab:sensors-start}#
7    periodicSensors: #\label{vocab:periodicsensors-start}#
8     TemperatureSensor
9        generate  tempMeasurement : TempStruct; #\label{vocab:periodicsensors-generate}#    
10        sample period 1000 for 6000000; #\label{vocab:periodicsensors-sampleperiod}# 
11    eventDrivenSensors:  #\label{vocab:eventdrivensensors-start}#
12     SmokeDetector                
13        generate smokeMeasurement: SmokeStruct;  #\label{vocab:eventdrivensensor-generate-start}#
14        onCondition smokeValue>650PPM; #\label{vocab:eventdrivensensor-onCondition-end}#
15    requestBasedSensors: #\label{vocab:requestbasedsensor-start}#
16        YahooWeatherService   
17            generate weatherMeasurement: TempStruct accessed
18					by locationID: String;#\label{vocab:requestbasedsensor-end}#
19  tags:   #\label{vocab:tag-start}#
20      BadgeReader             #\label{vocab:badgereader-tag-start}#     
21        generate badgeDetectedStruct: BadgeStruct; #\label{vocab:badgereader-tag-end}#  
22  actuators:      #\label{vocab:actuators-start}#
23      Heater       #\label{vocab:heater}#
24        action Off();    #\label{vocab:action-end}#
25        action SetTemp(setTemp:TempStruct);	 #\label{vocab:actuators-end}#
26  storages:        #\label{vocab:storages-start}#
27     ProfileDB  #\label{vocab:profile-start}#       
28        generate profile: TempStruct accessed-by badgeID:
29					          String; #\label{vocab:storages-generate-end}#
30        action InsertProfileData(profileData:ProfileStruct); #\label{vocab:storages-end}#
\end{lstlisting}


\section{Architecture Specification}\label{sec:arch}

\begin{figure}[!ht]
\centering
\includegraphics[width=1.0\textwidth]{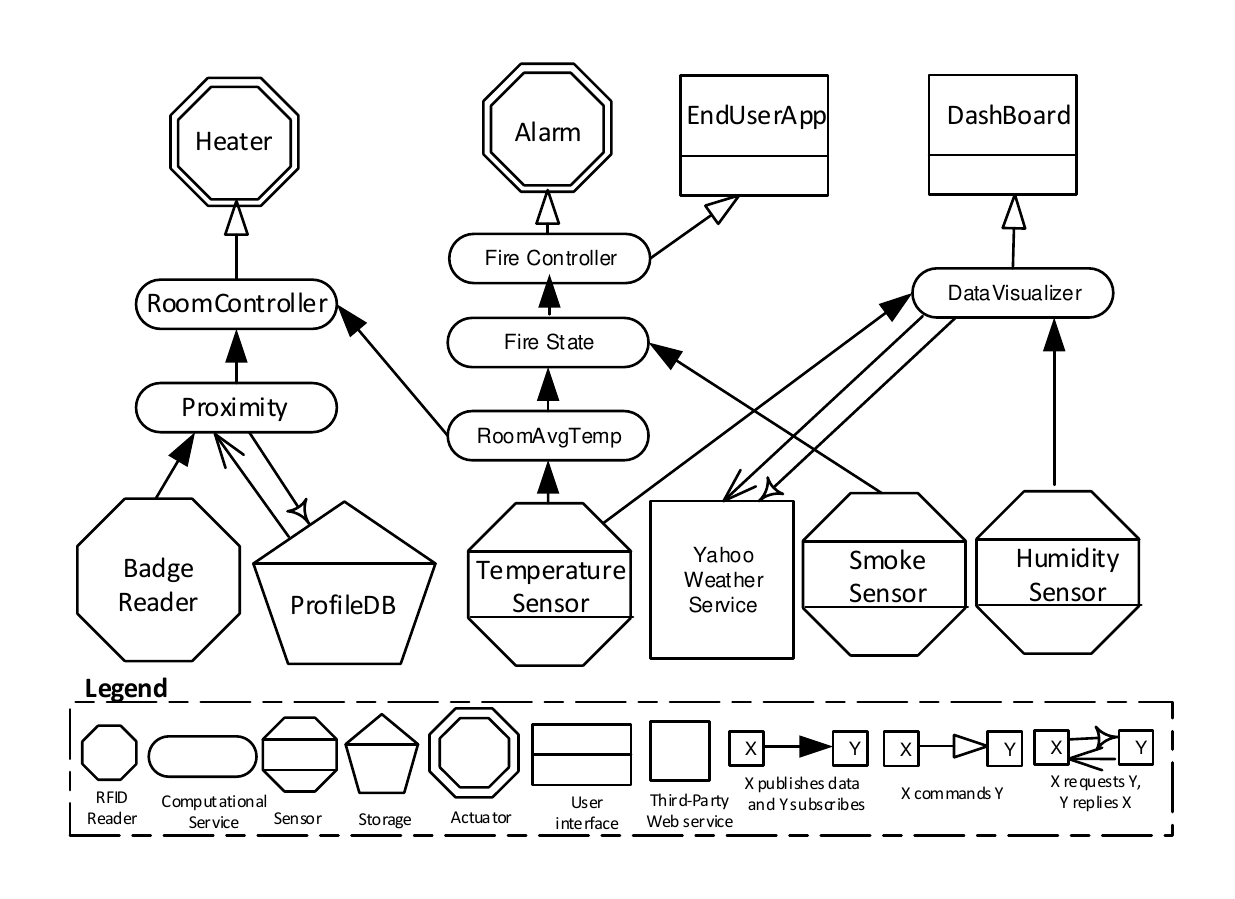}
\caption{Dataflow of Smart Home Application.}
\label{fig:dataflowOfScenario}
\end{figure}

Referring the concepts defined in the domain specification, the software designer  specifies an architecture specification. It is described as a set of computational services. It consists of two types of computational services: (1)  {\em Common} components specify common operations (e.g., \texttt{average}, \texttt{count}, and \texttt{sum}) in the application logic. For instance, \texttt{RoomAvgTemp} component consumes 5 
temperature measurements (Listing~\ref{list:archdsl}, line~\ref{arch-roomavgtemp-consume}), apply average by sample operation~(Listing~\ref{list:archdsl}, line~\ref{arch-roomavgtemp-compute}), and generates room average temperature measurements (Listing~\ref{list:archdsl}, line~\ref{arch-roomavgtemp-generate}). (2) {\em Custom} specifies an application-specific logic. For instance, \texttt{Proximity} component is a custom
component (in Figure~\ref{fig:dataflowOfScenario}) that coordinates events from \texttt{BadgeReader} with the content from \texttt{ProfileDB}.

Each computational service is described by a set of inputs and outputs. We describe them below:

\fakeparagraph{Consume and Generate} They represent a set of subscriptions~(or consume) and publications~(or generate) expressed by a computational service. For instance, \texttt{RoomAvgTemp} consumes \texttt{tempMeasurement} (Listing~\ref{list:archdsl}, line~\ref{arch-roomavgtemp-consume}), calculates an average temperature (Listing~\ref{list:archdsl}, line~\ref{arch-roomavgtemp-compute})  and generates \texttt{roomAvgTempMeasurement}~(Listing~\ref{list:archdsl}, line~\ref{arch-roomavgtemp-generate}).

\fakeparagraph{Request} It is a set of requests issued by a computational service to retrieve data. For instance, to access user's profile, \texttt{Proximity}~(Listing~\ref{list:archdsl}, line~\ref{arch-proximity:proximity-request}) sends a request message containing profile information as an access parameter to storage \texttt{ProfileDB}~(Listing~\ref{vocabDSL}, lines~\ref{vocab:profile-start}-\ref{vocab:storages-end}).

\fakeparagraph{Command} 
It is a set of commands,  issued by a computational service to trigger actions.
The software designer can pass arguments to a command depend on action signature. 
For instance, the \texttt{RoomController} issues a \texttt{SetTemp} command(Listing~\ref{list:archdsl}, line~\ref{arch-controller:controller-command}) with a \texttt{settemp} as an argument to \texttt{Heater}~(Listing~\ref{vocabDSL}, 
lines~\ref{vocab:heater}-\ref{vocab:actuators-end}).

Figure~\ref{fig:dataflowOfScenario} shows a layered architecture of the application discussed in 
Section~\ref{sec:case-study}. Listing~\ref{list:archdsl} 
describes a part of Figure~\ref{fig:dataflowOfScenario}. It revolves around the actions of the \texttt{Proximity} service (Listing~\ref{list:archdsl}, lines 
\ref{arch-proximity:proximity-start}-\ref{arch-proximity:proximity-generate}),
which coordinates events from the \texttt{BadgeReader} with the
content of \texttt{ProfileDB} storage service.  To do so,
the \texttt{Proximity} composes information from two sources, one for
badge events, and one for requesting the
user's temperature profile from \texttt{ProfileDB}. The output of the \texttt{Proximity} and \texttt{RoomAvgTemp} are
consumed by the \texttt{RoomController} service~(Listing~\ref{list:archdsl},
lines~\ref{arch-controller:regulatetemp}-\ref{arch-controller:proximity}).
This service is responsible for taking decisions that are carried
out by invoking command (Listing~\ref{list:archdsl}, line~\ref{arch-controller:controller-command}).

\lstset{emph={generate, computationalServices, command, from, to, 
consume, request, Custom, Common, COMPUTE, AVG_BY_SAMPLE }, emphstyle={\color{blue}\bfseries\emph} }

\lstset{caption={A code snippet of architecture spec.}, escapechar=\#, label=list:archdsl}
\begin{lstlisting}
1 computationalServices:
2 Common:  #\label{arch-inbuit-component}#
3	RoomAvgTemp#\label{arch-roomavgtemp}#
4	   consume tempMeasurement from TemperatureSensor;#\label{arch-roomavgtemp-consume}# 
5	   COMPUTE(AVG_BY_SAMPLE,5);#\label{arch-roomavgtemp-compute}#
6	   generate roomAvgTempMeasurement:TempStruct;#\label{arch-roomavgtemp-generate}#
7 Custom: #\label{arch-custom-component}#
8	Proximity #\label{arch-proximity:proximity-start}#
9	   consume badgeDetected from BadgeReader;#\label{arch-proximity:proximity-badgedetected-consume}#
10	   request profile(badgeID) to ProfileDB;#\label{arch-proximity:proximity-request}#
11	   generate tempPref: UserTempPrefStruct; #\label{arch-proximity:proximity-generate}#
12  RoomController #\label{arch-controller:regulatetemp-start}#
13	   consume roomAvgTempMeasurement from RoomAvgTemp; #\label{arch-controller:regulatetemp}#
14	   consume tempPref from Proximity; #\label{arch-controller:proximity}#
15	   command SetTemp(setTemp) to Heater;  #\label{arch-controller:controller-command}# 
\end{lstlisting}

\section{User-interactions Specification}\label{sec:spec-user-interactions}

The user interactions specification defines {\em what} interactions are required by an application. We design a set of abstract \emph{interactors} that denotes data exchange between an application and a user. The following are abstract interactors that are specified using User Interaction Language~(UIL). 


{Command} It denotes information flow from a user to an application. It is declared using the \texttt{command} keyword~(Listing~\ref{userInteractionDSL}, lines~\ref{ui:userinteraction-command}-\ref{ui:userinteraction-commandwithparameter}). For instance, a user can control an actuator by triggering a \texttt{Off()} command~(Listing~\ref{userInteractionDSL}, line~\ref{ui:userinteraction-command}). Command could be parametrized too. For instance, a user can set a temperature of heater by triggering a \texttt{SetTemp()} command~(Listing~\ref{userInteractionDSL}, line~\ref{ui:userinteraction-commandwithparameter}). 

{Notify} It denotes information flow from an application to a user. For instance, an application notifies a user in case of fire. It is declared using the \texttt{notify} keyword~(Listing~\ref{userInteractionDSL}, lines~\ref{ui:enduserapp-notify}-\ref{ui:enduserapp-notify1}). The application notifies users with the fire information specified in the data structure~(Listing~\ref{userInteractionDSL}, lines~\ref{ui:userinteraction-FireStateStruct-Start}-\ref{ui:userinteraction-FireStateStruct-end}).

{Request} It denotes information flow round-trip between an application and a user, but initiated from the user. For instance, a user can retrieve data by requesting to data storage. This is declared using the \texttt{request} keyword. 
The user requests \texttt{ProfileDB} for data~(Listing~\ref{userInteractionDSL}, line~\ref{ui:userinteraction-request}) and the \texttt{ProfileDB} responses
back with data to the user.

\lstset{emph={resources, command, request, action, to, userInteractions, EndUserApp, notify, DashBoard, structs, double, String, from}, emphstyle={\color{blue}\bfseries\emph}, caption={Code snippet of user interaction spec.}, escapechar=\#, 
 label=userInteractionDSL}	
 
\begin{lstlisting}
1 structs:
2 FireStateStruct    #\label{ui:userinteraction-FireStateStruct-Start}#
3  fireValue:String;  
4  timeStamp:String;	#\label{ui:userinteraction-FireStateStruct-end}#    
5 resources :
6 userInteractions:   #\label{ui:userinteraction-keyword}#
7  EndUserApp    #\label{ui:userinteraction-name}#	
8   command Off() to Heater;  #\label{ui:userinteraction-command}#
9   command SetTemp(setTemp) to Heater;  #\label{ui:userinteraction-commandwithparameter}#
10   request profile to ProfileDB;	  #\label{ui:userinteraction-request}#		
11   notify FireNotify(fireNotify:FireStateStruct) #\label{ui:enduserapp-notify}#
12					from FireController;  #\label{ui:enduserapp-notify1}#
13  DashBoard    #\label{ui:userinteraction-dashboard}#
14   notify DisplaySensorMeasurement(sensorMeasurement:Visu-
15			alizeStruct) from DisplayController;#\label{ui:userinteraction-dashboard-notify}# 
\end{lstlisting}

\section{Deployment Specification}\label{sec:network}
The deployment specification describes a device and its properties in a
target deployment. It includes properties such as \emph{location} that defines where 
a device is deployed, \emph{resource} defines component(s) to be deployed 
on a device, \emph{language-platform} is used to generate an appropriate package for
a device, \emph{protocol} specifies a run-time system installed on a device to interact with other devices. Listing~\ref{networkspecification} shows a small code snippet to
illustrate these concepts. \texttt{TemperatureMgmt-Device-1} is located in room\#1~(line~\ref{deployd1:room-location}), \texttt{TemperatureSensor} and \texttt{Heater} are attached with the device~(line~\ref{deployd1:android-resources}) and the device driver code for these two components is in NodeJS~(line~\ref{deployd1:android-platform}), MQTT runtime system is installed on \texttt{TemperatureMgmt-Device-1} device~(line~\ref{deployd1:android-protocol}). 

A storage device contains the \textbf{database} field that specifies the installed database. This field is used to select an appropriate storage driver.
For instance, \texttt{DatabaseSrv--- \\ Device-2} (lines~\ref{DatabaseSrv-Device-2}-\ref{deployd2:javaSE-mysql}) runs \texttt{ProfileDB} component implemented in \texttt{MySQL} database. 

When an application is deployed, the network manger decides what user interaction components need to be deployed on devices. The announcement of user interaction components in the deployment specification is now an integral part of the  specification. This announcement is used to deploy the generated UI framework and UI code~(Step~\circled{5} in Figure~\ref{developmentcycle}) on a device.  Listing~\ref{networkspecification} illustrates a code snippet to describe this concept.  \texttt{SmartPhone-Device-3} announces that \texttt{EndUserApp}~(specified in user interaction specification) should be deployed on \texttt{Android} device~(Lines~\ref{deployd3:android-start}-\ref{deployd3:android-platform}).

\lstset{emph={resources, softwarecomponents,  devices, location, protocol, description,database, language,platform}, emphstyle={\color{blue}\bfseries\emph},caption={Code snippet of deployment spec.}, label= networkspecification}
\begin{lstlisting}
1 devices: 
2    TemperatureMgmt-Device-1:
3        location:    #\label{deployd1:android-start}#
4            Room: 1; #\label{deployd1:room-location}#
5        resources: TemperatureSensor, Heater; #\label{deployd1:android-resources}#
6        language-platform: NodeJS;  #\label{deployd1:android-platform}#
7        protocol : mqtt; #\label{deployd1:android-protocol}#
8    DatabaseSrv-Device-2:  #\label{DatabaseSrv-Device-2}#
9        resources: ProfileDB;   #\label{DatabaseSrv-Device-2:profileDB}#
10        database: MySQL;   #\label{deployd2:javaSE-mysql}#
11    SmartPhone-Device-3:   #\label{deployd3:android-start}#
12        resources: EndUserApp; #\label{deployd3:android-resources}#
13        language-platform: Android;    #\label{deployd3:android-platform}#
14      ...  
        
\end{lstlisting}

\section{Chapter Summary}
\hspace*{0.2in} This chapter presents an implementation of proposed framework called as IoTSuite- A toolsuite for programming Internet of Things, Service and People~(IoTSP) applications. It separates different aspect of application development into domain, functional, platform, and deployment with integration of high-level languages to specify these aspects. To specify high-level specifications, we presented editor with the use of Xtext. The proposed framework consists of IoTSuite compiler to parse these high-level specifications. It takes high-level specification as input and generates source code as output.

\chapter{Application development using IoTSuite}
\label{IoTSuite}
This chapter describes each step of Internet of Things, Service, and People~(IoTSP) application development process using IoTSuite. Application development using IoTSuite is a multi-step process and focuses on design, implement, and deployment phases to develop IoTSP applications~\cite{chauhan2016user}.\\
\hspace*{0.2in}We take an example of smart home application~(Refer Figure~\ref{fig:dataflowOfScenario}) to demonstrate application development using IoTSuite.

\section{Create IoTSuite Project} \label{sec:project}
To develop a smart home application using IoTSuite, developers carry out the following steps.
	\subsubsection{Download IoTSuite-Eclipse-Plugin}
	\begin{itemize}
	\item Download IoTSuite-Eclipse-Plugin\footnote{IoTSuite-Eclipse-Plugin is available at \url{https://github.com/chauhansaurabhb/IoTSuite-Eclipse-Plugin}}.
	\item Go to Downloads folder.
	\item Extract downloaded IoTSuite-Eclipse-Plugin into C:$\backslash$ drive.
	\item Go to C:$\backslash$IoTSuite-Eclipse-Plugin$\backslash$Template.
	\item Extract IoTSuite-TemplateV9-master.rar in to same directory~(C:$\backslash$IoTSuite-Eclipse-Plugin$\backslash$Template).
		\end{itemize}
 \section{Open IoTSuite-Eclipse}
	\begin{itemize}	
	\item Go to C:$\backslash$IoTSuite-Eclipse-Plugin.
	\item Right click on IoTSuite-Eclipse.exe as shown in Figure~\ref{fig:openeclipse} (Step 1), and select Open option (Step 2).
	\begin{figure}[!ht]
\centering
\includegraphics[width=0.9\textwidth]{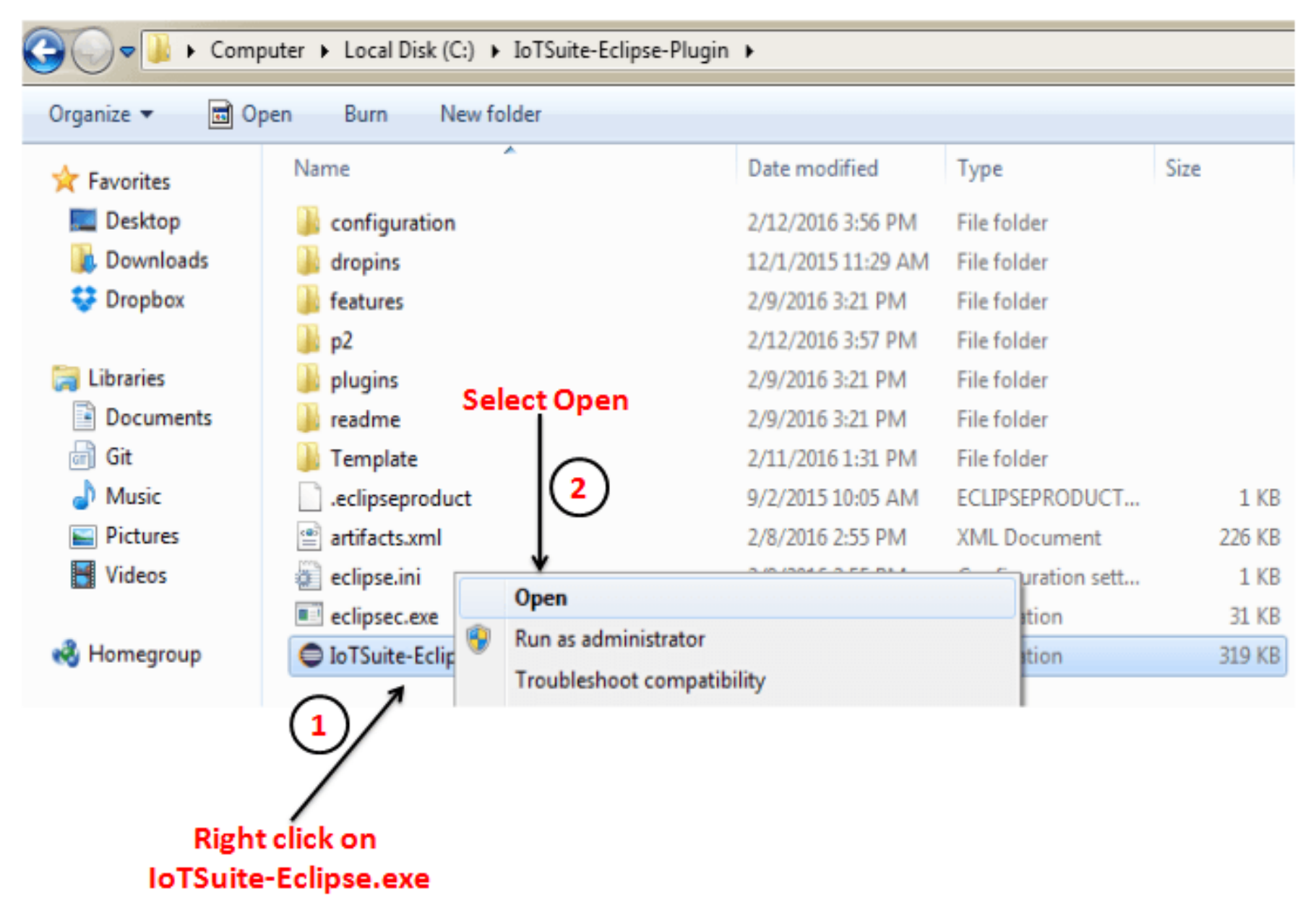}
\caption{Open IoTSuite-Eclipse}
\label{fig:openeclipse}
\end{figure}

\item When IoTSuite-Eclipse is open, usually developers have to select workspace but here developers don’t need to go to C:$\backslash$IoTSuite-Eclipse-Plugin$\backslash$Template in order to select workspace, it is already provided as default workspace.
\item Check on “Use this as the default and do not ask again”  (Step 1), and click on OK button (Step 2)  (Refer Figure~\ref{fig:workspace}).

\begin{figure}[!ht]
\centering
\includegraphics[width=0.9\textwidth]{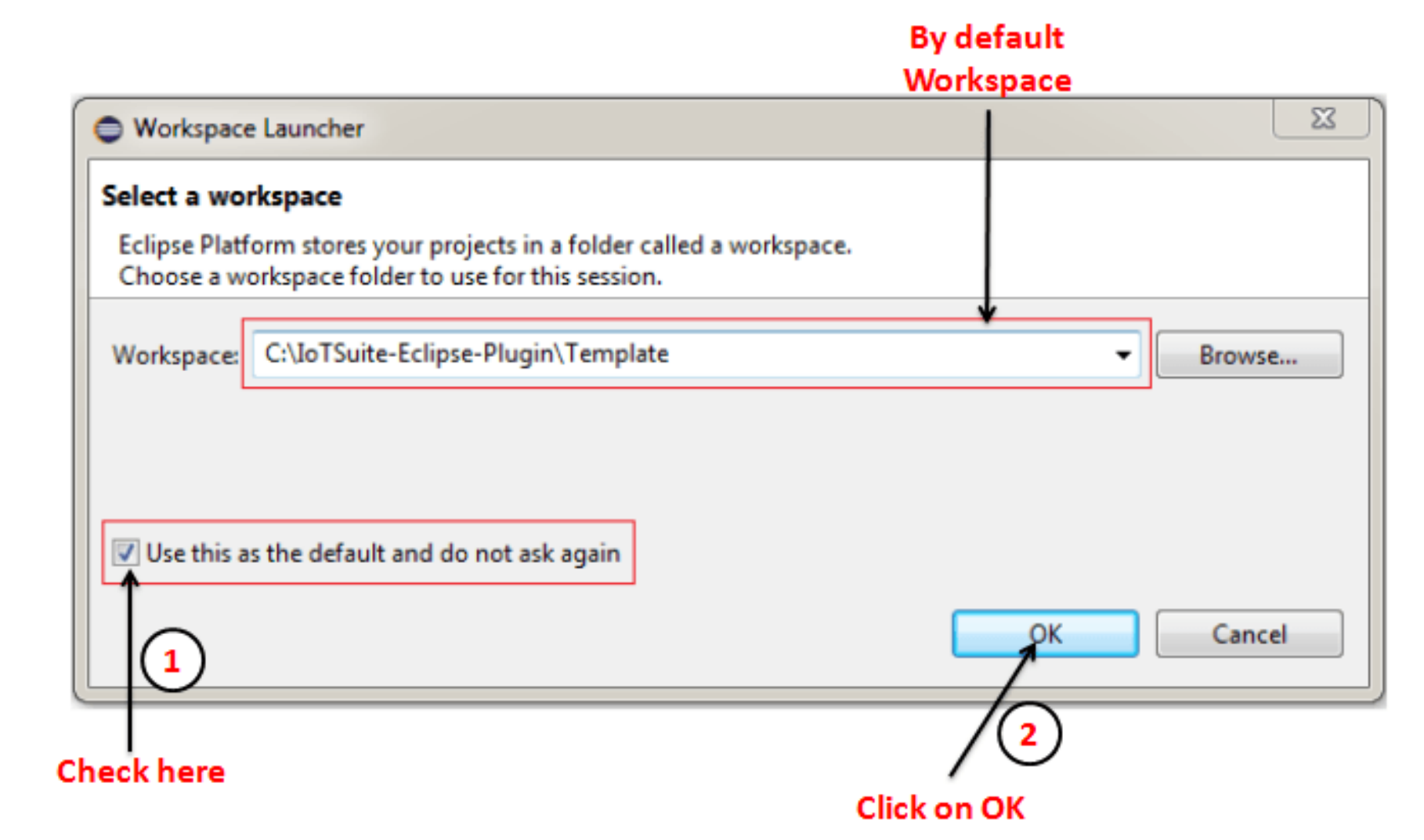}
\caption{Select default Workspace}
\label{fig:workspace}
\end{figure}
\end{itemize}
\newpage	
	\section{Create a new IoTSuite Project}	
\begin{itemize}
	\item To create a new IoTSuite project using IoTSuite-Eclipse, click on \textbf{File>New>Pr-oject..}, choose IoTSuite Project from the list~(Refer Figure~\ref{fig:iotproject}).
\begin{figure}[!ht]
\centering
\includegraphics[width=0.7\textwidth]{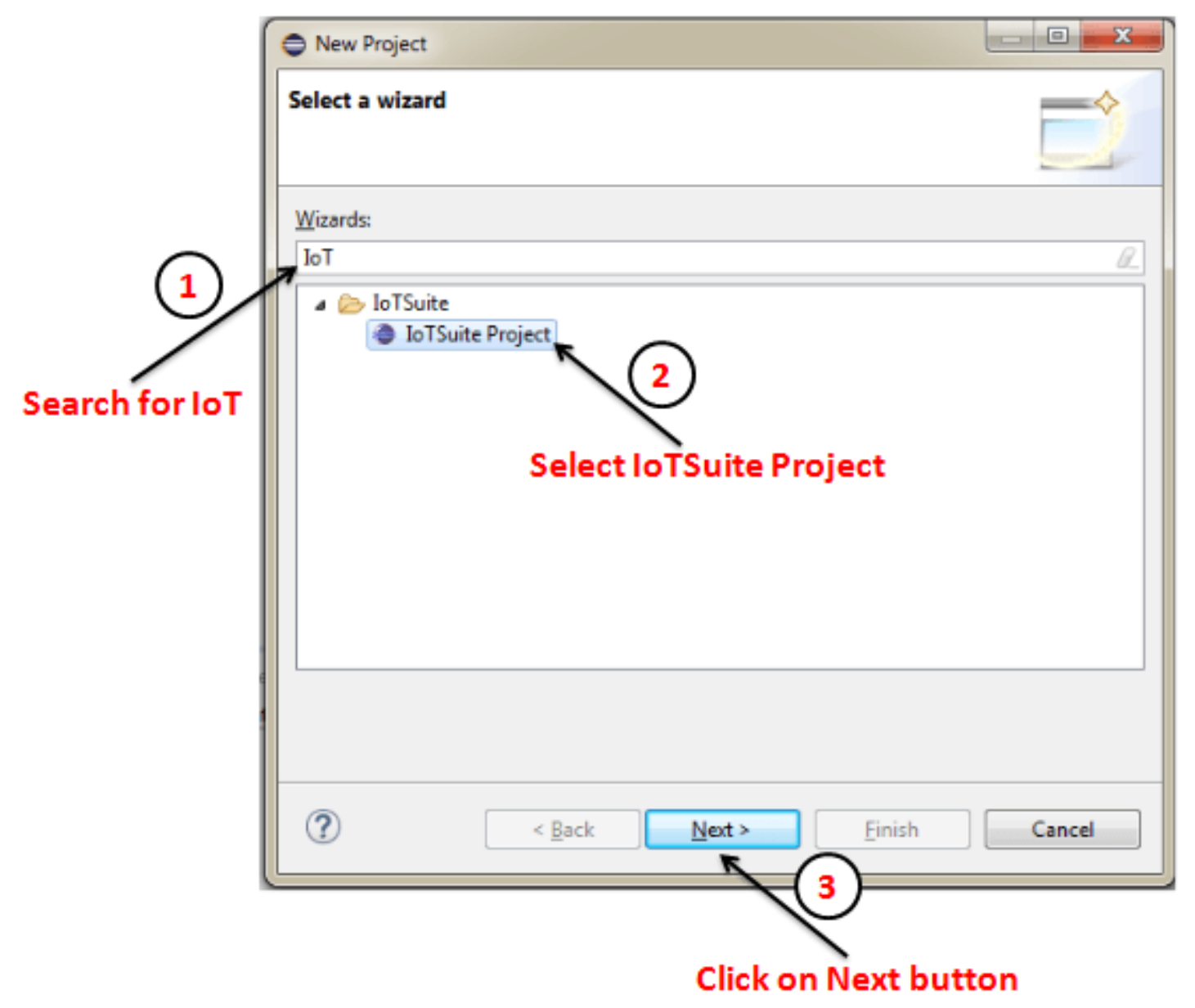}
\caption{IoTSuite Project Creation Wizard (1/2)}
\label{fig:iotproject}
\end{figure}

\newpage 
\item In the next screen, write the project name as IoTSuiteSpecification~(Step 1), check Use default location~(Step 2), and click on Finish button~(Step 3)~(Refer Figure~\ref{fig:projectname}).

\begin{figure}[!ht]
\centering
\includegraphics[width=0.8\textwidth]{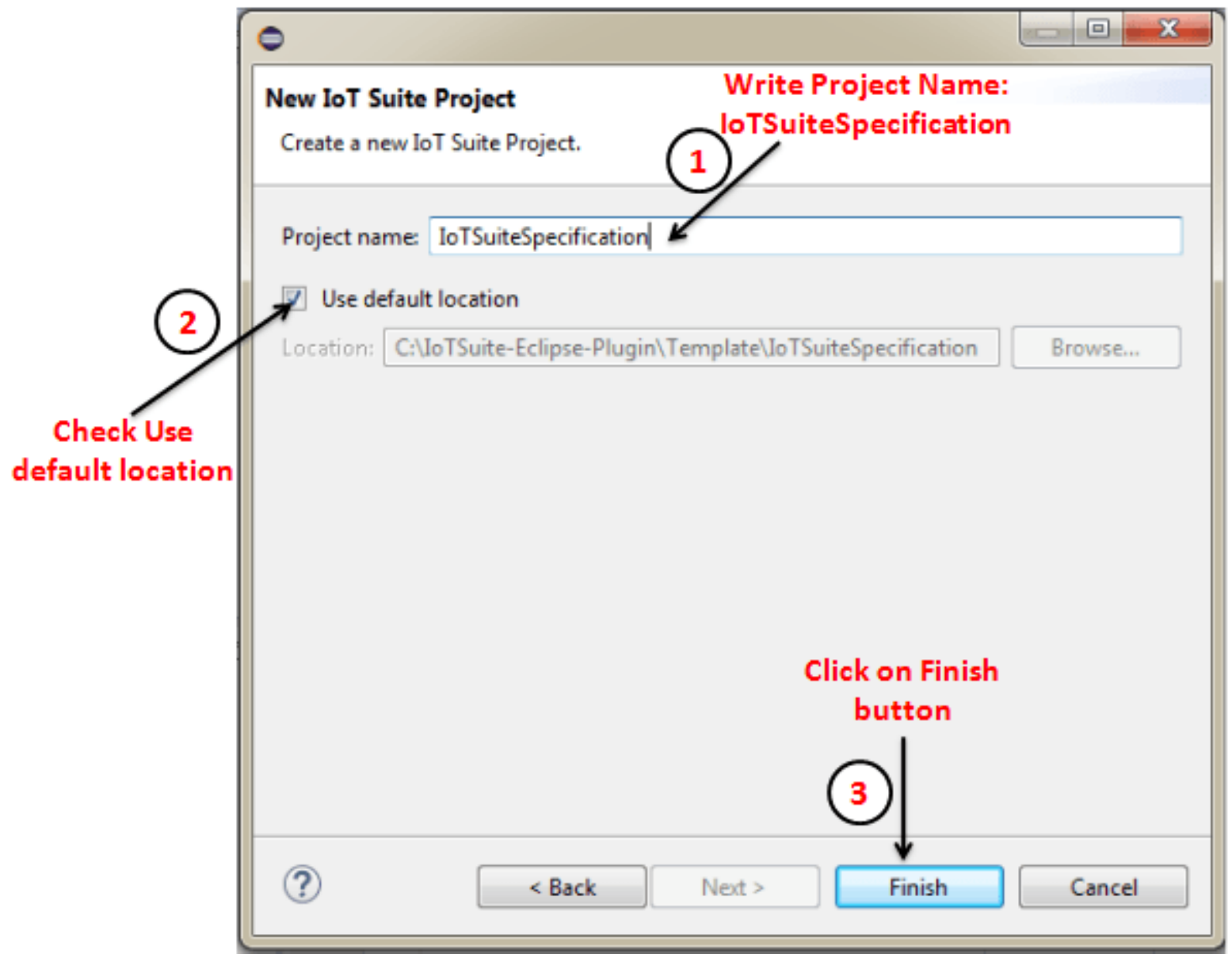}
\caption{IoTSuite Project Creation Wizard (2/2)}
\label{fig:projectname}
\end{figure}

As result, new IoTSuite project is created, and the corresponding IoTSuite specification files will be opened in Xtext mode. By default, they have predefined contents in order to guide developers. The structure of created project is shown in Figure~\ref{fig:iotprojectstructure}. Once the IoTSuiteProject is created, next step is to specify high-level specification~(Refer Section~\ref{high-level spec.}).
 
\begin{figure}[!ht]
\centering
\includegraphics[width=0.4\textwidth]{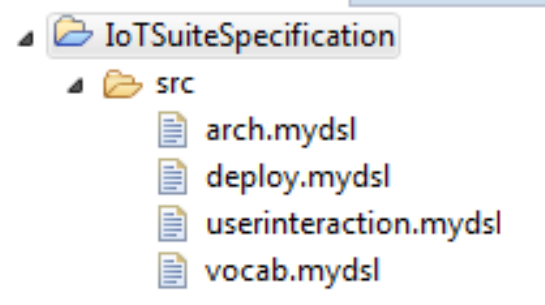}
\caption{IoTSuite Project Structure}
\label{fig:iotprojectstructure}
\end{figure}

\end{itemize}
\section{Specifying high-level specifications}\label{high-level spec.}
\begin{itemize}
\item  Specifying high-level specification using editor: To write these specifications, we present Xtext for a full fledged editor support with features such as \textit{syntax coloring, error checking, auto completion, rename re-factoring, outline view and code folding}.

\item We implemented \textit{Outline/Structure view} feature which is displayed on top most right side of the screen~(Refer~Figure~\ref{fig:outline}). It displays an outline of a file highlighting its structure. This is useful for quick navigation. 

\item As shown in Figure~\ref{fig:outline}, vocab.mydsl file contains large number of  structures, sensors,  and actuators, than from outline view by just clicking on  particular structures (e.g., TempStruct) it  navigates to TempStruct definition in the vocab.mydsl. So, developers don’t need to look in to entire file.

\begin{figure}[!ht]
\centering
\includegraphics[width=1.1\textwidth]{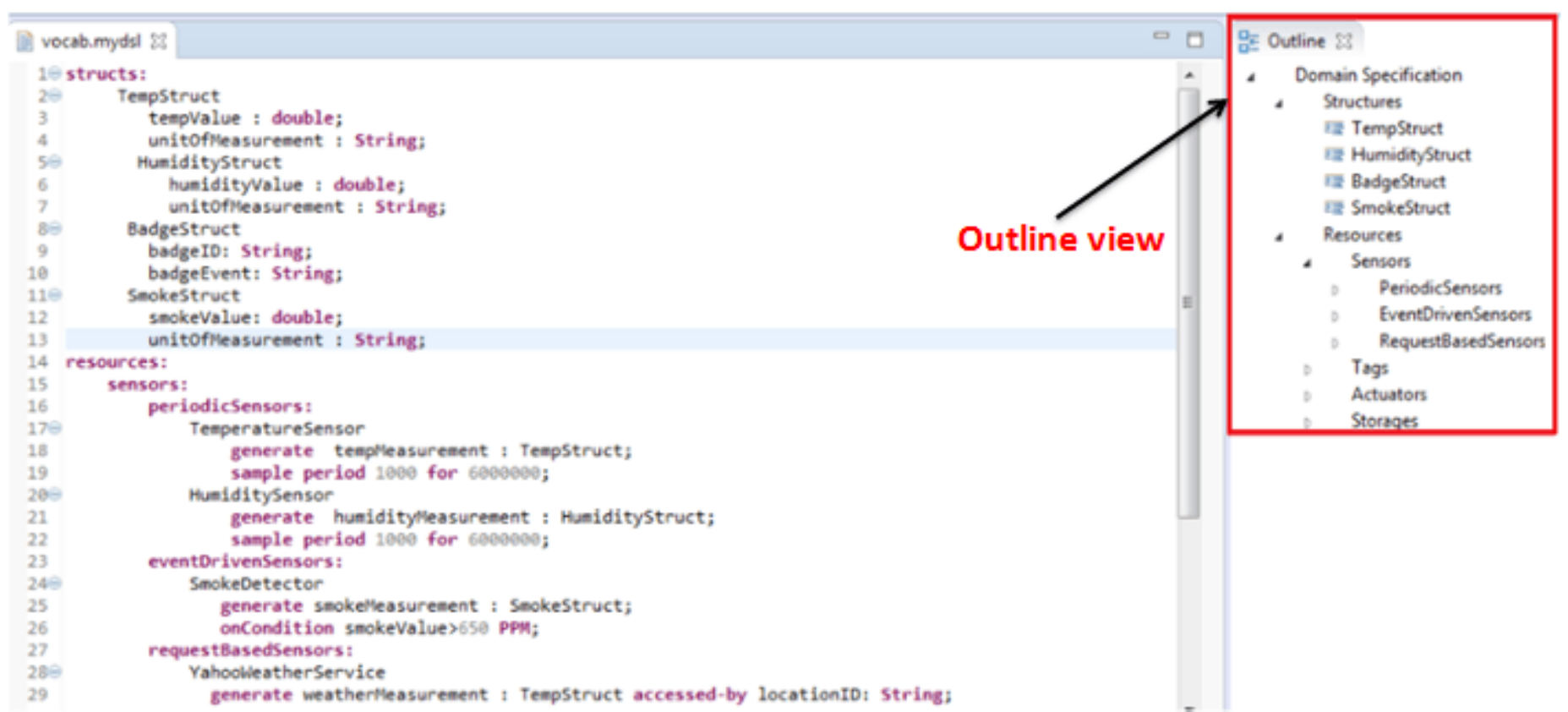}
\caption{IoTSuite editor feature: Outline view}
\label{fig:outline}
\end{figure}

\item Using \textit{syntax coloring} feature, keywords are appeared in colored text~(Refer~Figure~\ref{fig:coloring}).

\begin{figure}[!ht]
\centering
\includegraphics[width=0.7\textwidth]{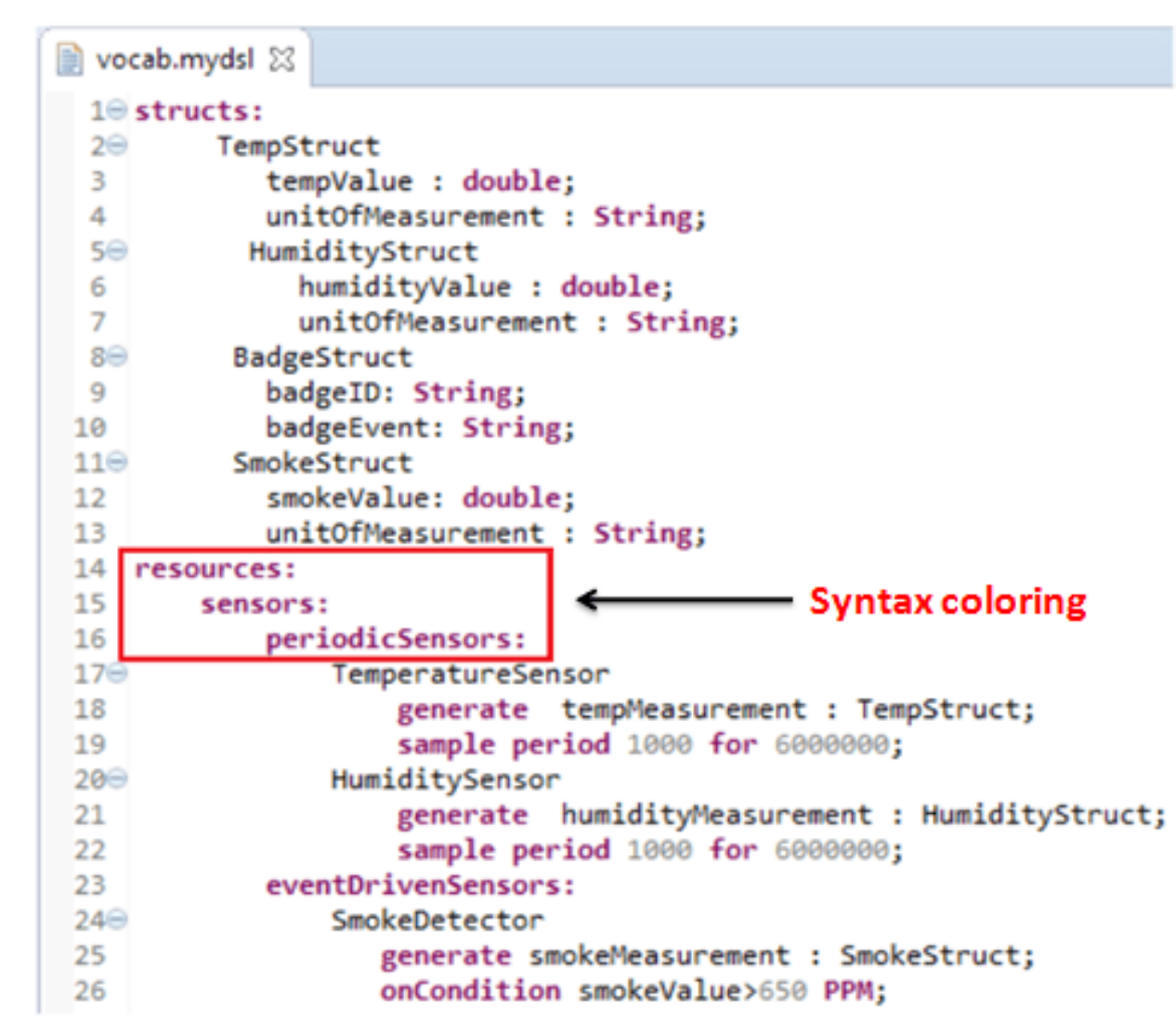}
\caption{IoTSuite editor feature: Syntax coloring}
\label{fig:coloring}
\end{figure}

\item Using \textit{code folding}, developer can collapse parts of a file that are not important for current task. In order to implement code folding, click on dashed sign located in left most side of the editor. When developer clicked on dashed sign, it will fold code and sign is converted to plus. In order to unfold code, again click on plus sign.
As shown in Figure~\ref{fig:folding}, the code is folded for TempStruct, HumidityStruct, and BadgeStruct.

\begin{figure}[!ht]
\centering
\includegraphics[width=0.8\textwidth]{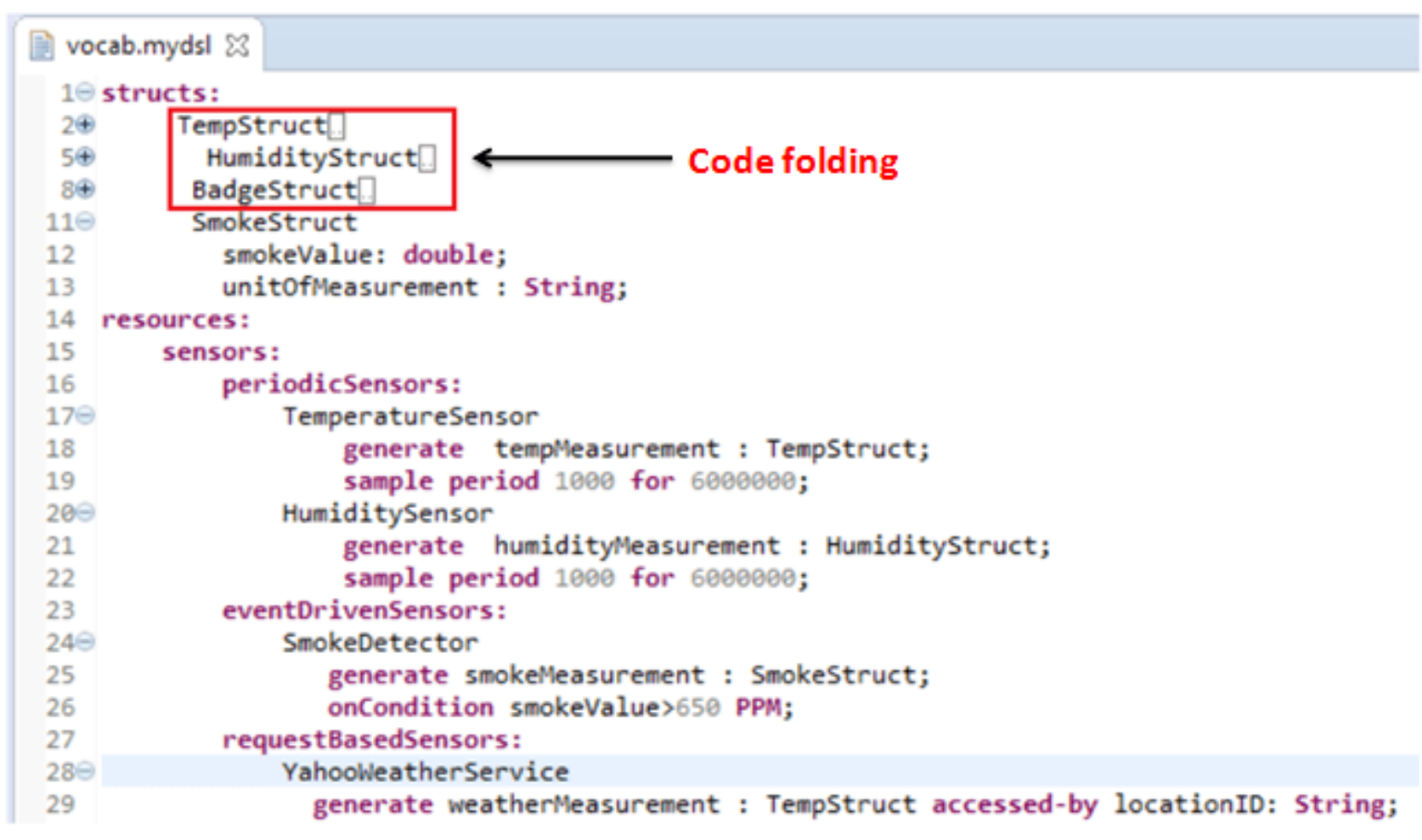}
\caption{IoTSuite editor feature: Code folding}
\label{fig:folding}
\end{figure}

\item The \textit{error checking} feature guides developer if any error is there.  General error in the file is marked automatically e.g., violation of the specified syntax or reference to undefined elements. Error checking indicates if anything is missing/wrong  in particular specification file.

\item \textit{Auto completion}  is used to speed up writing text in specification files. In order to use auto completion feature, developer need to press ctrl+space key at current cursor position,  so it will provide suggestion. Here~(Refer~Figure~\ref{fig:error}), in TemperatureSensor definition if developer write T and press ctrl+space than it will suggest developer to write TempStruct.

\begin{figure}[!ht]
\centering
\includegraphics[width=0.8\textwidth]{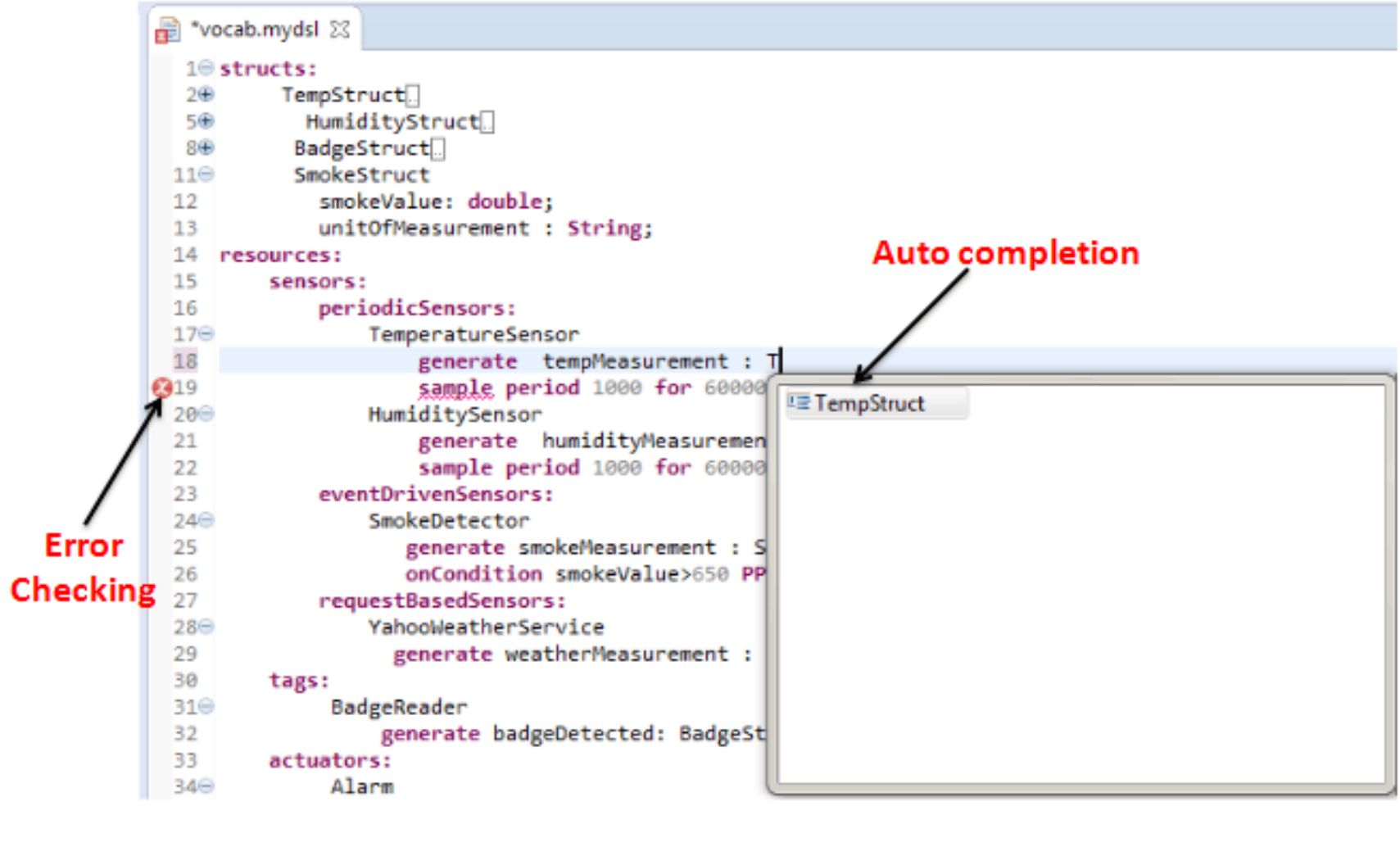}
\caption{IoTSuite editor features: Error checking \& Auto completion}
\label{fig:error}
\end{figure}

\begin{itemize}
	\item \emph{Specifying  domain specification} \label{vocab} 	
	
Developer specifies domain specification using vocab.mydsl file. To do this, double click on \textit{vocab.mydsl}~(Step 1), and specify domain specification using IoTSuite editor~(Step 2) as shown in Figure \ref{fig:vocab}.

\begin{figure}[!ht]
\centering
\includegraphics[width=0.8\textwidth]{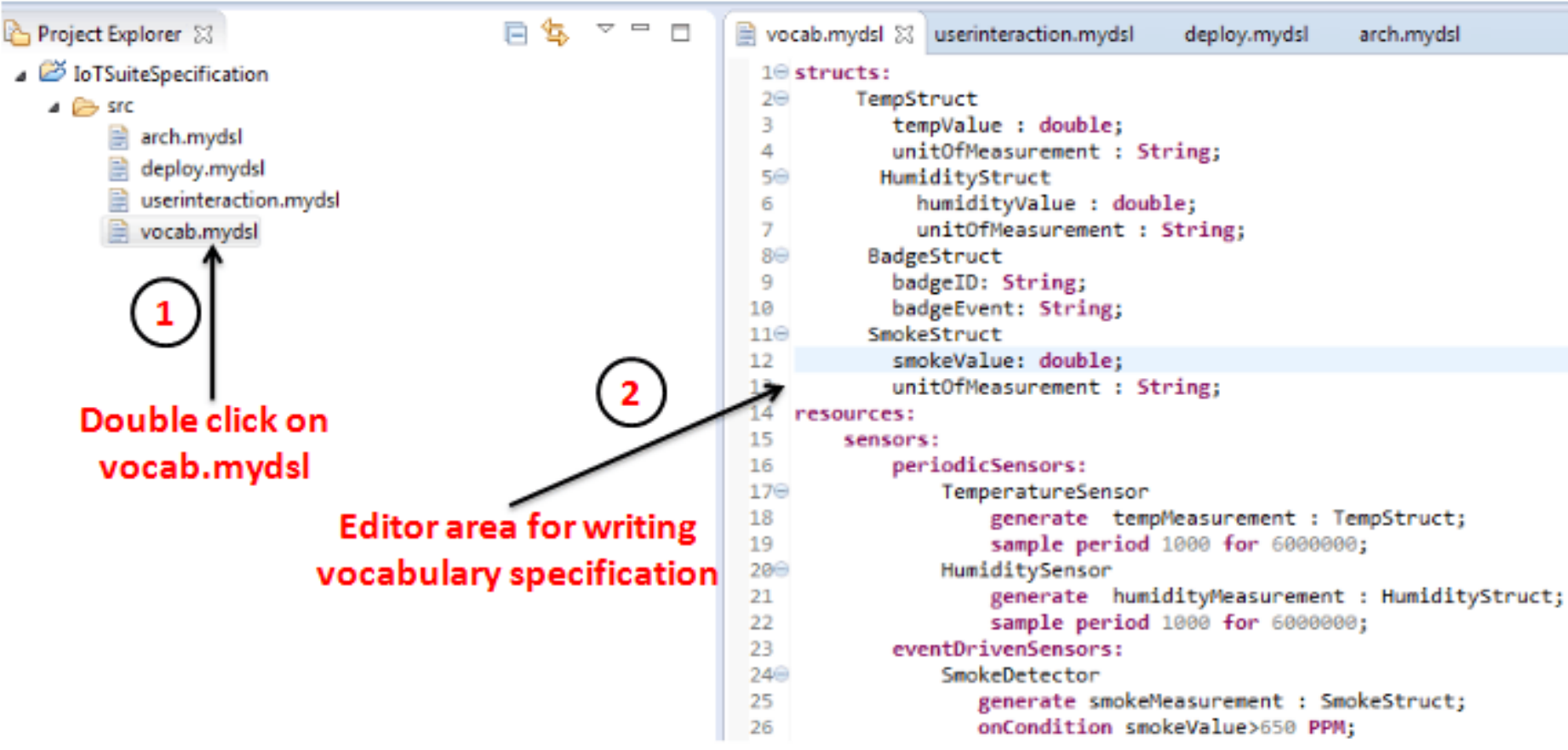}
\caption{Domain Specification}
\label{fig:vocab}
\end{figure}

\newpage

	\item \emph{Specifying arch specification} \\
Developer specifies architecture specification using arch.mydsl. To specify architecture specification, double click on  \textit{arch.mydsl}~(Step 1), and specify architecture specification using  IoTSuite editor~(Step 2) as shown in Figure~\ref{fig:arch}.

\begin{figure}[!ht]
\centering
\includegraphics[width=1.0\textwidth]{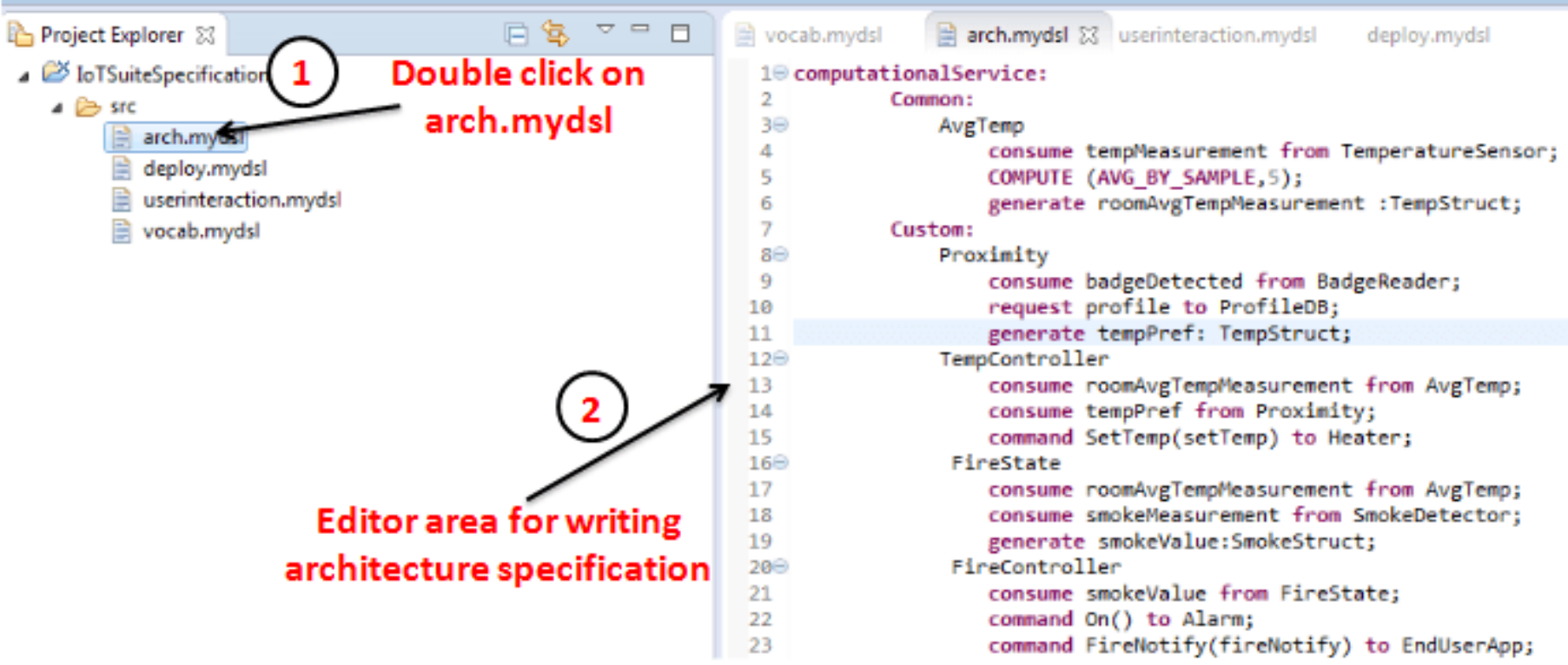}
\caption{Architecture Specification}
\label{fig:arch}
\end{figure}

\item \emph{Specifying user-interaction specification} \\

	To specify user-interaction specification, developer double clicks on \textit{userinteraction.mydsl}, and specify user-interaction specification using  IoTSuite editor~(Step 2) as shown in Figure~\ref{fig:ui}.
	
\begin{figure}[!ht]
\centering
\includegraphics[width=1.0\textwidth]{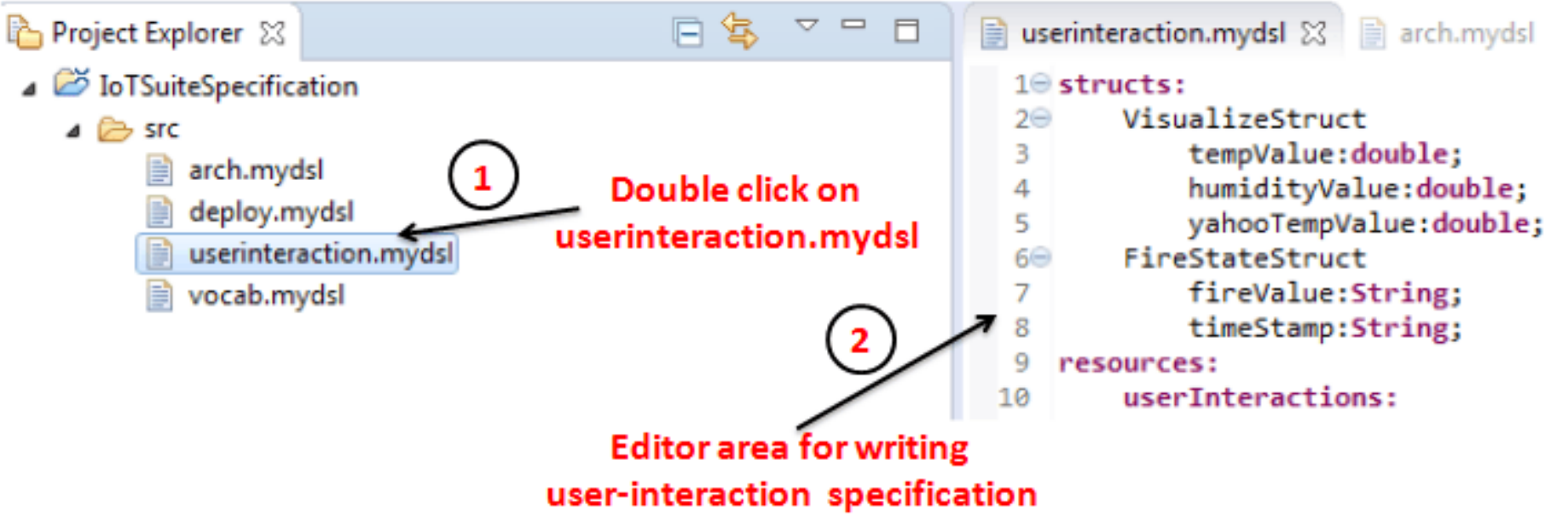}
\caption{User-Interaction Specification}
\label{fig:ui}
\end{figure}	

\newpage
	\item \emph{Specifying deployment specification}	\\
	Developer specifies deployment specification using deploy.mydsl. To do this, double click on \textit{deploy.mydsl}, and specify deployment specification using IoTSuite editor~(Step 2) as shown in Figure~\ref{fig:deploy}.

	\begin{figure}[!ht]
\centering
\includegraphics[width=1.0\textwidth]{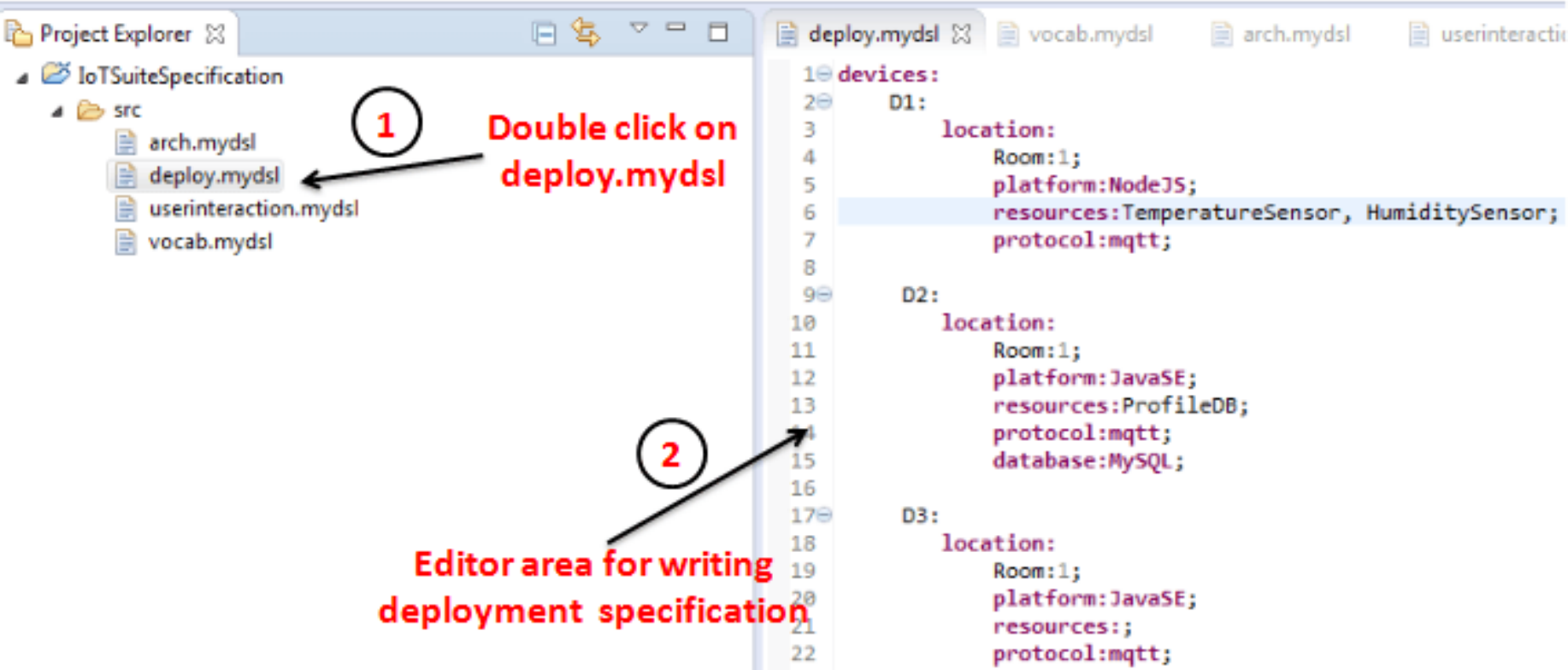}
\caption{Deployment Specification}
\label{fig:deploy}
\end{figure}		
\end{itemize}
\end{itemize}

\section{Compilation of IoTSuite Project} \label{sec:compile}
Now, developer compiles the high-level specification written in Section \ref{high-level spec.}. The compilation of high-level specification generates programming framework. Developer compiles high-level specification by performing following steps:
\subsubsection{Compilation of high-level specification}
\begin{itemize}
	\item \emph{Compilation of vocab specification}- Right click on vocab.mydsl file~(Step 1) and selecting "Compile Vocab"~(Step 2)~(Refer Figure~\ref{fig:compilevocab})  generates a vocabulary framework.
	\begin{figure}[!ht]
\centering
\includegraphics[width=0.5\textwidth]{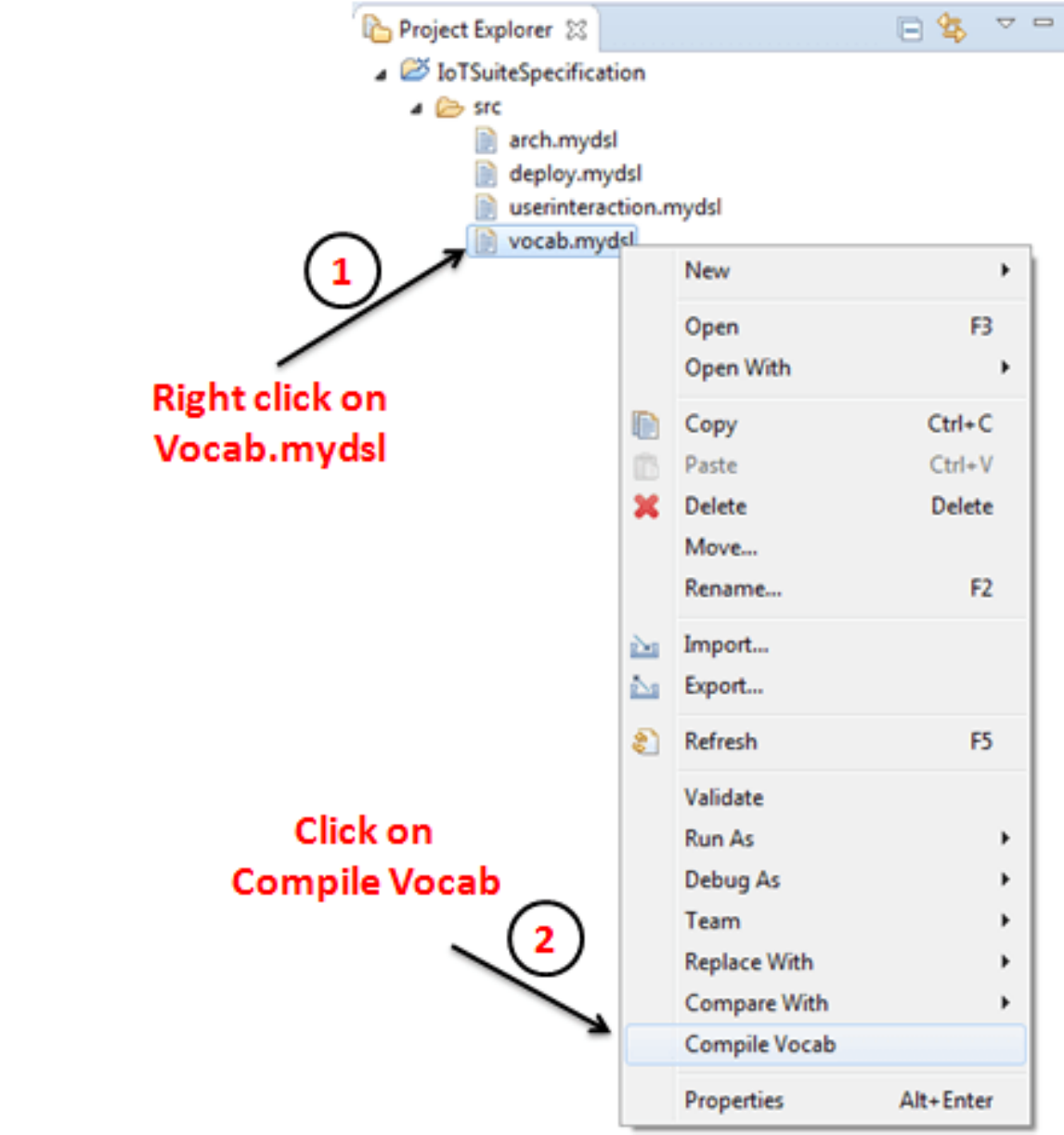}
\caption{Compilation of Vocab specification}
\label{fig:compilevocab}
\end{figure}

	\newpage
\item \emph{Compilation of architecture specification}- Right click on arch.mydsl file~(Step 1) and selecting "Compile Arch"~(Step 2)~(Refer Figure~\ref{fig:compilearch}) generates an architecture framework.
\begin{figure}[h]
\centering
\includegraphics[width=0.5\textwidth]{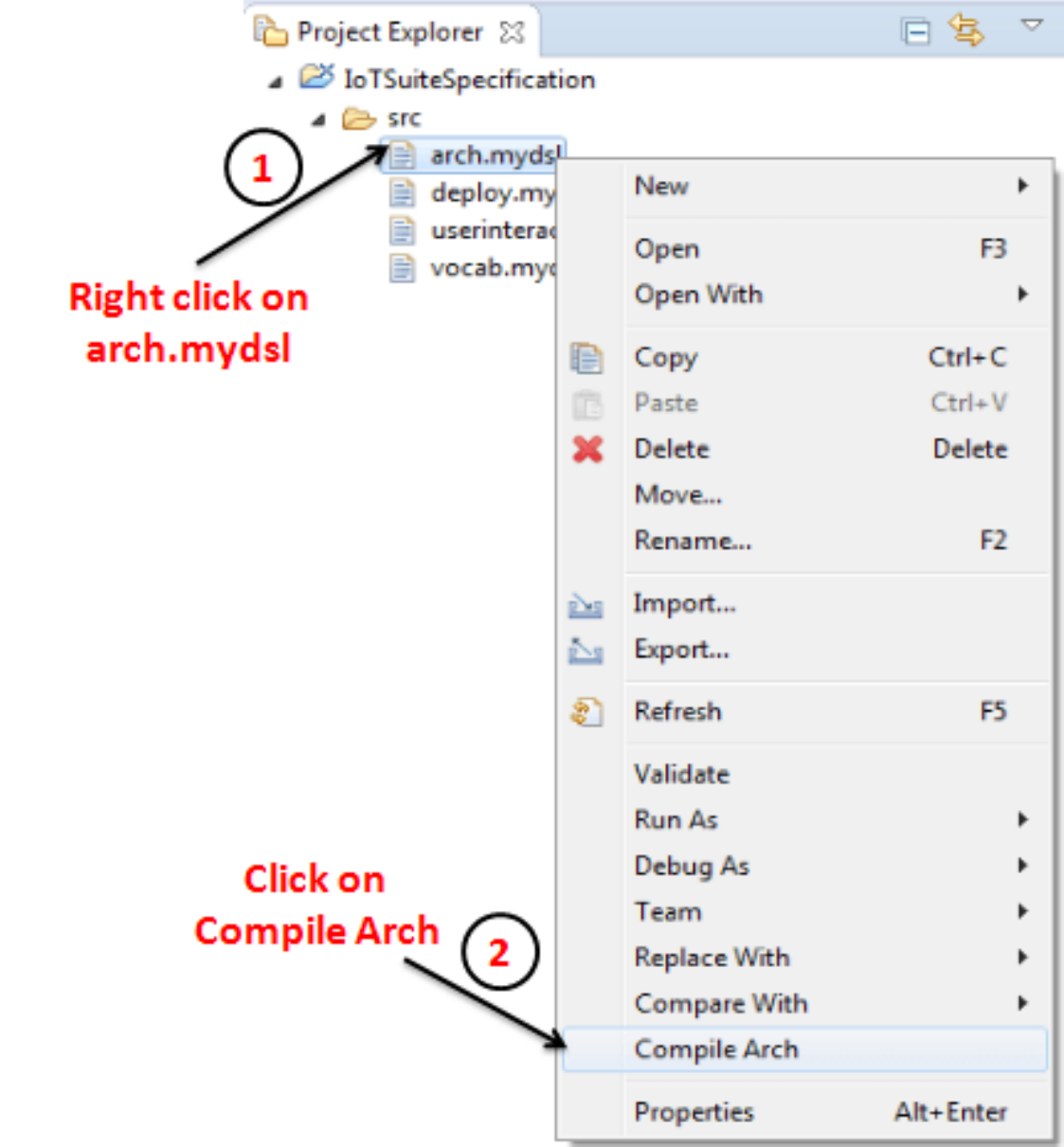}
\caption{Compilation of Architecture specification}
\label{fig:compilearch}
\end{figure}
 
\item \emph{Import application logic package}- To import application logic package, click on File Menu~(Step 1), and  select Import option~(Step 2) as shown in Figure~\ref{fig:importlogic}.
\begin{figure}[!ht]
\centering
\includegraphics[width=0.5\textwidth]{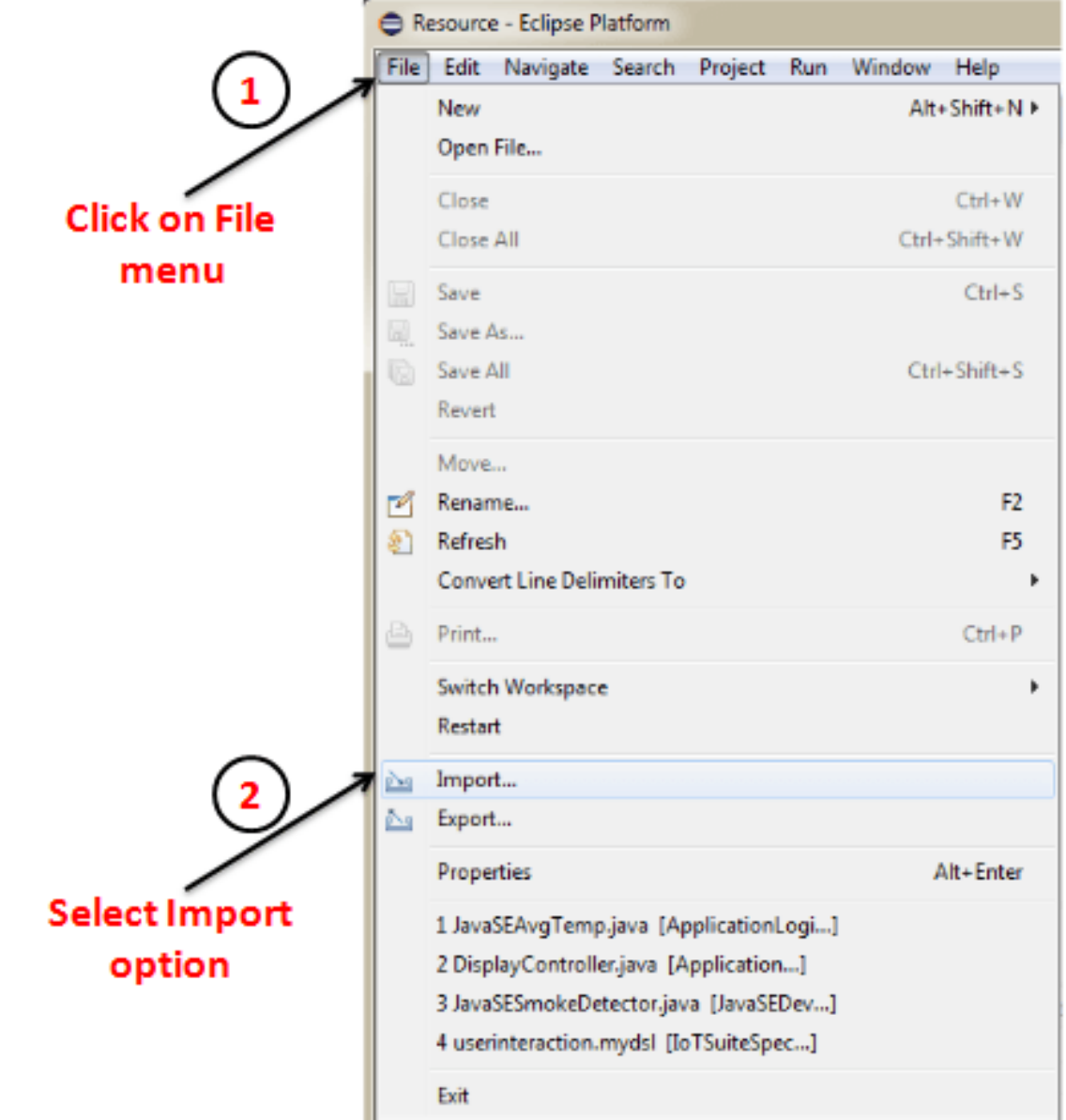}
\caption{Import application logic package }
\label{fig:importlogic}
\end{figure}
\item \emph{Locate application logic package}- To locate application logic package, browse to Template path~(Step 1), select application logic package~(Step 2), and click on Finish button~(Step 3) as shown in Figure~\ref{fig:importlogic2}. 
\begin{figure}[!ht]
\centering
\includegraphics[width=0.68\textwidth]{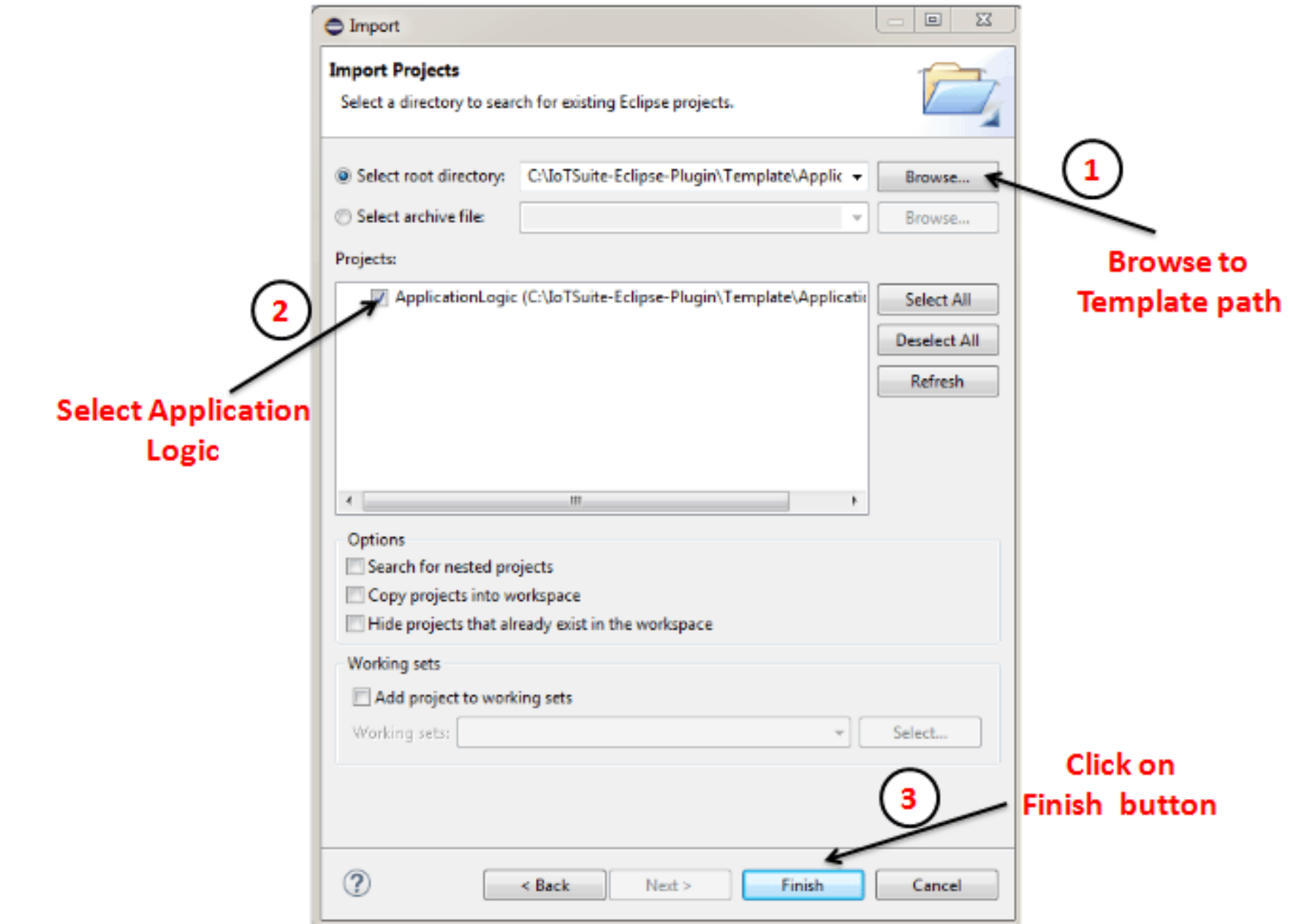}
\caption{Locate application logic package}
\label{fig:importlogic2}
\end{figure}
\item \emph{Implementing application logic}- The application logic project contains a generated framework that hides low-level details  from a developer and allows the developer to focus on the application logic. The developer writes application specific logic in logic package~~\cite[p.~12-13]{Patel201562}. To implement application logic of DisplayController, developers have to implement the generated abstract methods as shown in Figure~\ref{fig:implementlogic}. The DisplayController component receives the data from temperature, humidity Sensor, and yahoo weather service and coordinates with Dashboard component. To implement this logic, three methods have to implemented. In similar ways, developer implements application logic of other components.

\begin{figure}[!ht]
\centering
\includegraphics[width=0.8\textwidth]{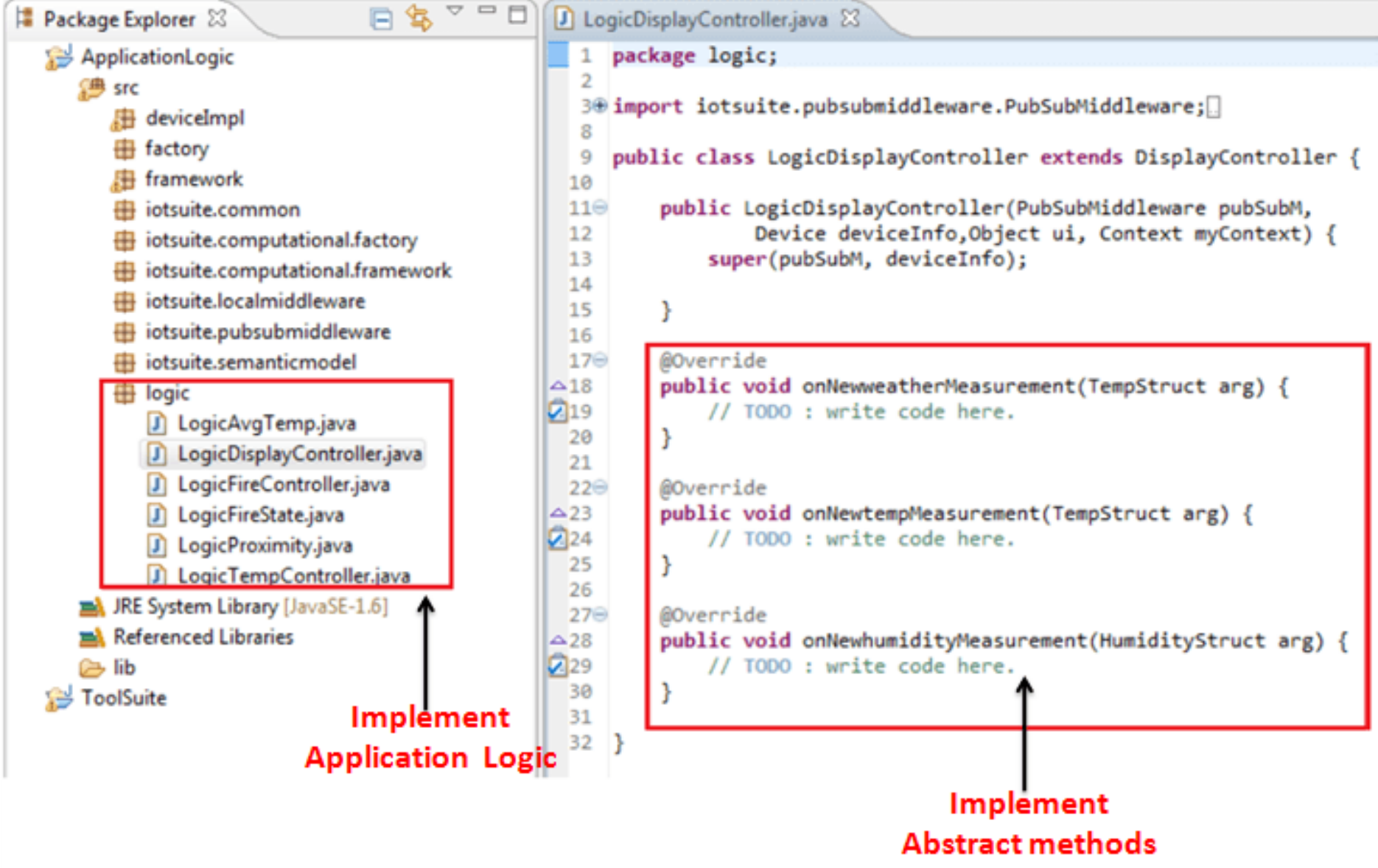}
\caption{Implementation of application logic}
\label{fig:implementlogic}
\end{figure}

\item \emph{Compilation of user-interaction specification}- Right click on userinteraction.mydsl file and selecting "Compile UserInteraction"~(Refer Figure~\ref{fig:compileui}) generates a User Interaction~(UI) framework.
\begin{figure}[!ht]
\centering
\includegraphics[width=0.5\textwidth]{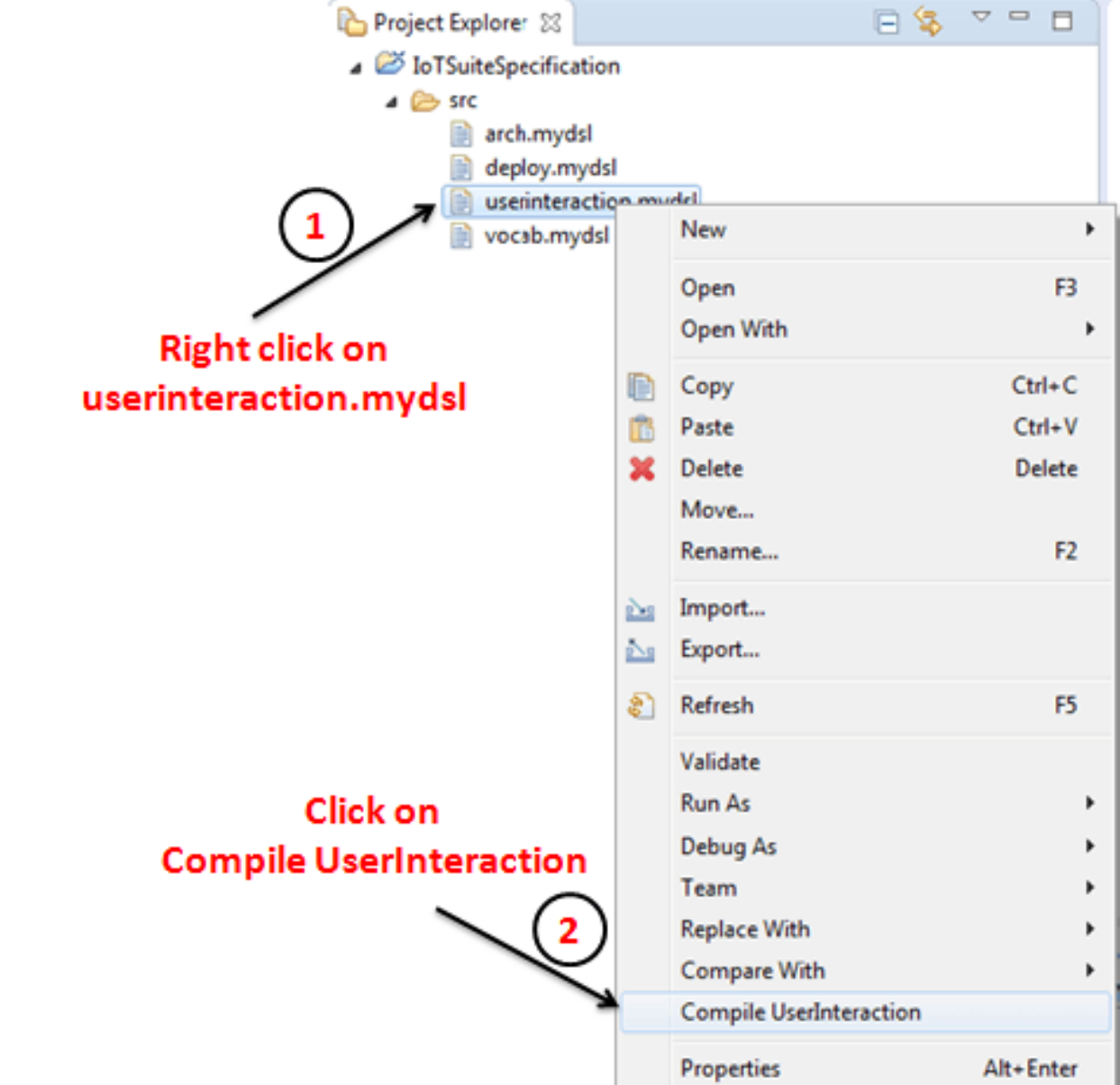}
\caption{Compilation of User-Interaction specification}
\label{fig:compileui}
\end{figure}
\newpage

\item \emph{Compilation of deployment specification}- Right click on deploy.mydsl file and selecting "Compile Deploy"~(Refer Figure \ref{fig:compiledeploy}) generates a deployment packages.
\begin{figure}[!ht]
\centering
\includegraphics[width=0.5\textwidth]{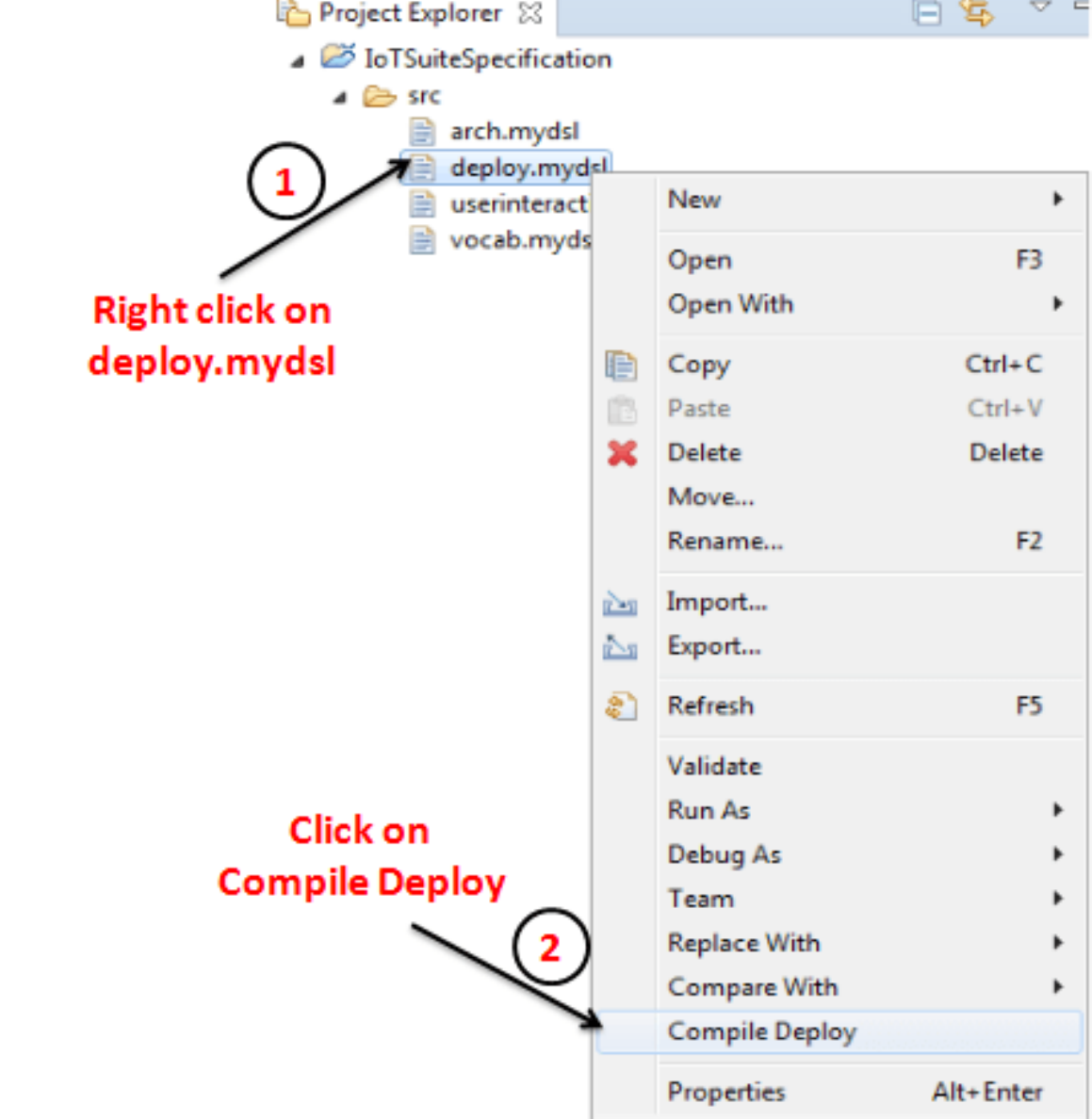}
\caption{Compilation of Deployment specification}
\label{fig:compiledeploy}
\end{figure}

\item \emph{Import user-interface project }- To import user-interface project, click on File Menu~(Step 1), and  select Import option~(Step 2) as shown in Figure~\ref{fig:importlogic1}.
\begin{figure}[!ht]
\centering
\includegraphics[width=0.5\textwidth]{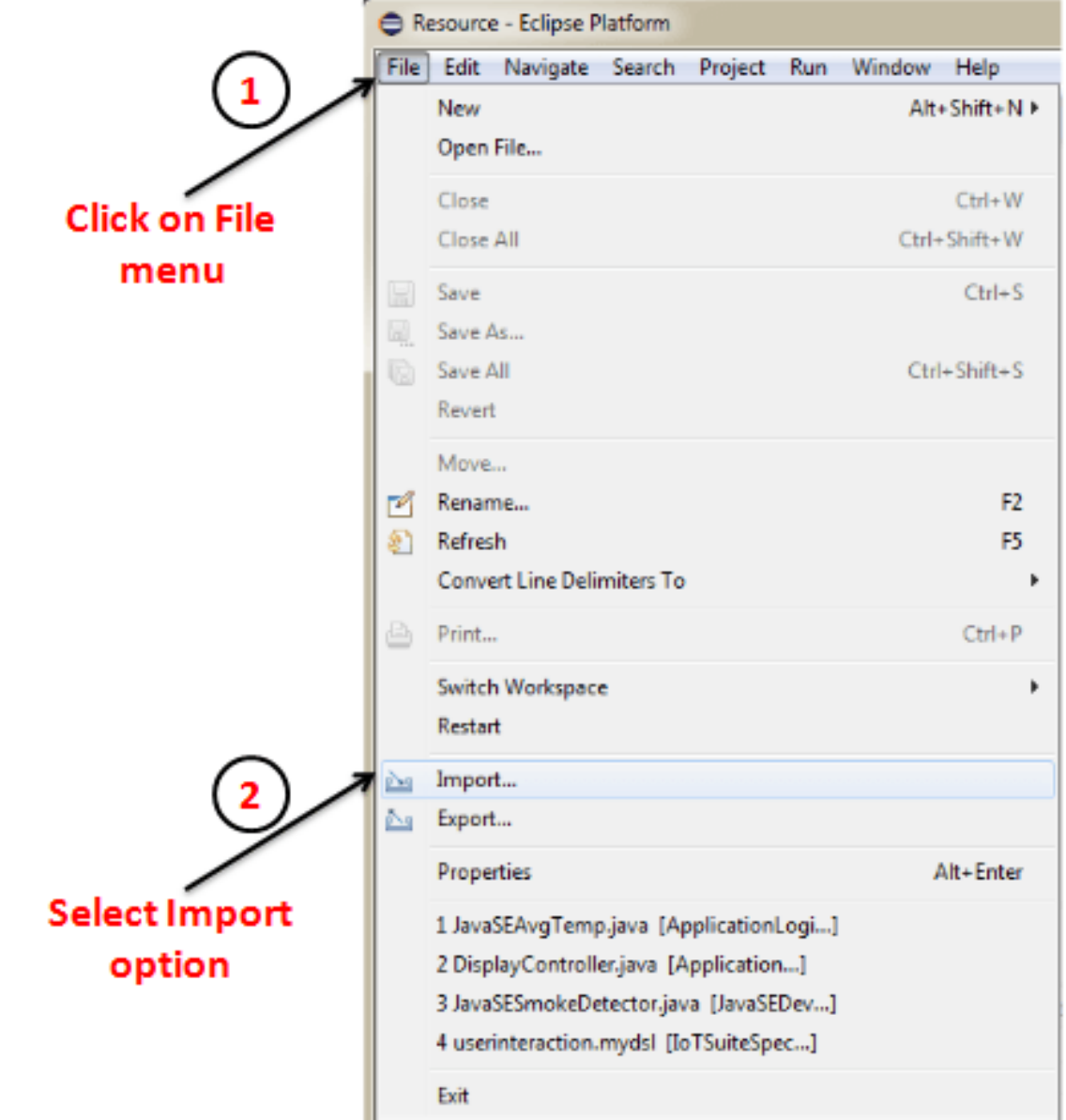}
\caption{Import user-interface project}
\label{fig:importlogic1}
\end{figure}
\newpage

\item \emph{Locate user-interface project}- To locate user-interface project, browse to CodeForDeployment folder in Template path~(Step 1), select project specified in the use-interaction specification~(Step 2), and click on Finish button~(Step 3) as shown in Figure~\ref{fig:importui}. 
\begin{figure}[!ht]
\centering
\includegraphics[width=0.7\textwidth]{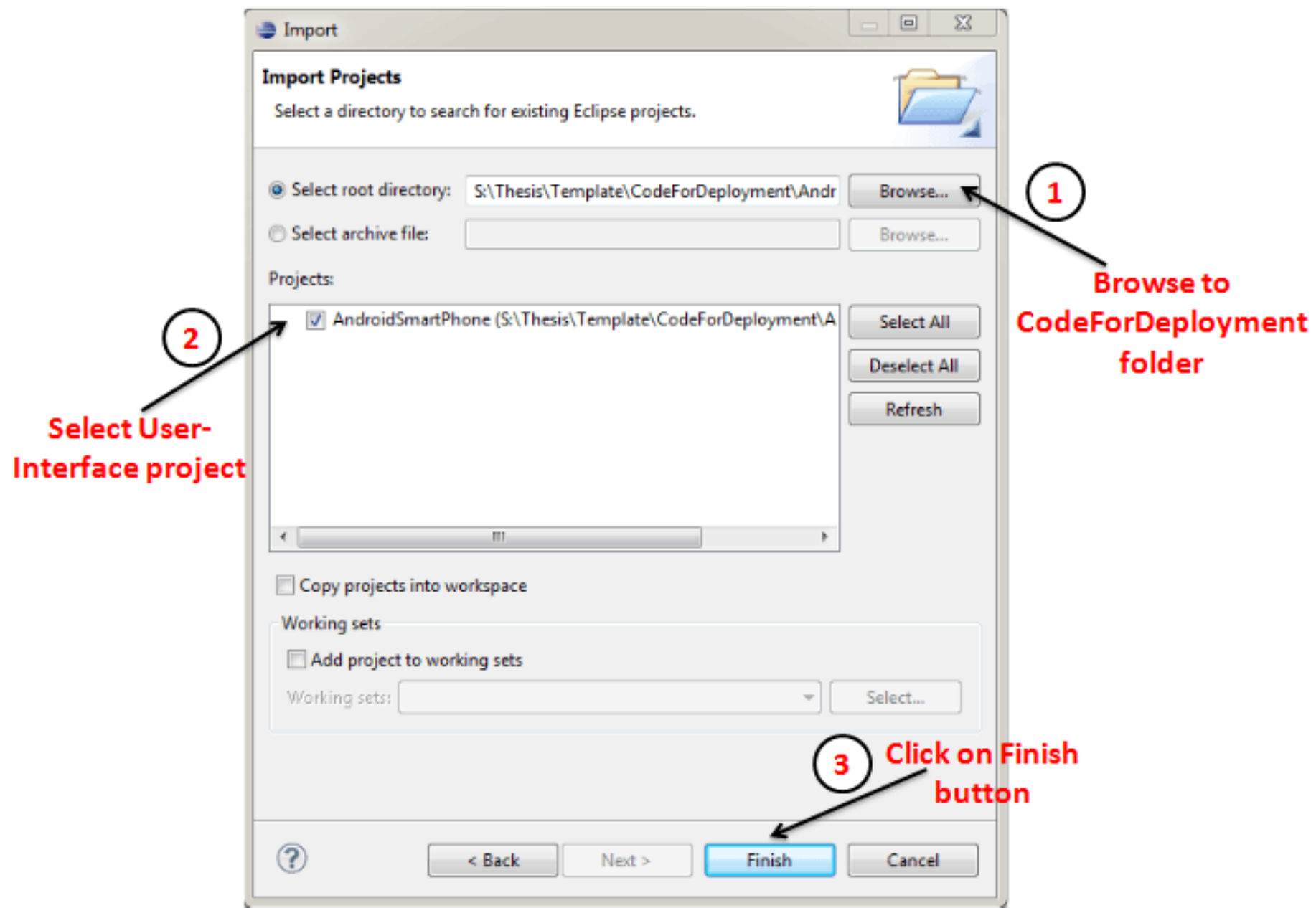}
\caption{Locate user-interface project}
\label{fig:importui}
\end{figure}

\item \emph{Implementing user-interface code}- In this step, developer implements user-interface code generated by compilation of user-interaction specification. Developer implements user-interface code in deviceImpl package~\cite[p.~15-16]{Patel201562}. The implementation of user-interface code involves the use of drag-and-drop functionality provided by form-widget using activity\_main.xml as shown in Figure~\ref{fig:implementui}. The developer connects this interaction with generated framework in the AndroidEndUserApp.java file. 

\begin{figure}[!ht]
\centering
\includegraphics[width=1.0\textwidth]{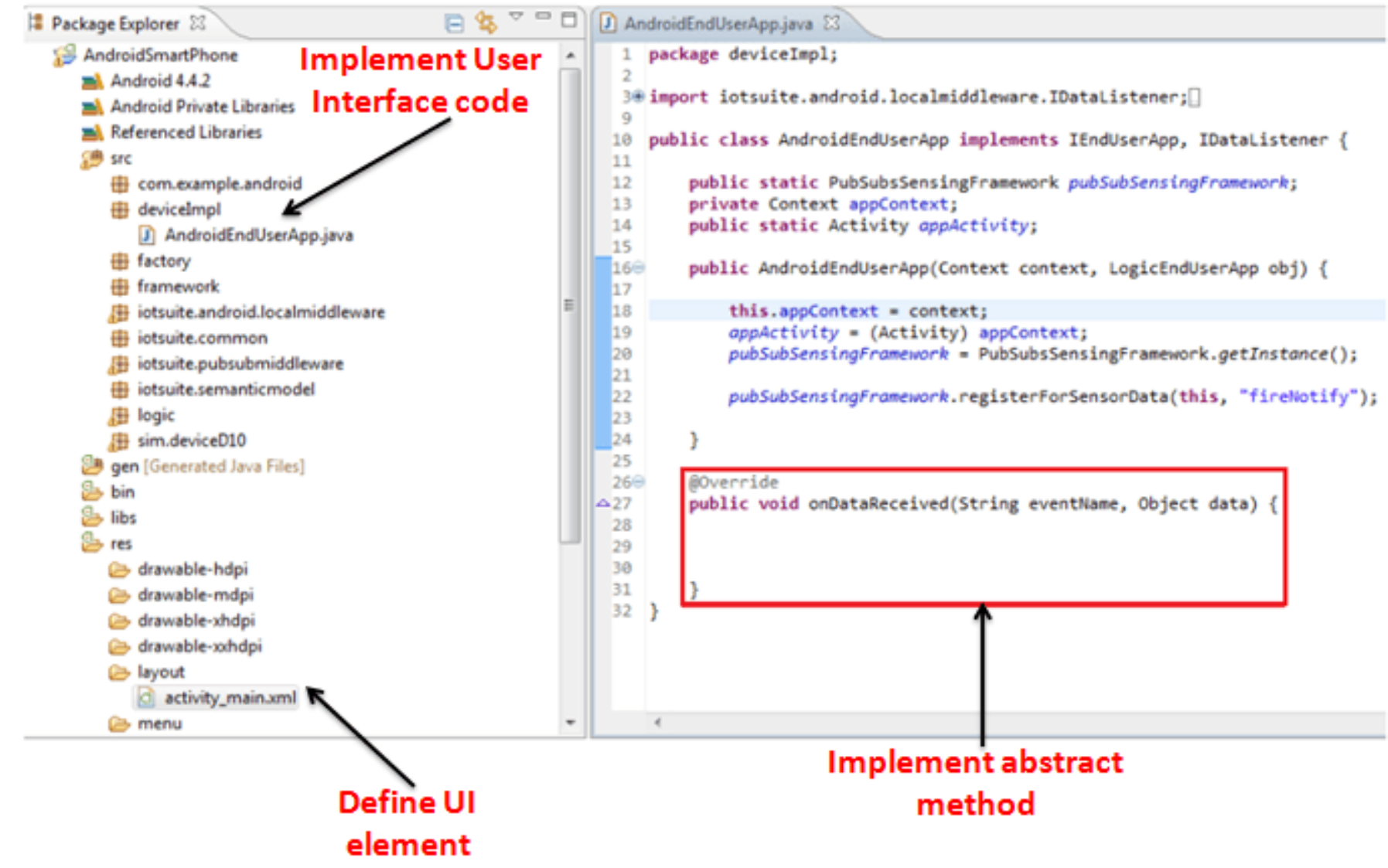}
\caption{Implementation of  user-interface code}
\label{fig:implementui}
\end{figure}
\end{itemize}
\section{Deployment of generated packages} \label{sec:deploy}
\begin{itemize}
\item The output of compilation of deployment specification produce a set of platform specific project/packages as shown in Figure~\ref{fig:deployPackages} for devices, specified in the deployment specification~(Refer Figure~\ref{fig:deploy}). These projects compiled by device-specific compiler designed for the target platform. The  generated packages integrate the run-time system, it enables a distributed execution of IoTSP applications. 
\begin{figure}[]
\centering
\includegraphics[width=0.8\textwidth]{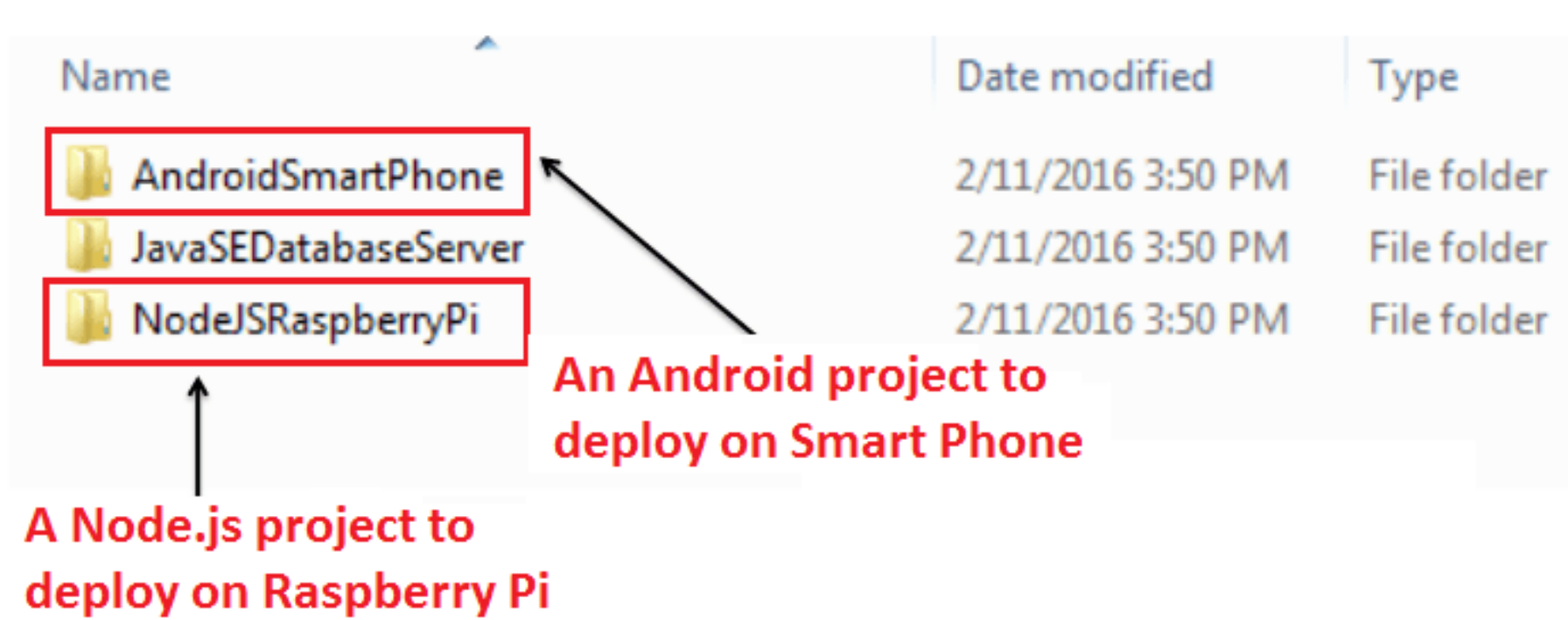}
\caption{Packages for target devices specified in the deployment specification}
\label{fig:deployPackages}
\end{figure}
\end{itemize}

\section{Chapter Summary}
\hspace*{0.2in}This chapter focuses on usability aspects of developer to implement IoTSP application using IoTSuite.  We have exposed IoTSuite as Eclipse plug-in which enables rapid development of IoTSP application using IoTSuite-Eclipse-Plugin. We have also provided support starting from creating an IoTSuite project to generating deployment ready packages and IoTSuite editor to specify high-level specification with features such as code folding, auto completion, rename/refactoring etc. The system is equipped with Eclipse plug-in such that it also provides support to enable single click compilation of high-level specifications.

\chapter{Evaluation} \label{evaluation}
The goal of this chapter is to describe how well the proposed approach eases the IoTSP application
development compared to the existing approaches~(discussed in Section~\ref{literature}). We implemented the 
smart home application~(discussed in Section~\ref{sec:case-study}) using existing 
approaches. 

\section{Lines of Code~(LoC) \& Cyclomatic Complexity} \label{LOCandCyclomatic}
To evaluate the proposed approach, we employed the {\em lines of code~(LoC)} and {\em Cyclomatic Complexity} as  metrics to measure the development effort. We are aware that LoC is not a precise metrics and it depends on programming languages, styles and stakeholders' programming skills. However, it provides an approximate measurement of development effort. The Cyclomatic Complexity is used to measure the structural complexity of code with respect to number of execution paths. It also focus on maintainability aspect of the code, code with more Cyclomatic Complexity is less maintainable~(more complex) as it likely contains more bugs. The value of Cyclomatic Complexity is between 1 to 10, code with lesser value is easy to maintain compare to code with higher value.  We measured LoC and Cyclomatic Complexity (Java code) using Eclipse Metrics 1.3.6 plug-in\footnote{\url{http://metrics.sourceforge.net/}}. We also measured Cyclomatic Complexity of GPL and Node-RED using JS Complexity\footnote{\url{http://jscomplexity.org/}}. To count LoC, we used Eclipse Metrics 1.3.6 plug-in, which counts actual Java statement as LoC and does not consider blank lines or lines with comments. 

\begin{table}[!ht]
\centering 
\scriptsize
\begin{tabular}{|>{\centering\arraybackslash}p{3.30cm} |>{\centering\arraybackslash} p{4.6cm} | >{\centering\arraybackslash} 
p{2.5cm} |>{\centering\arraybackslash}p{2.2cm} |}
 \toprule
\textbf{Entities}&\textbf{Component(model)}&\textbf{Interaction mode}  & \textbf{Runs on} \\ \midrule 
\multirow{3}{*}{Sensor}& Temperature(AM2302)  & Periodic  & RaspberryPi \\
                       & Humidity(AM2302)    & Periodic  & RaspberryPi \\
											 & Smoke(MQ2)     & Event  & Arduino \\ \midrule
\multirow{2}{*}{Actuator}& \parbox{3.1cm}{Heater(using LCD)}   & Cmd  & RaspberryPi \\
                       & \parbox{3.1cm}{Alarm(using buzzer)}  & Cmd  & RaspberryPi \\ \midrule
\multirow{1}{*}{Tag}& BadgeReader(RFID-RC522) & Event & RaspberryPi \\ \midrule	
\multirow{1}{*}{\parbox{1.4cm}{WebService}}& Yahoo Weather    & Req./Resp.  & Yahoo Server \\  \midrule	
\multirow{3}{*}{\parbox{1.4cm}{End-user Interaction}}& EndUserApp  & Notify  & AndroidPhone \\ 
& DashBoard   & Notify  & Desktop \\ \\ \midrule	
\multirow{1}{*}{Storage}& ProfileDB(MySQL) & Req./Resp.  & Microsoft Cloud \\ \midrule	
\multirow{1}{*}{Computation}& Proximity \& others  & Event, Cmd, Req./Resp.  & Desktop \\ \bottomrule
\end{tabular}
\caption{Experiment setup: Smart home application} 
\label{table:experimentsetup} 
\end{table}

We selected the smart home application as a case study for the evaluation, 
implemented it with the following existing approaches, and compared them
with our approach: (1) GPL: We have implemented application-specific functionality using general-purpose programming languages such as \texttt{Node.js}, \texttt{HTML}, \texttt{Android}, and \texttt{JavaScript}. (2) Macro Programming: 
As a representative tool for macro prog., we selected Node-RED. 
It is a widely popular visual tool for wiring together hardware devices, APIs and online services. 
It provides a flow editor where developers can drag-and-drop nodes and configure them 
using dialog editor to specify appropriate properties. To calculate the LoC using Node-RED, we counted each configuration specification as one line of code. To calculate the Cyclomatic Complexity using Node-Red, we exported code for each node and measured it.

We set up a {\em real} execution environment that is made up
of entities, exhibiting heterogeneity discussed in Section~\ref{sec:challenges}. 
Table~\ref{table:experimentsetup} shows entities used to perform the evaluation. All three approaches have been implemented and tested on this setup.
Table~\ref{table:locapp} shows the LoC required to develop the case study using all three
approaches. Using GPL, the stakeholders have to write more than twice number of LoC compared to Node-RED to
implement the same application~(\texttt{57\%} effort  reduction compared to GPL). The primary reason 
of effort reduction is that Node-RED provides high-level constructs that hides low-level details. 

\begin{table}[!ht]
\centering 
\scriptsize
\begin{tabular}{ |>{\centering\arraybackslash} p{2.1cm} |>{\centering\arraybackslash}p{0.8cm}>{\centering\arraybackslash}p{0.8cm}>{\centering\arraybackslash}p{0.8cm} >{\centering\arraybackslash}p{0.8cm}>{\centering\arraybackslash}p{0.8cm}>{\centering\arraybackslash}p{0.8cm}>{\centering\arraybackslash}p{0.8cm}|>{\centering\arraybackslash}p{0.8cm} |}
\toprule
\textbf{Approch.}&\textbf{S}&\textbf{A}&\textbf{T}&\textbf{WS}
&\textbf{EU} & \textbf{ST} & \textbf{Comp.} & \textbf{Total}   \\ \midrule
GPL&51&40&9&19&211 &36& 267 & 633 \\ \midrule
Node-RED&51*&40*&9*&0&39 & 14  &118 & 271 \\ \midrule
MDD& \multicolumn{7}{c|}{40(Vocab. Spec.)+29(Arch. Spec.)+} & 188\\ 
(IoTSuite) &\multicolumn{7}{c|}{43(Deploy. Spec.)+14(User Interaction Spec.)} & \\ 
 &\multicolumn{7}{c|}{+26(App. Logic code)+36(User Interface Code)} & \\ 
\bottomrule
\end{tabular}
\caption{Comparison of existing approaches: Lines of code required to develop smart home application. \textbf{S}~(Sensor), \textbf{A}~(Actuator), \textbf{T}~(Tag), \textbf{WS}~(External Web Service), \textbf{EU}~(End user Application),
\textbf{ST}~(Storage),\textbf{Comp.}~(Computational Services).} 
\label{table:locapp} 
\end{table}

Macro prog. is a viable approach compared to the GPL. It reduces the development effort
by providing cloud-based APIs to implement common functionality. However, one of Node-RED drawbacks is  its {\em node-centric} approach. The stakeholders have to write code to implement 
platform-specific functionality such as reading values from sensors, and actuating actuators. 
For Node-RED,  nodes for sensors and actuators are contributed at public repository 
\footnote{http://flows.nodered.org/}. However, many nodes are not available in 
the library till \today, marked as * in Table~\ref{table:locapp}. This leads to a platform-dependent 
design and increases the development effort. 

Table~\ref{table:locapp} shows that stakeholders can  write the same application using our approach
in \texttt{188} LoC (\texttt{70\%}  effort reduction compared to GPL and \texttt{30\%} effort reduction 
compared to Node-RED). The reason is that  our approach provides the ability 
to specify an application at global-level rather than individual devices. For instance, 
the domain language provides abstractions to specify entities in platform-independent 
manner. The translation of this specification to platform-specific code is taken care by our approach.
So, the stakeholders do not have to write platform-specific code while developing an application.

\begin{table}[!ht]
\centering 
\scriptsize
\begin{tabular}{|>{\centering\arraybackslash} p{2.9cm} |>{\centering\arraybackslash}p{3.9cm} |}
\toprule
\textbf{Approch.}& \textbf{Avg. Cyclomatic Complexity}   \\ \midrule
GPL& 2.7319 \\ \midrule
Node-RED & 1.3846 \\ \midrule
MDD~(IoTSuite)&  1.9432\\ 
\bottomrule
\end{tabular}
\caption{Comparison of existing approaches: avg. Cyclomatic Complexity to develop smart home application.} 
\label{table:complexity} 
\end{table}
Table~\ref{table:complexity} shows the avg. Cyclomatic Complexity to develop the case study using all three
approaches. Cyclomatic Complexity for GPL is more than 2, where in case of Node-RED and MDD it is less than 2. The case study develop using Node-RED and MDD is less complex as compare to GPL, which helps the stakeholders to maintain the code in the efficient way and highlight maintainability aspect.

\section{Automation results using IoTSuite}\label{sec:automation}
The goal of this section is to  evaluate automation provided by IoTSuite. We focused on \textit{scalability} and \textit{re-usability} aspects to evaluate automation results using IoTSuite. For that we have implemented Personalized-HVAC application using IoTSuite. 
\subsection{Re-usability}
\begin{figure}[!ht]
\centering
\includegraphics[width=1.0\textwidth]{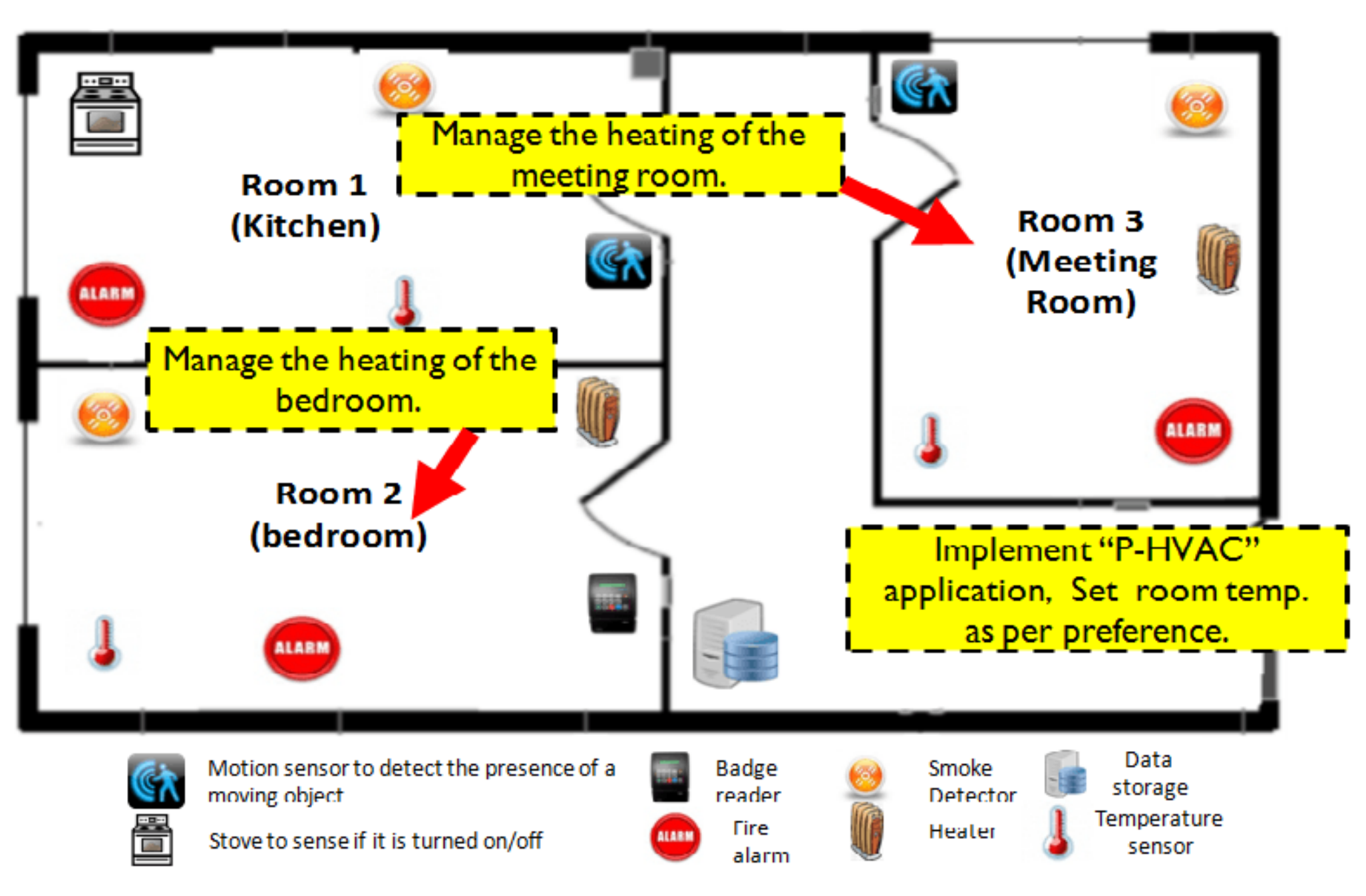}
\caption{A smart home with deployed Personalized-HVAC application in Room 2 and 3.}
\label{fig:phvac}
\end{figure}
To focus on re-usability aspect, we implemented Personalized-HVAC using IoTSuite as per experiment setup presented in Table~\ref{table:phvac} and deployed in bed room~(Room 2) and meeting room~(Room 3) as shown in Figure~\ref{fig:phvac}. 
\begin{table}[!ht]
\centering 
\scriptsize
\begin{tabular}{ |  p{1.6cm} |>{\centering\arraybackslash}p{1.6cm}|>{\centering\arraybackslash}p{1.6cm}|>{\centering\arraybackslash}p{1.6cm}|>{\centering\arraybackslash}p{1.6cm}|>{\centering\arraybackslash}p{1.6cm}|>{\centering\arraybackslash}p{1.6cm} |}
\toprule
\textbf{Entities}&\textbf{Component}&\textbf{Interaction mode}&\textbf{Platform (HVAC in meeting room)
}&\textbf{Deployment (HVAC in meeting room)
}
&\textbf{Platform (HVAC in bed room)
} & \textbf{Deployment (HVAC in bed room)
} \\ \midrule
Sensor&Badge Reader&Event-driven&Node.js&Raspberry Pi-1&Android & Smart Phone \\ \midrule
Storage&ProfileDB&Req/Res&MySQL&MySQL Server&Azure DB & Microsoft Azure Cloud   \\ \midrule
\multirow{ 2}{*}{Computation} &Proximity&Req/Res&\multirow{ 2}{*}{JavaSE}&Desktop-1&\multirow{ 2}{*}{JavaSE} & \multirow{ 2}{*}{Desktop-2} \\ 
& Temp. Controller&Cmd&&Laptop-1& &    \\ \midrule
Actuator&Heater&Cmd&Node.js&Raspberry Pi-2&JavaSE& Laptop-1   \\ \midrule
\end{tabular}
\caption{Experiment setup to implement Personalized-HVAC application in meeting room and bed room} 
\label{table:phvac} 
\end{table}
\begin{figure}[!ht]
\centering
\includegraphics[width=1.0\textwidth]{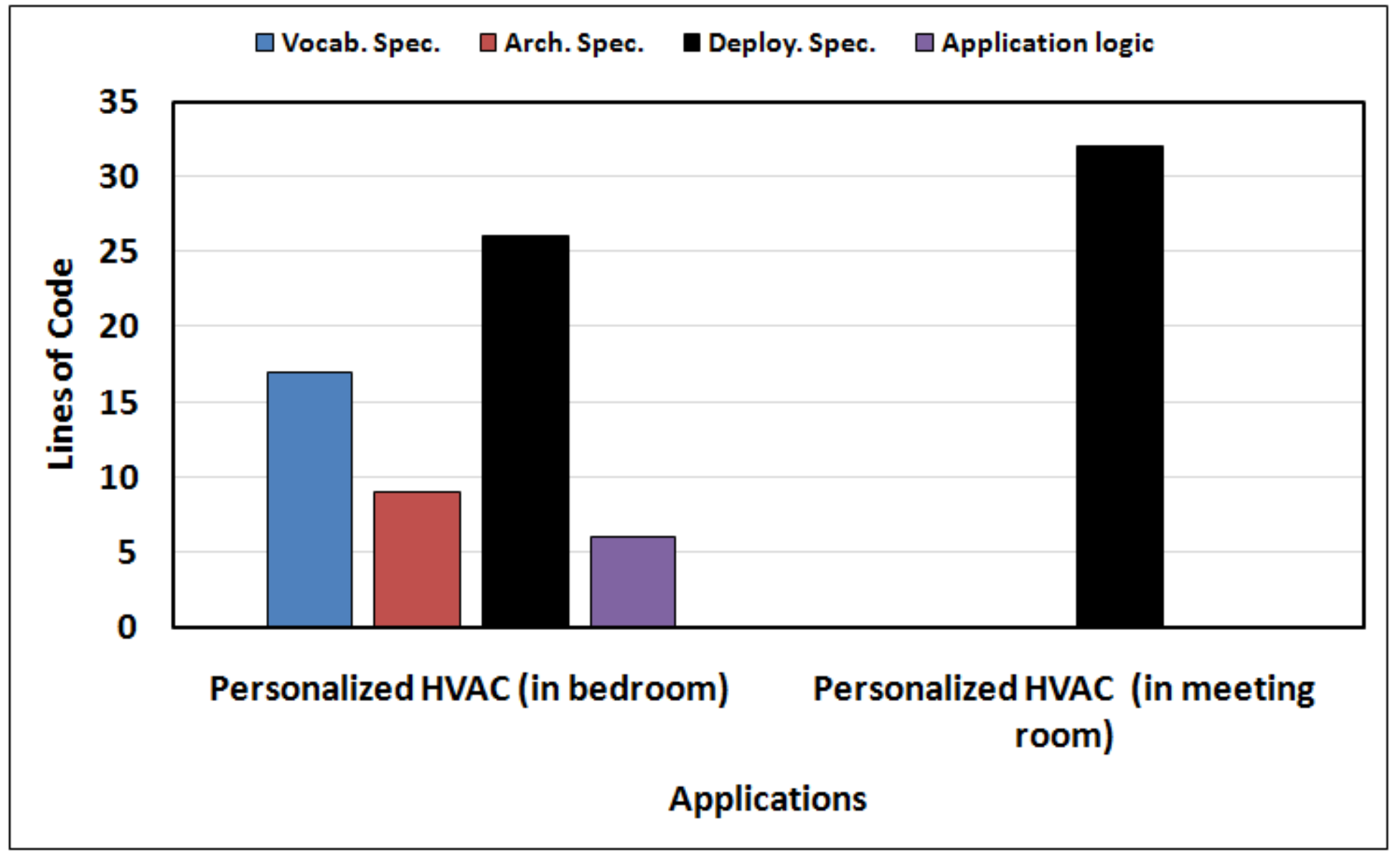}
\caption{Automation result: Re-usability in high-level specification}
\label{fig:reusable}
\end{figure}
If developer wants to deploy Personalized-HVAC application in bed room as per deployment scenario presented in~Table~\ref{table:phvac}, than developer has to write high-level specification~(vocab, arch, and deploy spec.) using IoTSuite followed by implementing application logic as shown in Figure~\ref{fig:reusable}. Now, if need arises to deploy Personalized-HVAC application in meeting room as per deployment scenario presented in~Table~\ref{table:phvac}, in such case developer can re-use high-level specification ~(except deploy spec.) as well as application logic of Personalized-HVAC application that is deployed in bed room. Here only deployment scenario is changed and developer has to implement it. This focus on re-usability aspect of application development using IoTSuite.
\subsection{Scalability}

\begin{figure}[!ht]
\centering
\includegraphics[width=1.0\textwidth]{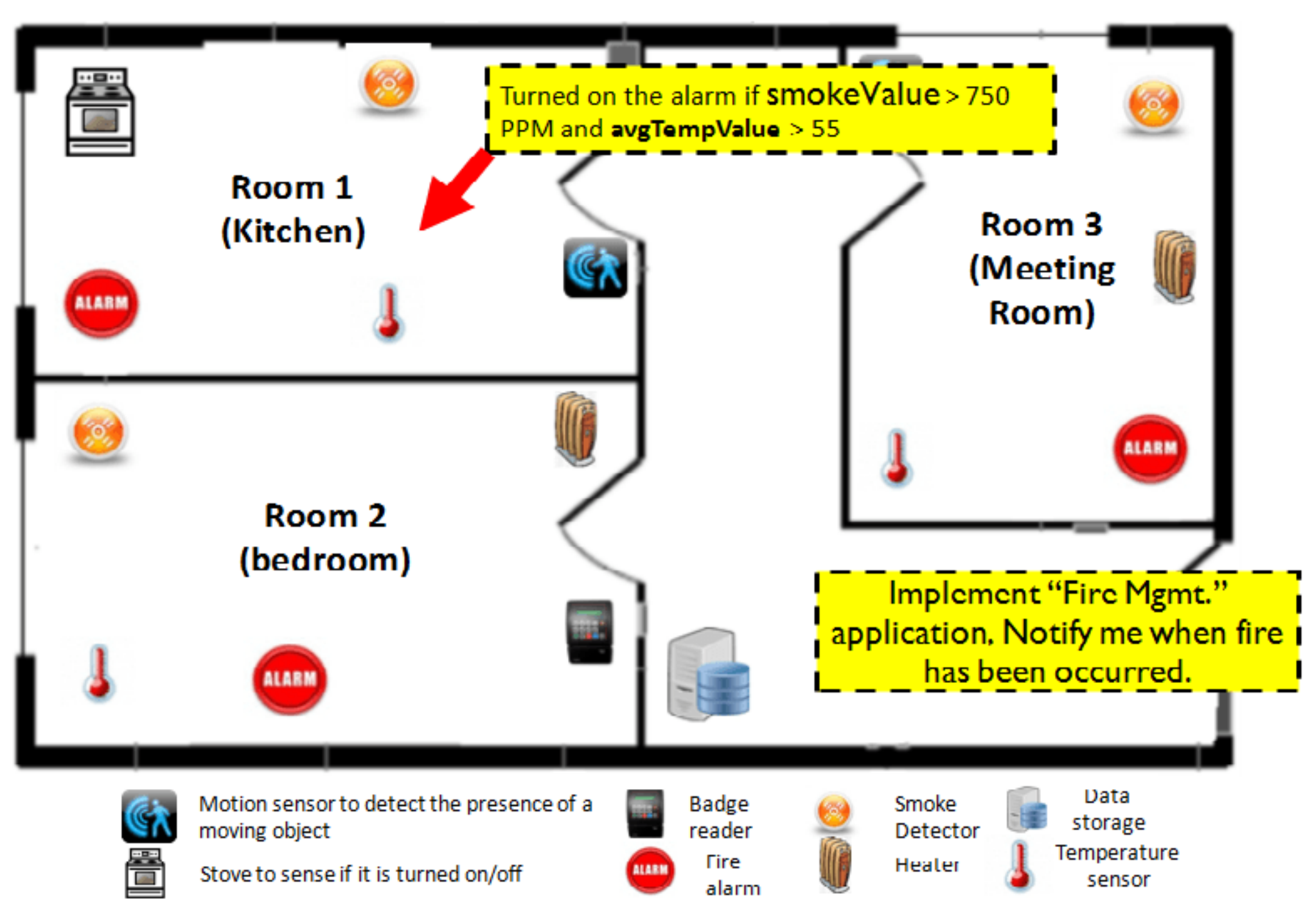}
\caption{A smart home with deployed Fire Management app.}
\label{fig:firemgmt}
\end{figure}
\begin{figure}[!ht]
\centering
\includegraphics[width=0.9\textwidth]{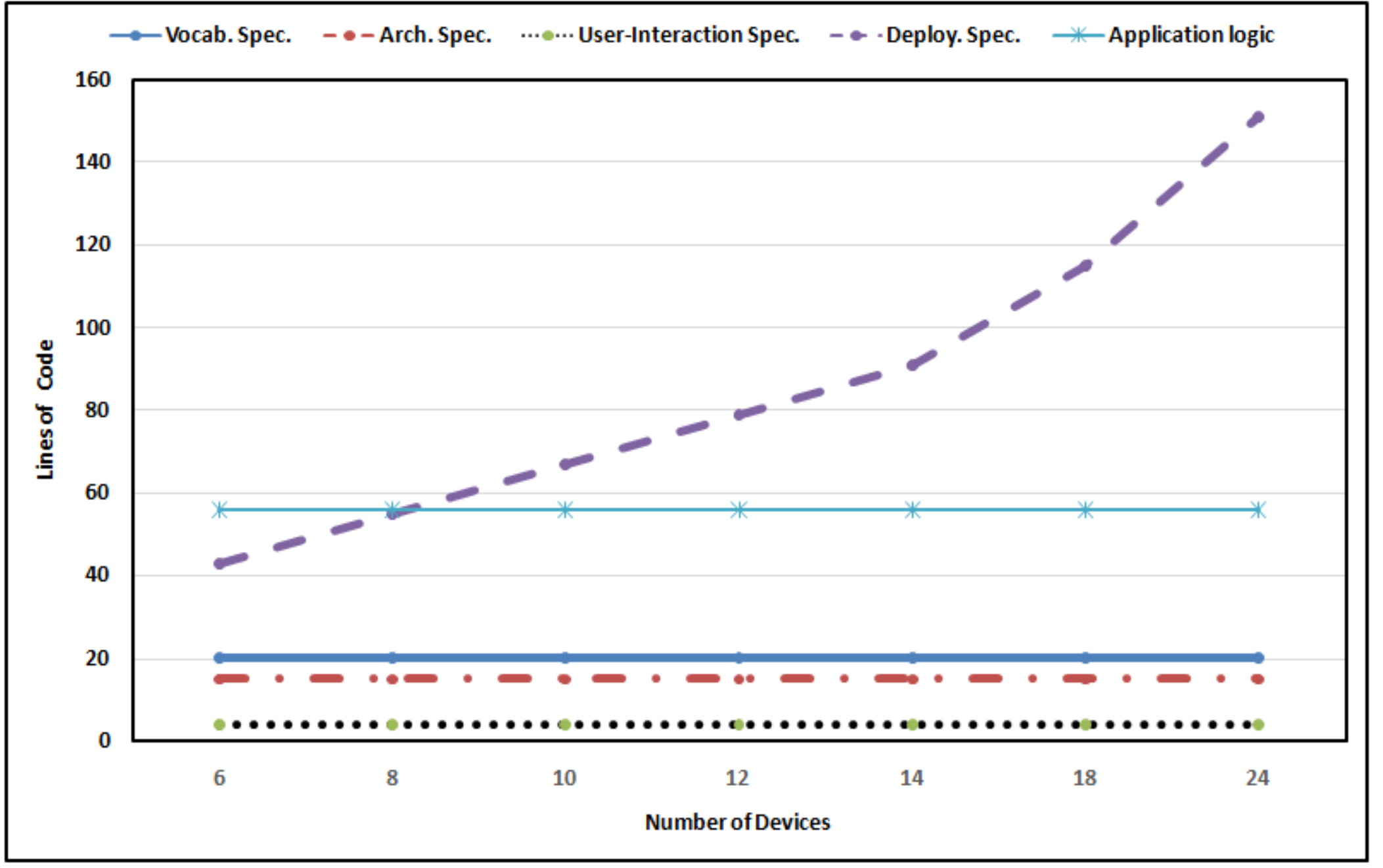}
\caption{Automation result: Scalability in high-level specifications}
\label{fig:scalability}
\end{figure}

To focus on scalability aspect, we implemented Fire Management application using IoTSuite as per experiment setup presented in Table~\ref{table:firemgmt} and deployed in Kitchen (Room 1) as shown in Figure~\ref{fig:firemgmt}. To implement Fire Management application, developer has to write high-level specifications (vocab, arch, user-interaction, and deploy spec.) followed by implementation of application logic. To evaluate scalability aspect of IoTSuite, we deploy Fire Management app. among large number of devices as shown in Figure~\ref{fig:scalability}. First, we implemented application with 6 devices exhibiting heterogeneity and measured development effort in term of LoC. In subsequent experiments, we kept increasing the number of sensors, actuators and end-user devices. We have measured LoC to specify high-level specifications followed implementing application logic during each experiment. We have noticed that with increase in number of devices, LoC remain constant to specify vocab, arch, user-interaction and application logic. The reason is that these high-level specifications are independent of the deployment scenario. As increase number of devices in deploy spec., we have to specify each devices individually in deploy specification. We can consider this as limitation of deployment language~(DL) of IoTSuite. 

\begin{table}[]
\centering 
\scriptsize
\begin{tabular}{ |>{\centering\arraybackslash} p{1.9cm} |>{\centering\arraybackslash}p{2.6cm}|>{\centering\arraybackslash}p{1.9cm}|>{\centering\arraybackslash}p{1.6cm}|>{\centering\arraybackslash}p{1.6cm}|}
\toprule
\textbf{Entities}&\textbf{Component}&\textbf{Interaction mode}&\textbf{Platform}&\textbf{Deployment} \\ \midrule
\multirow{2}{*}{Sensor} &Temperature Sensor (AM2302)& Periodic & Node.js & Raspberry Pi-1 \\ 
 &Smoke Detector (MQ2)& Event-driven & Node.js & Raspberry Pi-2 \\ \midrule
\multirow{2}{*}{Computation} &Room Avg. Temp& Event-driven & JavaSE & Dektop-1 \\ 
  &Fire Controller& Cmd & JavaSE & Dektop-2 \\ \midrule
Actuator&Alarm&Cmd&Node.js&Arduino   \\ \midrule
End User Interaction&End User App&Notify&Android&Android Phone   \\ \midrule
\end{tabular}
\caption{Experiment setup to implement Fire Management app.} 
\label{table:firemgmt} 
\end{table}

\section{Chapter Summary}
\hspace*{0.2in}
 This chapter evaluates proposed approach that reduces development efforts to develop IoTSP application. It also evaluates implementation of smart home application exhibits heterogeneous entities implemented on real devices. We evaluated proposed approach using \textit{Lines of Code~(LoC)} and \textit{Cyclomatic Complexity} to measure development effort. Our experimental results demonstrate that our approach drastically reduces development effort to develop smart home application using existing approaches. This chapter also discusses the automation provided by IoTSuite, in that we focused on re-usability and scalability aspects to evaluate automation. By examining results, we identify limitation of IoTSuite from Deployment Language~(DL) perspective as it should be independent of no. of devices in deployment specification.

\chapter{Conclusion and Future work}\label{sec:conclusionandfuturework}
We have built upon existing framework and evolved it into a framework for developing IoTSP applications, with substantial additions and enhancements such as 
describing heterogeneous entities, third party services, and end-user interactions. We present a comparative evaluation results with existing approaches. This provides the IoTSP community for further benchmarking.  The evaluation is carried out on real devices 
exhibiting IoTSP heterogeneity. Our experimental analysis and results demonstrates that our approach drastically reduces development effort for IoTSP application development compared to existing approaches. \\
Our immediate future work will be:
\begin{itemize}
	\item To evaluate the usability  of this development framework. This will help us to assess the potential for transferring it to industrial environment for IoTSP domain.
	\item Provide drag-and-drop based editor which further reduces development effort as compare to our current editor for writing high-level specification.
	\item Provide scalability in deployment specification as it should be independent of no. of devices. Add support to handle heterogeneous protocols in deployment specification.
\end{itemize}  

\section{Further Readings}

This document will answer this question: -- \emph{how developers can create an IoT application using IoTSuite?}. The video guide is available at URL\footnote{\url{https://www.youtube.com/watch?v=nS_Je7IzPvM}}.

For further reading, we suggest readers to refer our previous publications:

\begin{itemize} 
	
  \item In our IPSN 2016 demo paper\cite{7460669}, we  demonstrate IoTSuite to the research communities. This paper contains high-level architecture, a use-case exhibiting IoT application characteristics.

	\item In our ICSE 2016 paper\cite{Chauhan:2016:DFP:2897035.2897039}, we present comparative evaluation results with existing approaches such as general-purpose programming, macroprogramming and cloud-based platforms. The evaluation is carried out on real devices exhibiting characteristics of IoT applications. Our experimental analysis and results demonstrate that our approach drastically reduces development effort for IoT applications compared to existing approaches. 
	
	\item In our work~\cite{7467275}, we conduct tutorial using IoTSuite at 22nd Asia-Pacific Software Engineering Conference 2015.
	
	\item In our work~\cite{patel-comnet-iot15}, we documented the implementation details of IoTSuite. 
	
	\item In our work~\cite{patel:hal-00788366, DBLP:journals/corr/PatelLB16} we have documented our evaluation results.
	
	\item In our work~\cite{Patel201562, patel-icse14, patelhal00809438}, we revised and refined the IoTSuite.
   This work integrates a set of high-level languages to specify an IoT application. It had provided automation  techniques to parse the specifications written using these high-level languages and generate platform-specific code. The IoTSuite integrates different high-level modeling languages that abstract platform-specific complexity. It is supported by automation techniques such as a code-generator that generates platform-specific code by parsing the specification written using the supported high-level programming languages.

	\item In our Middleware 2011 paper~\cite{appdevIoTpatel2011}, we had started identifying concepts and associations among the concepts and proposed an early domain model for an IoT applications.

\end{itemize}

\bibliographystyle{unsrt}
\bibliography{ref}

\begin{appendices}
  \chapter{High-level specifications for smart home app.}
	\section{Vocabulary specification}
\lstset{emph={requestBasedSensors, sensors, String, generate, actuators, structs, double,  action, resources, long, storages, accessed, by, periodicSensors, sample, period, for, eventDrivenSensors, onCondition, true, tags, =}, emphstyle={\color{blue}\bfseries\emph}, caption={}, escapechar=\#}	
 
\begin{lstlisting}
1 structs:   
2   TempStruct
3    tempValue : double;
4    unitOfMeasurement : String;	
5   HumidityStruct
6  	  humidityValue : double;
7   	  unitOfMeasurement : String;      	  
8   BadgeStruct
9      badgeID: String;
10      badgeEvent: String; 
11  SmokeStruct
12	  smokeValue: double; 
13	  unitOfMeasurement : String;	
14 resources:
15  sensors:   
16    periodicSensors: 
17     TemperatureSensor
18       generate  tempMeasurement : TempStruct; 
19       sample period 1000 for 6000000; 
20	  HumiditySensor
21         generate  humidityMeasurement : HumidityStruct;    
22         sample period 1000 for 6000000;   
23    eventDrivenSensors:  
24     SmokeDetector                
25        generate smokeMeasurement: SmokeStruct;  
26        onCondition smokeValue>650PPM; 
27    requestBasedSensors:
28        YahooWeatherService   
29            generate weatherMeasurement: TempStruct accessed
30					by locationID: String;
31    tags:   
32      BadgeReader
33        generate badgeDetectedStruct: BadgeStruct;  
34    actuators:     
35      Alarm       
36        action On();  
37      Heater
38          action SetTemp(setTemp:TempStruct);	
39    storages:       
40     ProfileDB 
41        generate profile: TempStruct accessed-by badgeID:
42					          String; 
\end{lstlisting}
	\section{Architecture specification}
	
	\lstset{emph={computationalServices, Common, Custom, generate, consume, COMPUTE, request, command, from, to, =}, emphstyle={\color{blue}\bfseries\emph}, caption={}, escapechar=\#}	
 
\begin{lstlisting}
1 computationalServices:
2  Common:
3	AvgTemp
4	  consume tempMeasurement from TemperatureSensor;
5	  COMPUTE (AVG_BY_SAMPLE,5);
6	  generate roomAvgTempMeasurement :TempStruct;
7 Custom:
8	Proximity
9	  consume badgeDetected from BadgeReader;
10	  request profile to ProfileDB;
11	  generate tempPref: TempStruct;
12	TempController
13	  consume roomAvgTempMeasurement from AvgTemp;
14	  consume tempPref from Proximity;
15	  command SetTemp(setTemp) to Heater;
16	FireController
17	  consume roomAvgTempMeasurement from AvgTemp;
18	  consume smokeMeasurement from SmokeDetector;			 	
19	  command FireNotify(fireNotify) to EndUserApp;			 	
20	DisplayController
21	  consume tempMeasurement from TemperatureSensor;
22	  consume humidityMeasurement from HumiditySensor;
23	  consume weatherMeasurement from YahooWeatherService;
24        command DisplaySensorMeasurement(sensorMeasurement)
25					to DashBoard;
\end{lstlisting}
	\section{Use-interaction specification}
		\lstset{emph={structs, String, double, resources, userInteractions, notify, from, =}, emphstyle={\color{blue}\bfseries\emph}, caption={}, escapechar=\#}	
 
\begin{lstlisting}
1 structs:
2  FireStateStruct    
3   fireValue:String;  
4   timeStamp:String;	
5  VisualizeStruct
6   tempValue:double;
7   humidityValue:double;
8   yahooTempValue:double;    
9 resources :
10 userInteractions:   
11  EndUserApp    
12    notify FireNotify(fireNotify:FireStateStruct) 
13					from FireController;  
14  DashBoard   
15   notify DisplaySensorMeasurement(sensorMeasurement:Visu-
16			alizeStruct) from DisplayController; 
\end{lstlisting}
	
	\section{Deployment specification}
		\lstset{emph={devices, location, platform, resources, protocol, database, =}, emphstyle={\color{blue}\bfseries\emph}, caption={}, escapechar=\#}	
 
\begin{lstlisting}
1 devices:
2   D1:
3	 location:
4	    Room:1;
5	 platform:NodeJS;
6	 resources:TemperatureSensor, HumiditySensor;
7	 protocol:mqtt;
8   D2:
9	 location:
10		Room:1;
11	 platform:NodeJS;
12	 resources:BadgeReader;
13	 protocol:mqtt;
14  D3:
15	 location:
16	    Room:1;
17	 platform:NodeJS;
18	 resources:SmokeDetector;
19	 protocol:mqtt;
20  D4:
21	 location:
22		Room:1;
23	 platform:NodeJS;
24	 resources:YahooWeatherService;
25	 protocol:mqtt;								
26  D5:
27	 location:
28	 	Room:1;
29	 platform:JavaSE;
30	 resources:ProfileDB;
31	 protocol:mqtt;
32	 database:MySQL;
33  D6:
34	 location:
35	 	Room:1;
36	 platform:JavaSE;
37	 resources:;
38	 protocol:mqtt;
39  D7:
40	 location:
41	 	Room:1;	
42	 platform:JavaSE;
43	 resources:;
44	 protocol:mqtt;
45  D8:
46	 location:
47	 	Room:1;	
48	 platform:JavaSE;
49	 resources:;
50	 protocol:mqtt;
51  D9:
52	 location:
53	 	Room:1;	
54	 platform:JavaSE;
55	 resources:;
56	 protocol:mqtt;
57  D10:
58	 location:
59	 	Room:1;	
60	 platform:Android;
61	 resources:EndUserApp;
62	 protocol:mqtt;
63  D11:	
64	 location:
65	 	Room:1;
66	 platform:NodeJS;
67	 resources:Heater;
68	 protocol:mqtt; 
69  D12: 
70	 location:
71		Room:1;
72	 platform:NodeJS;
73	 resources:Alarm;
74	 protocol:mqtt;
75  D13:
76	 location:
77	 	Room:1;
78	 platform:NodeJS;
79	 resources:DashBoard;
80	 protocol:mqtt;	

	 
		
	 
	 
	 
	
\end{lstlisting}
  \chapter{Publications}
	\begin{itemize}
	\item Saurabh Chauhan, Pankesh Patel, Flavia Delicato, and Sanjay Chaudhary \textit{"A
Development Framework for Programming Cyber-Physical Systems"}, at
2nd International Workshop on Software Engineering for Smart Cyber-Physical
Systems (SEsCPS), Co-located with 38th International Conference on Software Engineering (ICSE), 2016.
\item Saurabh Chauhan, Pankesh Patel,  Flavia Delicato, and Sanjay
Chaudhary \textit{"Demonstration Abstract: IoTSuite-A Framework to Design, Implement, and Deploy IoT Applications"}, at 15th ACM-IEEE International Conference on Information Processing in
Sensor Networks (IPSN), 2016.
\end{itemize} 
\end{appendices}
\end{document}